\documentclass[aps,preprint,showkeys,
superscriptaddress,
preprint,
amsmath,amssymb,
aps,
]{revtex4-1}

\usepackage[latin1]{inputenc}
\usepackage{graphicx}
\usepackage{dcolumn}
\usepackage{bm}
\usepackage{color}

\usepackage{amsmath}
\usepackage{amssymb}
\usepackage{makeidx}
\usepackage{subfigure}
\usepackage{indentfirst}

\begin{document}

\date{\today}


\title{Partial control of chaos: how to avoid undesirable behaviors with small controls in presence of noise}

\author{Rub\'en Cape\'ans}
\email{ruben.capeans@urjc.es}
\affiliation{Departamento de F\'isica, Universidad Rey Juan Carlos, Tulip\'an s/n, 28933 M\'ostoles, Madrid, Spain}

\author{Juan Sabuco}
\email{juan.sabuco@urjc.es}
\affiliation{Departamento de F\'isica, Universidad Rey Juan Carlos, Tulip\'an s/n, 28933 M\'ostoles, Madrid, Spain}
\affiliation{Institute for New Economic Thinking at the Oxford Martin School, Mathematical Institute, University of Oxford, Walton Well Road, Eagle House OX2 6ED, Oxford, UK.}

\author{Miguel A. F. Sanju\'an}
\email{miguel.sanjuan@urjc.es}
\affiliation{Departamento de F\'isica, Universidad Rey Juan Carlos, Tulip\'an s/n, 28933 M\'ostoles, Madrid, Spain}
\affiliation{Department of Applied Informatics, Kaunas University of Technology, Studentu 50-415, Kaunas LT-51368, Lithuania}
\affiliation{Institute for Physical Science and Technology, University of Maryland, College Park, Maryland 20742, USA}

\begin{abstract}

The presence of a nonattractive chaotic set, also called chaotic saddle, in phase space implies the appearance of a finite time kind of chaos that is known as transient chaos. For a given dynamical system in a certain region of phase space with transient chaos, trajectories eventually abandon the chaotic region escaping to an external attractor, if no external intervention is done on the system. In some situations, this attractor may involve an undesirable behavior, so the application of a control in the system is necessary to avoid it. Both, the nonattractive nature of transient chaos and eventually the presence of noise may hinder this task. Recently, a new method to control chaos called \emph{partial control} has been developed. The method is based on the existence of a set, called the safe set, that allows to sustain transient chaos by only using a small amount of control. The surprising result is that the trajectories can be controlled by using an amount of control smaller than the amount of noise affecting it. We present here a broad survey of results of this control method applied to a wide variety of dynamical systems. We also review here all the variations of the partial control method that have been developed so far. In all the cases various systems of different dimensionality are treated in order to see the potential of this method. We believe that this method is a step forward in controlling chaos in presence of disturbances.

\end{abstract}

\maketitle

\section{Introduction}

One of the most challenging phenomena found in nonlinear dynamics is the presence of chaos. The discover of this physical phenomenon involved a completely revolution in physics and many other fields, since it was shown that even a deterministic system can behave unpredictably. Due to this new kind of dynamical behaviour, many efforts were directed to study the possibility of controlling chaos.

Over the last 20 years, a lot of work has been devoted to the study of the control
of chaotic systems. In the literature, there are two basic approaches: feedback
control and non-feedback control. In the feedback methods chaos is stabilized on
a desired unstable periodic orbit embedded in the chaotic attractor by applying
small temporal perturbations to an accessible parameter. The perturbation required
is computed at every instance and is proportional to the difference between the
actual state and the desired state. For the non-feedback control, the idea is that a
small periodic parametric perturbation can suppress chaos, being this perturbation
permanent.

The partial control method presented here is a feedback control
method, like the well known OGY control method \cite{OGY}. However the goal of the partial control is completely different. The OGY method was presented in 1990 and it was designed to avoid the chaotic behavior through applying small perturbations to stabilize a desired periodic orbit. Nevertheless,
recently it had been stressed the importance of chaos in some practical systems. In mechanics for example, chaos helps to prevent undesirable resonances  \cite{Oscillator}. In engineering, the thermal pulse combustor is more efficient in the chaotic regime \cite{Thermal}. In living organisms, chaotic dynamics in vital functions can make the difference between health and disease \cite{Perc}. In biology, it has been suggested that the disappearance of chaos may be the signal of pathological behavior \cite{Biological}. In all these cases, chaos is a desirable property that is worth preserving.

\begin{figure}
\includegraphics [trim=0cm 0cm 0cm 0cm, clip=true,width=0.7\textwidth]{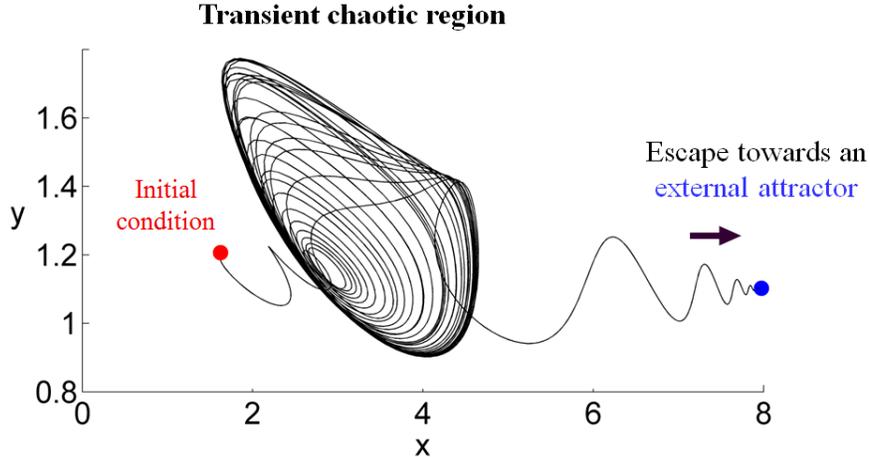}\\
\centering
\caption{\textbf{Transient chaos.} The figure illustrates the chaotic transient behavior of a trajectory. Starting in the red point, the trajectory falls in the chaotic region where it remains for a while.
After a finite amount of time, the trajectory escapes from the chaotic region towards an external
attractor (blue fixed point). The transient chaotic behavior is produced by the presence of a chaotic
saddle in phase space. This invariant set is a non-attractive chaotic set and this is the reason why
the trajectory eventually escapes. The goal of applying control is to sustain the chaotic behavior
forever, avoiding the escape of the trajectories.}
\label{tc}
\end{figure}

Sometimes the chaotic behavior is only transient in nature, and it is necessary to apply external perturbations to keep trajectories in the transient chaotic regime. Transient chaos is a characteristic dynamical behavior that occurs in a certain region of phase space, where chaotic orbits exist for a while, before escaping (after a crisis) to an external attractor (Fig.~\ref{tc}). This kind of behavior can be found in a broad variety of systems like the periodically driven $CO_2$ laser \cite{Laser}, voltage collapse in electrical power systems \cite{Dhamala}, or the Mcann-Yodzis ecological model \cite{McCann}, among many others. In many cases these escapes have catastrophic consequences for the system, for example in the thermal combustor example \cite{Thermal}  the crises involves the flame blowout making the device useless. Another undesirable behaviour occurs in the Mcann-Yodzis ecological model where the crisis conduct irreversibly to the extinction of one of the species.

The partial control method was proposed with the aim of sustaining the transient chaotic behaviour indefinitely and in consequence avoid undesirable escapes. With a similar objective, different control methods have been proposed in the literature \cite{Schwartz,Dhamala,Bertsekas,Bertsekasdos}. However, these methods differ from the partial control method in that they have been mainly designed to be applied in deterministic systems, while partial control is a robust method able to deal with random disturbances present in all realistic systems. The more remarkable feature of the partial control is the ability of keeping a control smaller than the disturbances. More precisely, this method ensures that the amount of control used will be smaller than the amount of disturbances affecting the system, which is a counterintuitive and surprising result. This is possible due to the presence of the chaotic saddle in the phase space which is responsible of the transient chaos. The partial control method benefits from the fractal structure of the chaotic saddle, to reduce the impact of the disturbances and at the same time, to enhance the effect of the control applied.

In the next sections, we will explain how the partial control method is defined and the algorithm to apply it. We will also show the different variations of the method created to deal with the several situations found in practice. In this sense the main algorithm was designed to apply the control on some variable of the system \cite{Automatic,Asymptotic}, however it is also possible, through minor modifications, to control some parameter of the system after minor modifications in the algorithm \cite{Parametric}. We will also show the application of the method in delay-coordinates maps \cite{Delayed}. Such maps are specially relevant in systems with memory or dynamical models reconstructed through the delay embedding technique. Finally, it will be introduced some general outlines to implement the method in experimental data that exhibits transient chaos.

\section{A general description of the partial control method}

Transient chaos is caused by the presence of a chaotic saddle in phase space. In contrast to a chaotic attractor that possesses a fractal structure only in the stable direction, the chaotic saddle is a nonattrative invariant set that is fractal in both, the stable and unstable directions. Due to the fractal structure in the unstable direction, infinite holes arise along the unstable manifold of the chaotic saddle. A trajectory that it is initially attracted along the stable direction for some finite amount of time, eventually escapes through one of the gaps present in the unstable direction.  The main mechanism to create a chaotic saddle is  when a chaotic attractor collides with the boundary of its own basin of attraction, causing a \textbf{boundary crisis}, and allowing the trajectories to reach other regions of the phase space, (see Fig.~\ref{tc}). In many occasions, this escape involves a highly undesirable state and therefore the application of some control scheme is required to prevent it.

\begin{figure}
\includegraphics [trim=0cm 0cm 0cm 0cm, clip=true,width=0.55\textwidth]{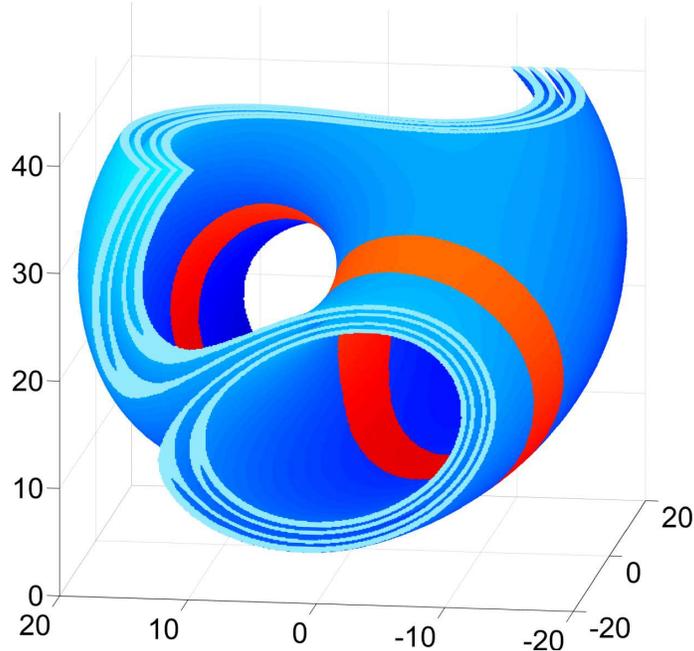}\\
\centering
\caption{\textbf{Example of the set needed to partially control the Lorenz system.} The figure shows an example of a safe set in the phase space computed for the partially controlled Lorenz system in the transient chaotic regime. The blue set represents the points of the phase space that satisfy the control condition defined by the partial control method. The red set is a subset of the blue set, and represents the asymptotic region where the controlled dynamics converges.}
\label{1z}
\end{figure}

The partial control method is a recently developed control strategy for preventing escapes associated with a transient chaotic behaviour.
It is particularly appropriate when it is desirable to keep the magnitude of the control small.  One of the main advantage of this method is the consideration of the random disturbance affecting the trajectories. In many experimental systems, the presence of disturbances may be unavoidable and must be considered, especially when it is necessary to keep the control as small as possible. Methods that perform well in systems in absence of disturbances can fail dramatically when disturbances appear. For this reason, it is reasonable to consider a term, that we call \textbf{disturbance}, that encloses all the uncertainty affecting the dynamics of the system, like modeling mismatches, finite precision in the measure of initial conditions or even systematic or random external disturbances. In most cases the amplitude of this disturbances can be limited, so we consider bounded disturbances. In addition, we also considered that the control available is also limited as in most real scenarios.

The intrinsic instability of the chaotic saddle together with the action of noise creates a difficult scenario where keeping the control small might seem impossible. However it is possible to keep the trajectories close to the chaotic saddle taking advantage of the horseshoe map presence in the phase space that produces the chaotic saddle. This geometrical action implies the existence of certain sets called safe sets that lie in the vicinity of the chaotic saddle. These sets are used to keep the trajectories controlled (close to the chaotic saddle) even when the control applied is smaller than the disturbance. Indeed, the control idea based on controlled set invariance, is pretty standard (see Refs.~\cite{Invariant,Dhamala,Hutson,Genesio,Kolmanovski}). It has also been shown that the shape of the invariant sets play an important role in the dynamics of the controlled system \cite{Nagumo,Invariant}. This situation is even more dramatic in the case of partial control, where the invariant set can be rather complex and it is only possible to find it using a numerical algorithm \cite{Automatic}. In Fig.~\ref{1z} an example of a safe set computed for the Lorenz system is shown.

We consider here dynamical systems of the form $q_{n+1}=f(q_n)$, where $q_{n} \in \mathbb{R}^n$. We assume that the map $f$ acts on a region $Q$ like a horseshoe map \cite{Asymptotic}. This implies that nearly all trajectories inside $Q$ (except a zero measure set) escape of it after some iterations. If we consider in addition the effect of disturbances, all trajectories eventually escape if no control is applied.

Thus, if we add the disturbances $\xi_n$ followed by a feedback control $u_n$, our model becomes $q_{n+1}=f(q_n)+\xi_n+u_n$, for $n=1,2,3..$
We assume in the following that $|\xi_n|\leq\xi_0$ , $|u_n|\leq u_0$ and $\xi_0>u_0>0$.

We define the safe set $Q_\infty \subset Q$ as the set of points $q$, so that the trajectory $q_{n+1}=f(q_n)+\xi_n+u_n$ stay in $Q_\infty$ forever.  The control $u_n$ is chosen with the knowledge of $f(q_n)+\xi_n$, and applied to place the trajectory again in the set $Q_\infty$. We say that trajectories found under these conditions are \textbf{admissible trajectories}.

\begin{figure}
\includegraphics [trim=3cm 0cm 0cm 0cm, clip=true,width=0.73\textwidth]{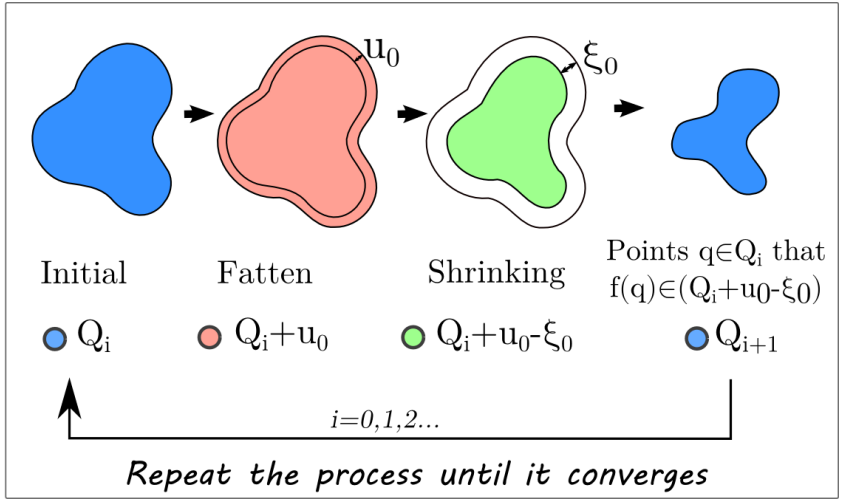}\\
\centering
\caption{\textbf{Graphical process used by the Sculpting Algorithm to obtain the safe set.} The denoted set $Q_i$ is fattened by the thickness $u_0$. The fattened set is displayed in red. Then, the new set is shrunk or contracted by a distance $\xi_0$, obtaining the set denoted $Q_i+u_0-\xi_0$ (in green). Finally we remove the grid points $q \in Q_i$ whose image $f(q)$ falls outside $Q_i+u_0-\xi_0$. Notice that $Q_{i+1} \subset Q_i$.}
\label{2z}
\end{figure}

One of the advantages of this method is that the set $Q_\infty$ can be determined computationally following an iterative process. The set $Q$ is represented by a grid stored in a computer. Beginning with the region $Q_0=Q$, in the first iteration we remove the grid points $q \in Q_0$ for which there are $\xi$ with $|\xi| \leq \xi_0$ such that $f(q)+\xi$ cannot be moved back inside $Q_0$ using a $u$ for which $|u| \leq u_0$. As a result of this first pruning, a new region $Q_1\subset Q_0$ is obtained. Applying the same process to $Q_1$, we obtain a smaller set $Q_2 \subset Q_1 \subset Q_0$. Repeating this process until it converges, the final set denoted $Q_\infty$ is found. This set is known as the \textbf{safe set}. Based on this idea, we developed an algorithm called the \textbf{Sculpting Algorithm} \cite{Automatic}, for computing the successive regions $Q_n$ until the safe set is finally found. We illustrate the procedure of finding the safe set in Fig.~\ref{2z}. We are given the bound $u_0$ and $\xi_0$ and the region $Q_0=Q$. The $i^{th}$ step can be summarized as follows:

\begin{enumerate}
  \item  Morphological dilation of the set $Q_i$ by $u_0$, obtaining the set denoted by $Q_i+u_0$.
  \item  Morphological erosion of set $Q_i+u_0$ by $\xi_0$, obtaining the set denoted by $Q_i+u_0-\xi_0$.
  \item  Let $Q_{i+1}$ be the points $q$ of $Q_i$, for which $f(q)$ is inside the set denoted $Q_i+u_0-\xi_0$.
  \item  Return to step $1$, unless $Q_{i+1}=Q_i$, in which case we set $Q_\infty=Q_i$. We call this final region, the \textit{safe set}. Note that if the chosen $u_0$ is too small, then $Q_\infty$ may be the empty set, so that a bigger value of $u_0$ is required to control the trajectories.
\end{enumerate}

Some practical considerations have to be done. In order to compute  the safe set $Q_\infty$, a finite grid covering $Q_0$ has to be used, since it is not possible to compute the infinite number of points in $Q_0$. We will call the grid resolution as the distance between two adjacent points $q$.  Higher resolutions give a more accurate safe set. In this sense, we have found that beyond a critical resolution of the grid of $Q$ and $\xi$, the safe set remains practically unchanged. Due to the complex shape of the chaotic saddle underlying the chaotic dynamics, the derivation of a rigorous proof of the convergence of the algorithm would be extremely difficult. However, from a computational view, it is easy to demonstrate that the algorithm converges in a finite number of steps, since the grid used is composed of a finite amount of points.  From a practical point of view, we recommend to compute the safe set with the algorithm proposed with increasing resolutions until finding the critical value for which the shape of the safe set found remains unchanged. That one will be a very good approximation of the real safe set.

Due to the The safe set obtained using the algorithm just described, is a positively invariant set \cite{Invariant}, since  all the points belonging to the safe set fall again in the safe set under the controlled dynamics $q_{n+1}=f(q_n)+\xi_n+u_n$. That is, by the safe set definition, if the controlled system's state is at some time inside the safe set, then it will also be contained again in this set in the future.

\section{The partial control method applied on variables}

  The first attempt to show the application of the partial control scheme was carried out in dynamical systems where some variables of the system are accessible for the controller. We present here the most relevant systems studied so far.

 \subsubsection{H\'enon map}

  We consider the H\'enon map with a choice of parameters close to the boundary crisis, which occurs for
\begin{equation}\label{Henon_map}
\begin{split}
\begin{split}
&x_{n+1}=2.16-0.3y_{n}-x^{2}_{n}\\
&y_{n+1}=x_{n}.
\end{split}
\end{split}
\end{equation}

For this choice of parameters, we observe that almost all of the initial conditions escape from the square $Q = [-5,5] \times [-5,5]$ after a finite number of iterations. The presence of a disturbance in
the system typically complicates the survival probability of the orbits inside the
square, since a small disturbance can drive the orbit outside the square. If this
happens, the orbit would go into the infinity very fast.

To apply the algorithm to the H\'enon map, we have chosen this $Q$ as the region of the phase space from which we want to avoid the escapes. This square completely covers the chaotic saddle formed in the parametric region which is close to the boundary crisis.

\begin{figure}
\begin{center}
\addtolength{\subfigcapskip}{-152pt}
\addtolength{\subfigtopskip}{0pt}
\addtolength{\subfiglabelskip}{-9pt}
\subfigure[]{\includegraphics[width=0.31 \textwidth]{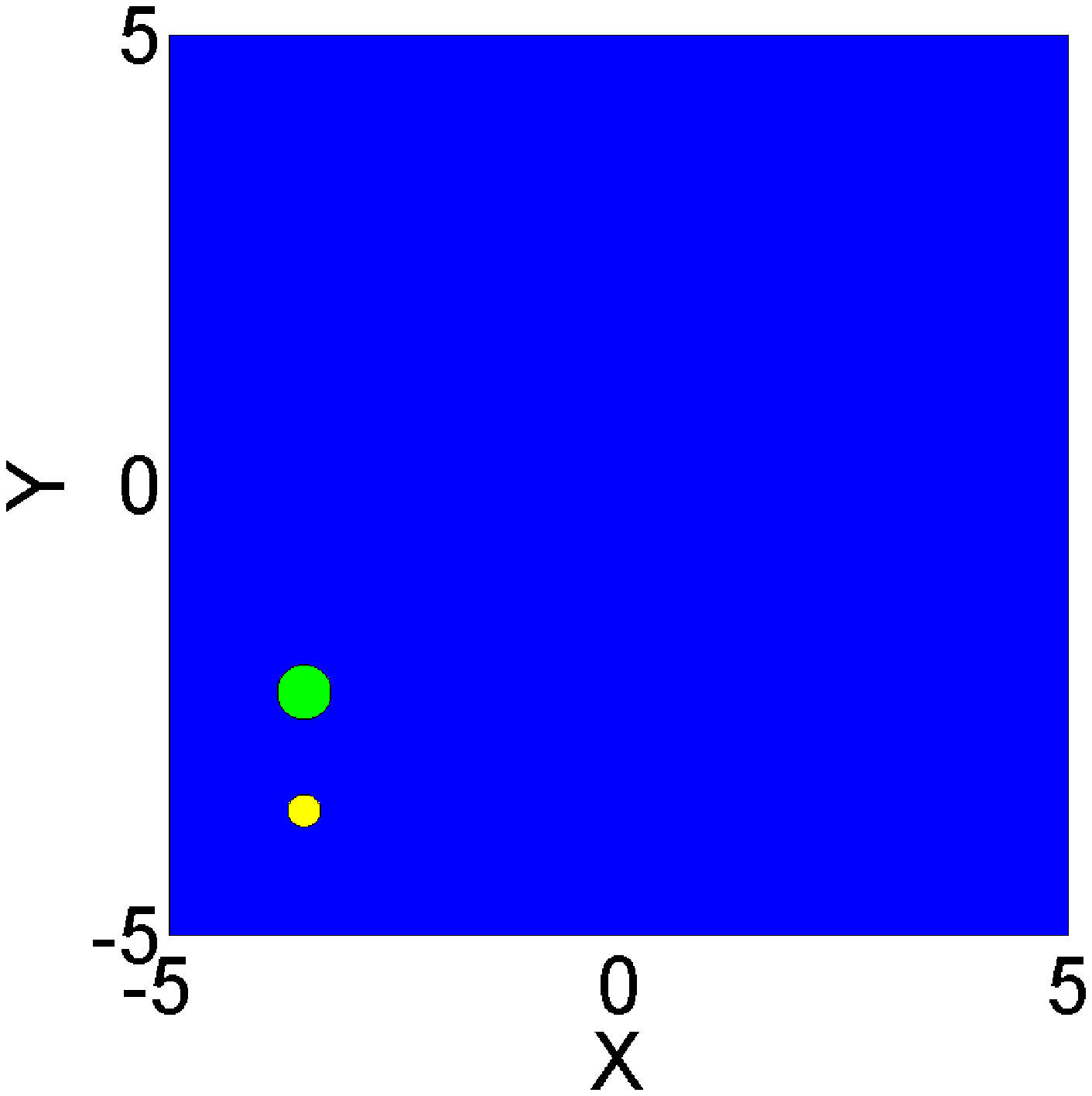}}
\subfigure[]{\includegraphics[width=0.31 \textwidth]{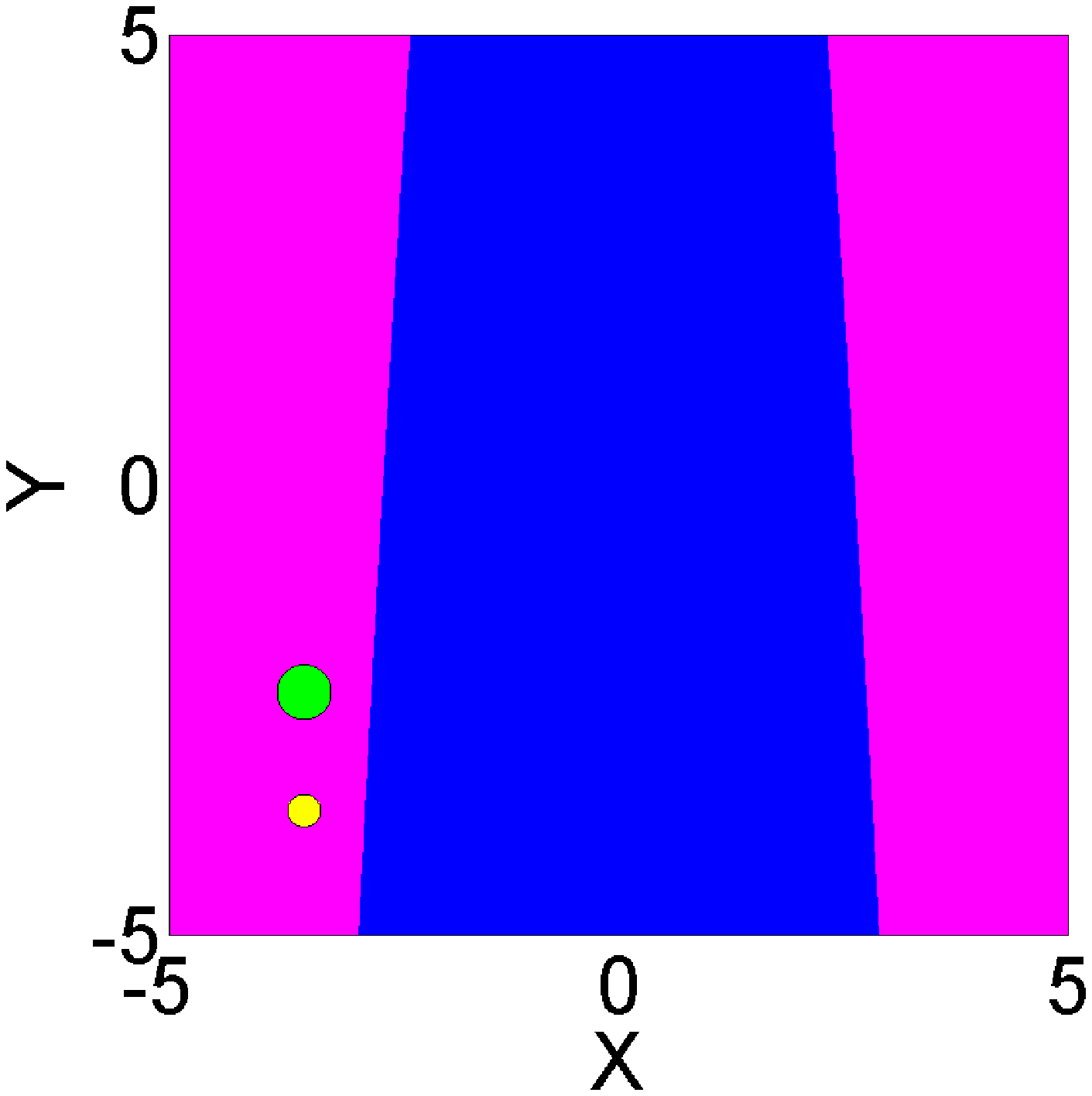}}
\subfigure[]{\includegraphics[width=0.31 \textwidth]{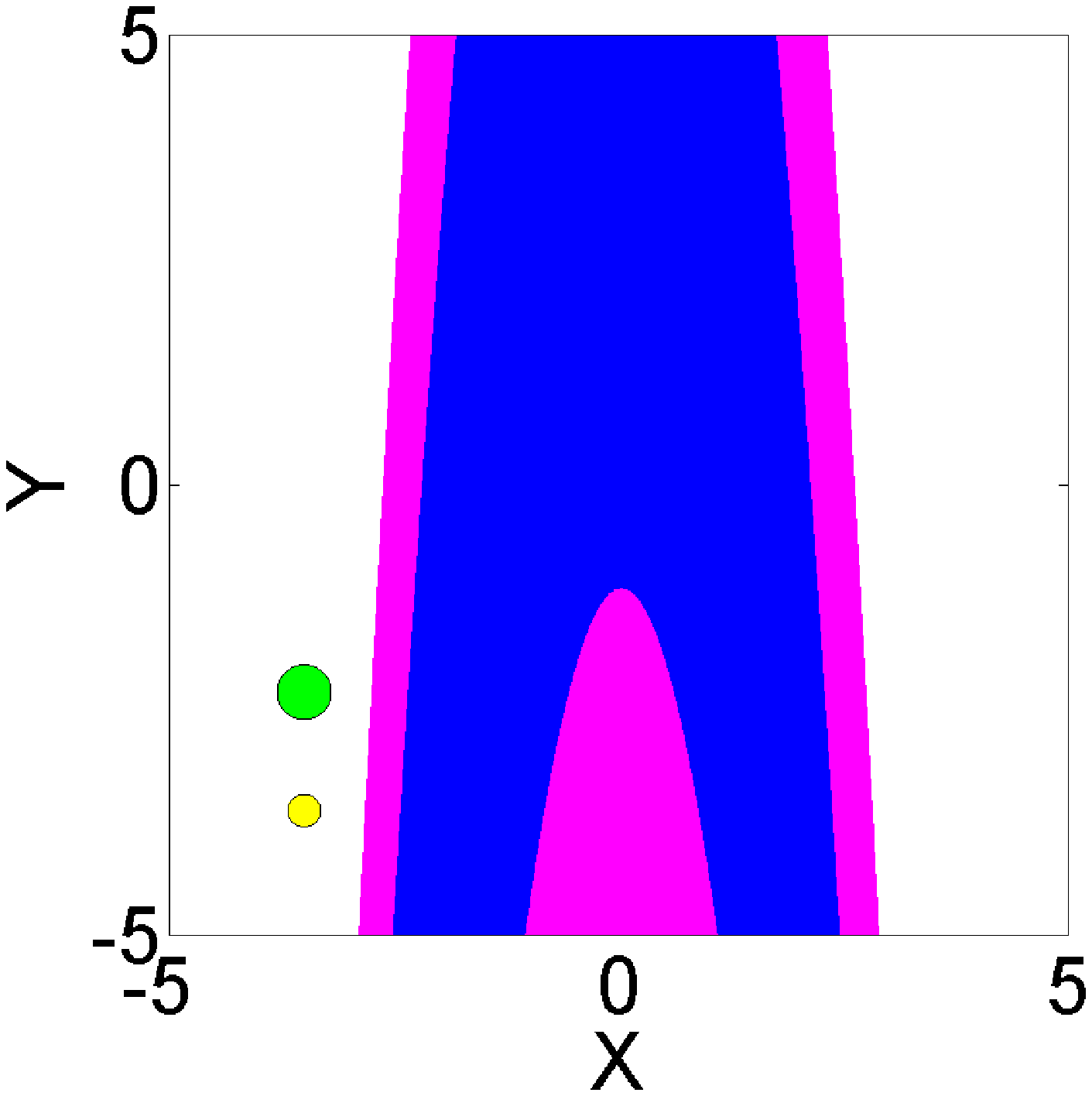}}
\subfigure[]{\includegraphics[width=0.31 \textwidth]{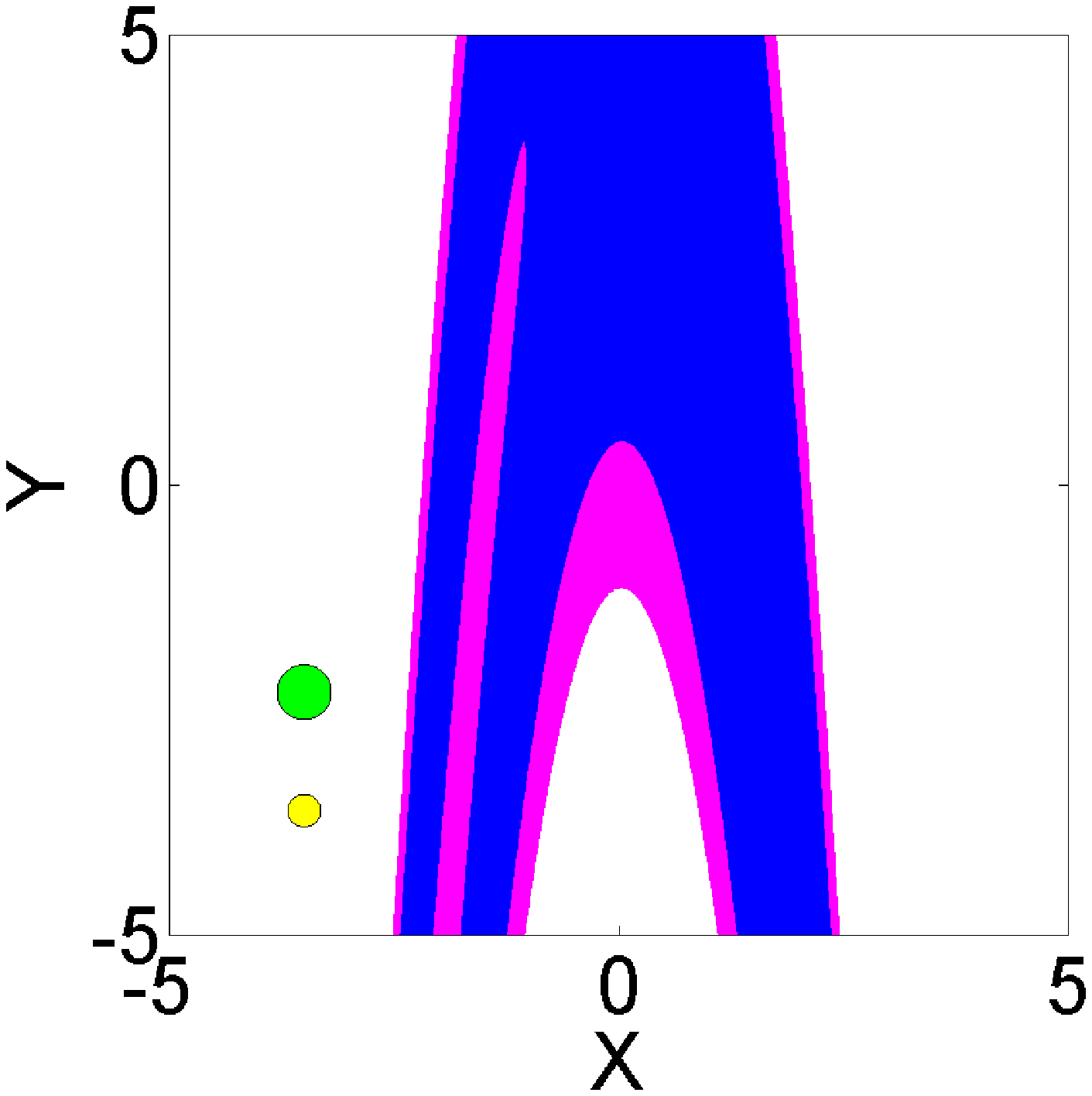}}
\subfigure[]{\includegraphics[width=0.31 \textwidth]{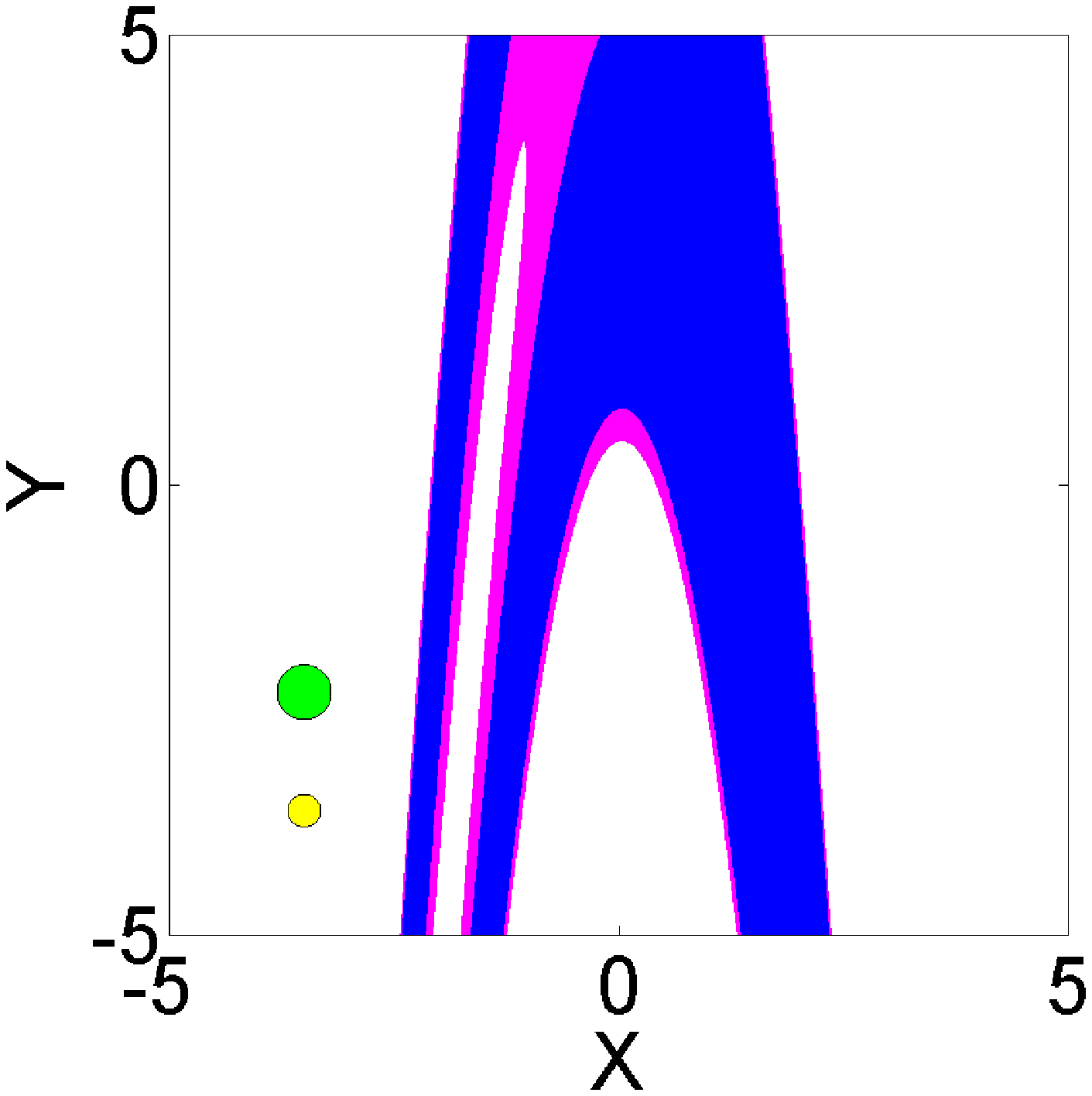}}
\subfigure[]{\includegraphics[width=0.31 \textwidth]{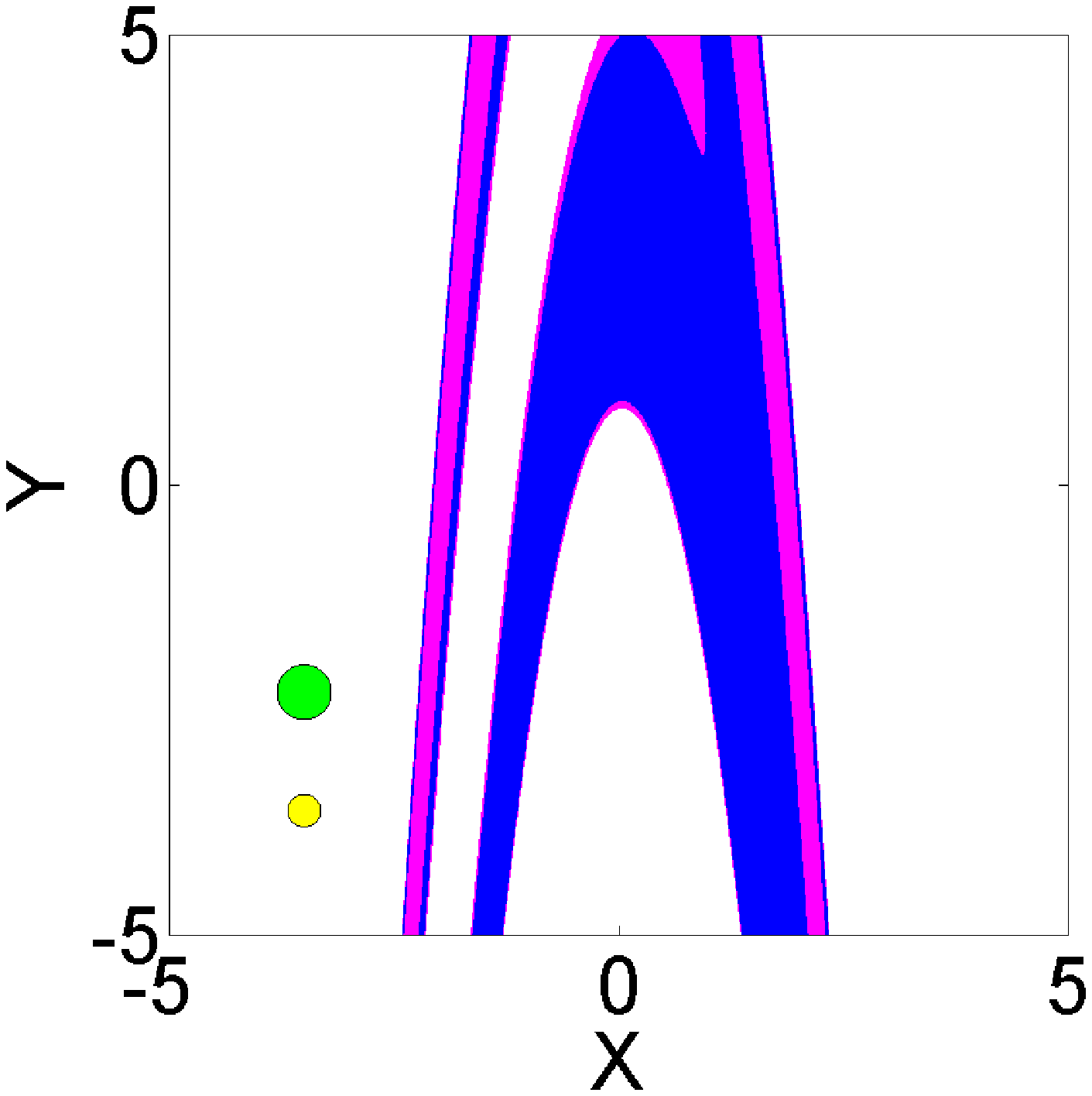}}
\subfigure[]{\includegraphics[width=0.31 \textwidth]{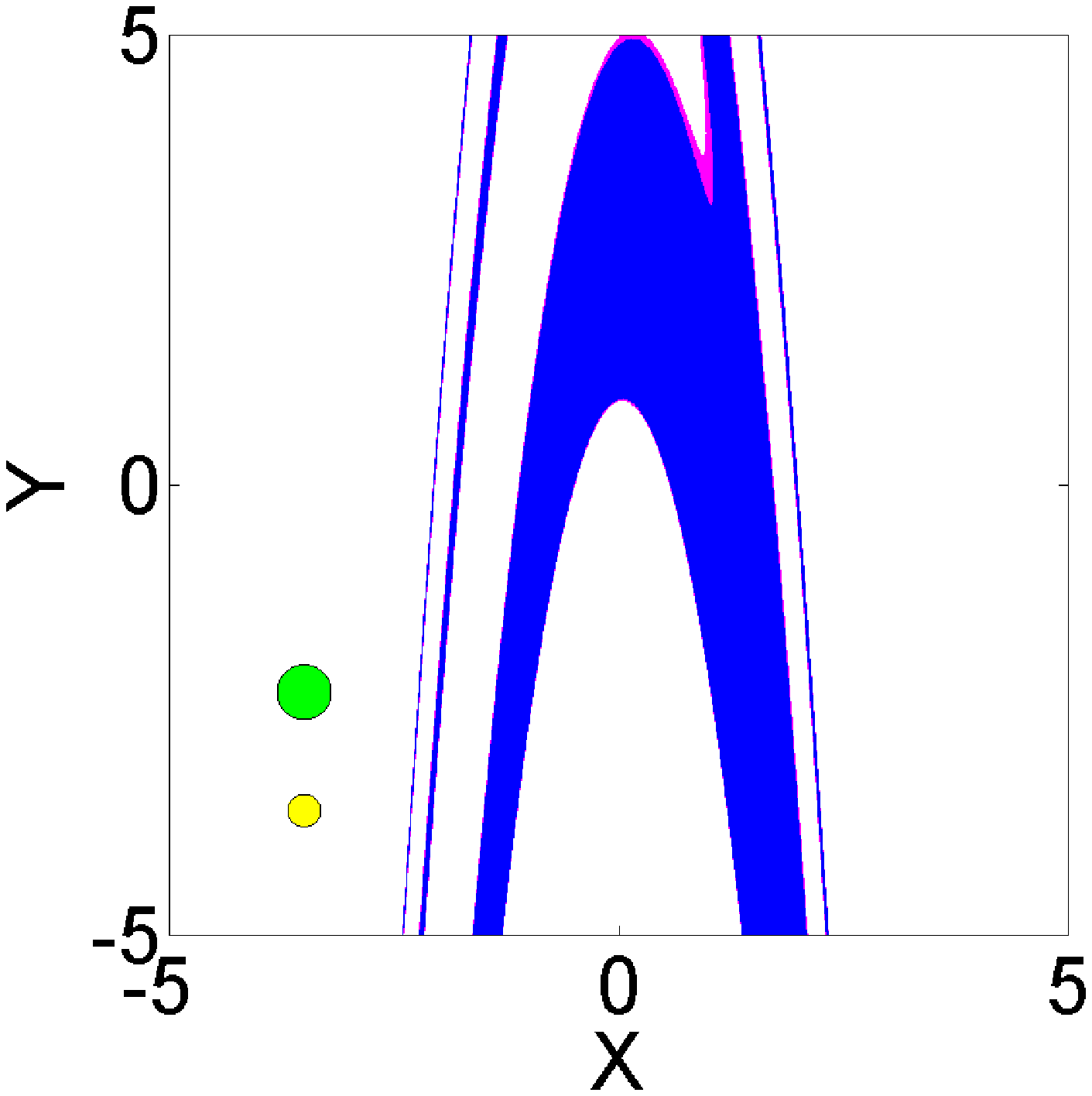}}
\subfigure[]{\includegraphics[width=0.31 \textwidth]{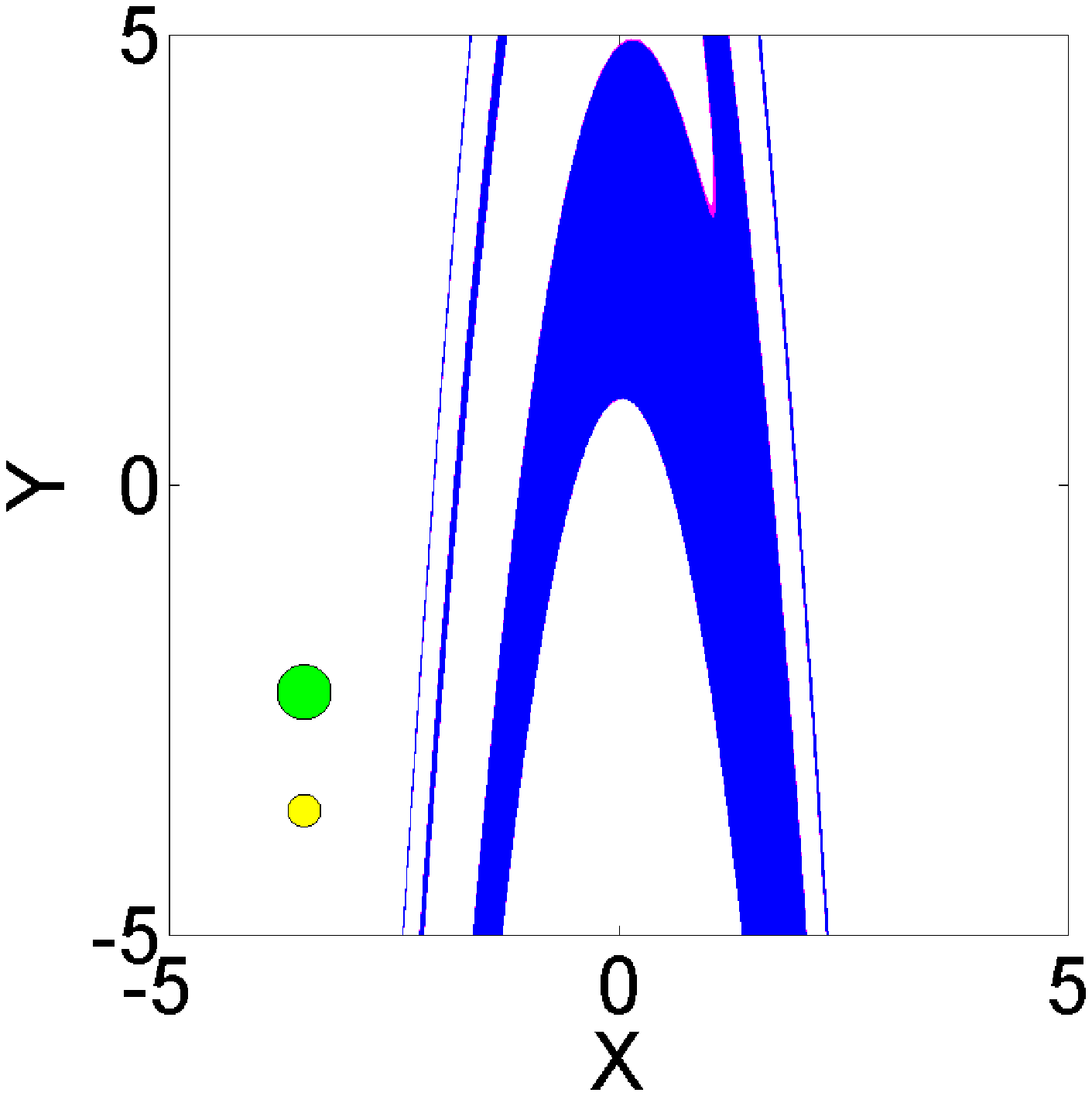}}
\subfigure[]{\includegraphics[width=0.31 \textwidth]{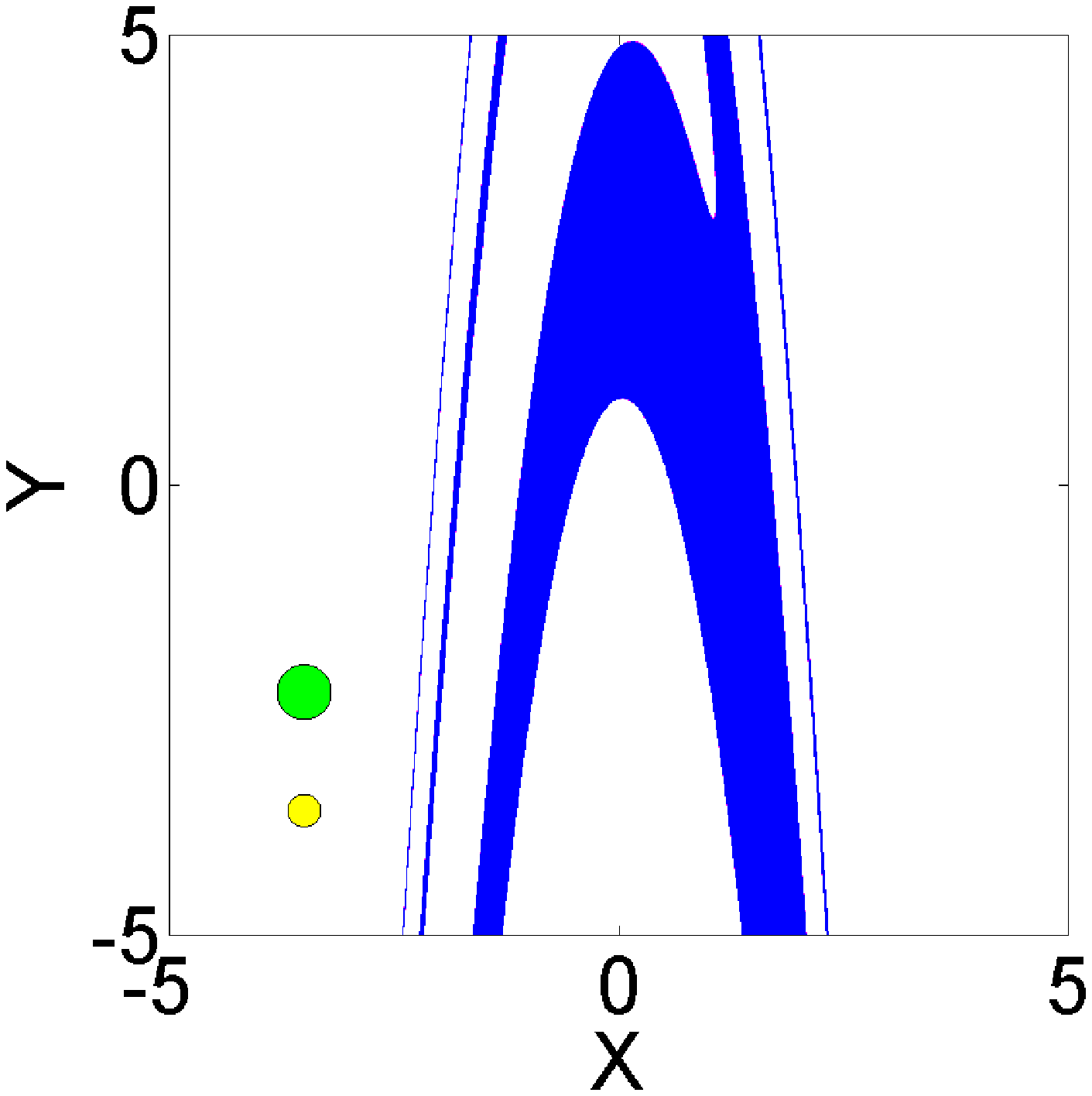}}
\subfigure[]{\includegraphics[width=0.31 \textwidth]{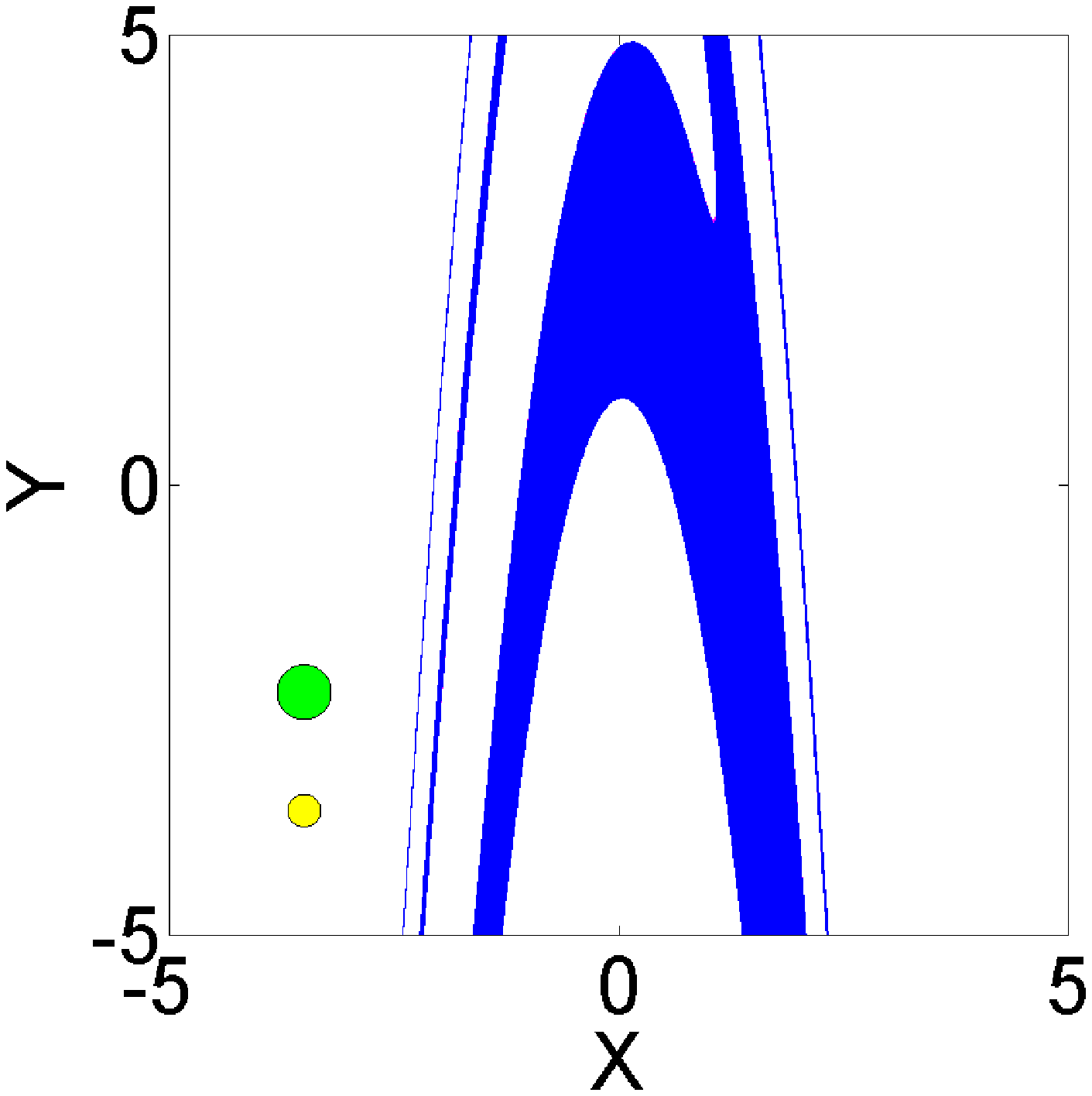}}
\subfigure[]{\includegraphics[width=0.31 \textwidth]{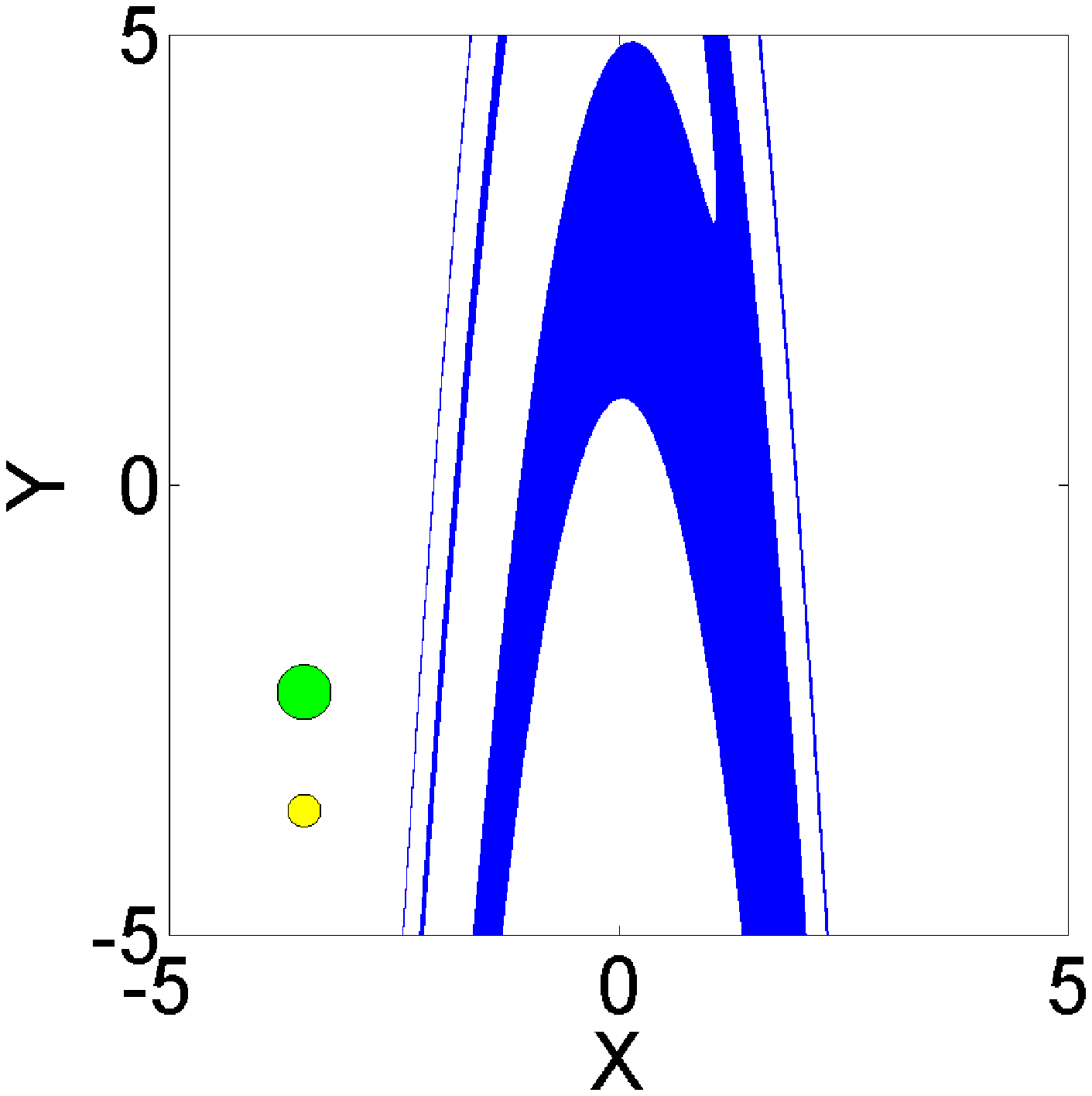}}
\subfigure[]{\includegraphics[width=0.31 \textwidth]{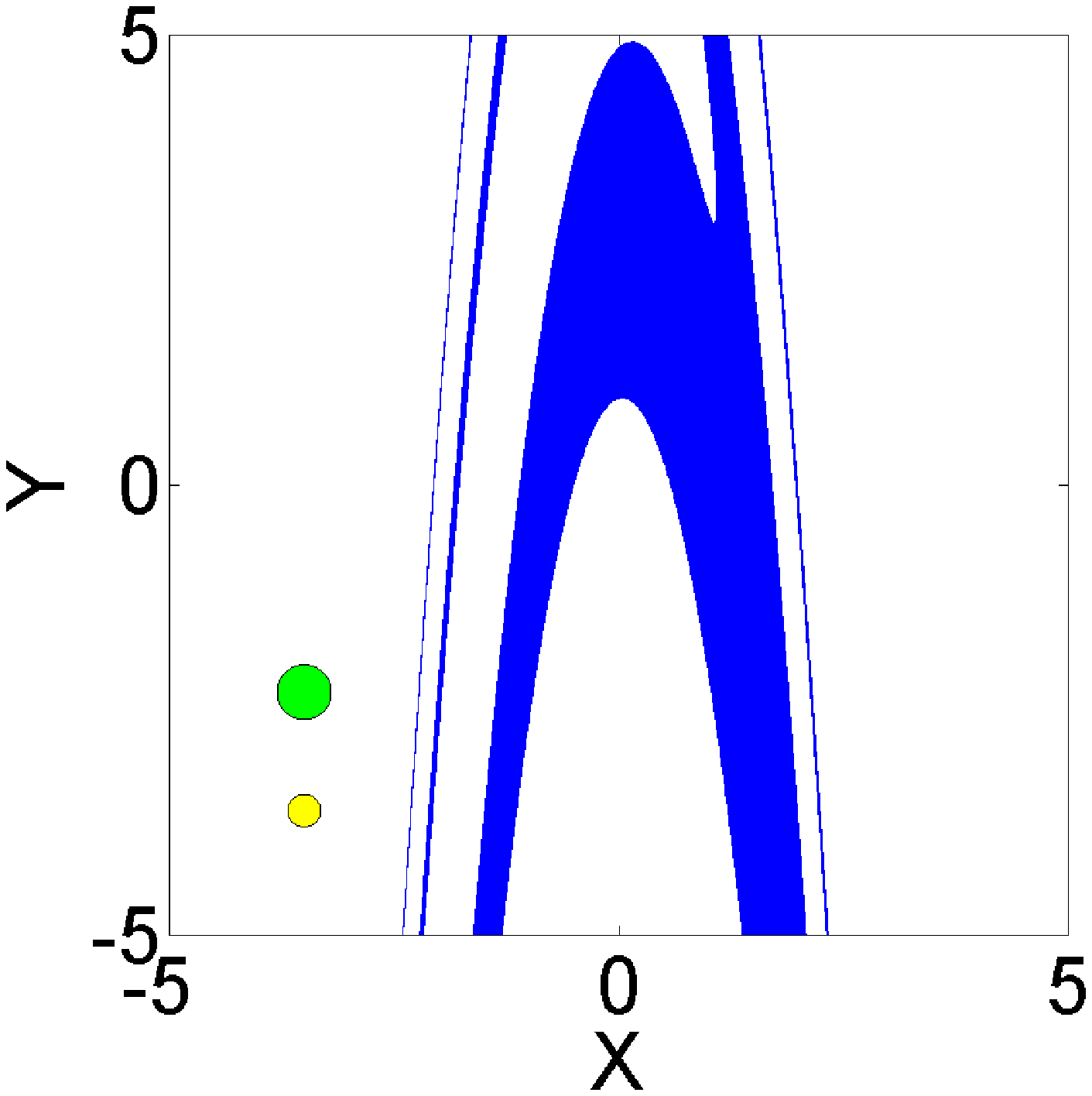}}
\caption{\textbf{Sequence for computing the safe set in the H\'enon map}. The map ($x_{n+1}=2.16-0.3y_{n}-x^{2}_{n};~ y_{n+1}=x_{n}$). The initial region $Q_0$ is the square $[-5,5]\times[-5,5]$, (a grid of $3000\times 3000$ points),  and the values $\xi_0=0.3$ (green circle) and $u_0=0.18$ (yellow circle). At each step, part of $Q_n$ is removed (magenta region) while the blue region remains. After $12$ steps the safe set converges.}
\label{3z}
\end{center}
\end{figure}

\begin{figure}
\begin{center}
\includegraphics[width=0.4 \textwidth]{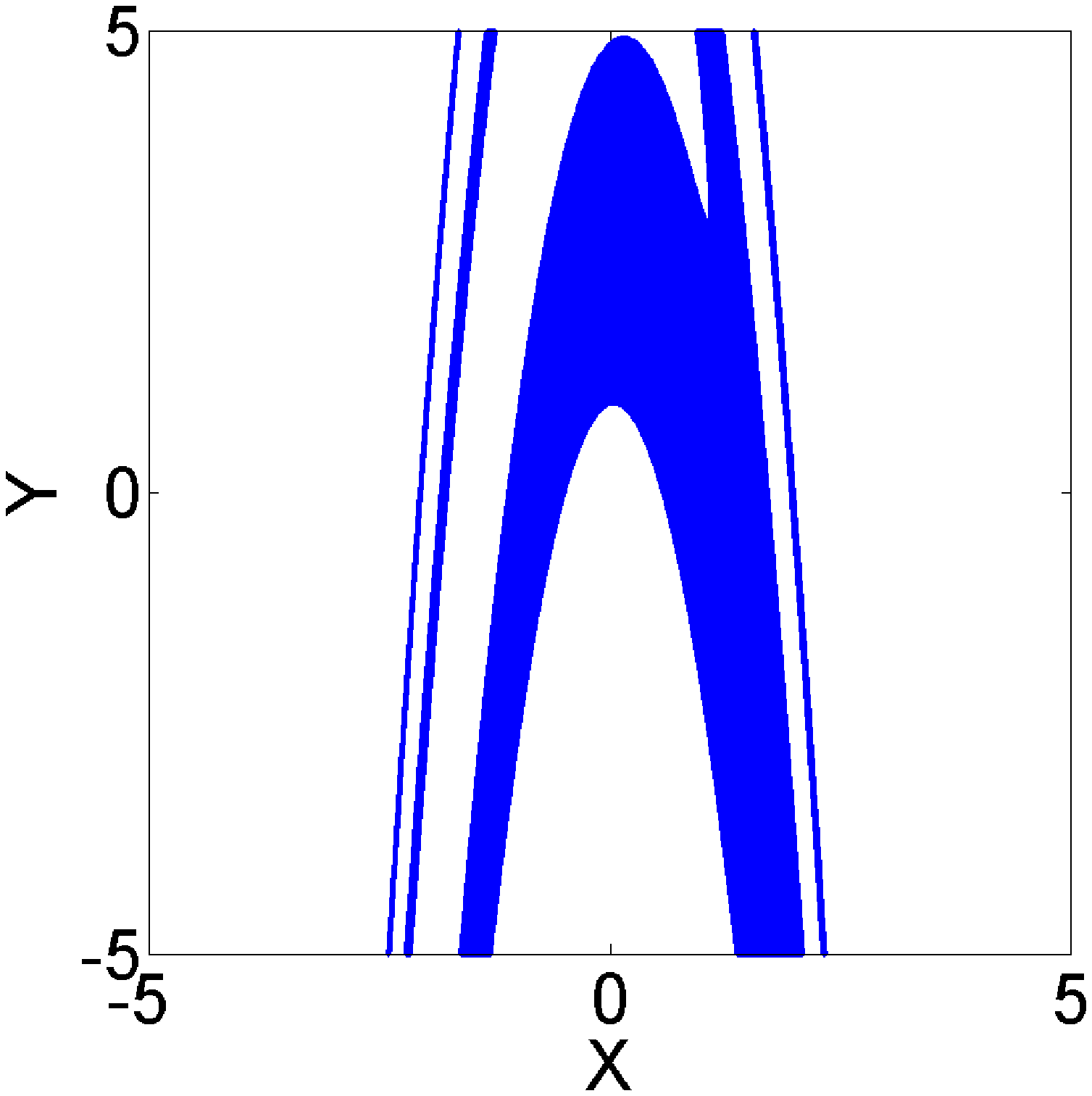}
\caption{\textbf{Safe set of the H\'enon map}. In this figure we can see the result of applying the Sculpting Algorithm to the H\'enon map,
$x_{n+1}=2.16-0.3y_{n}-x^{2}_{n}; y_{n+1}=x_{n}$ . The safe set appears in blue. The minimum control allowed so that it exits a safe set is $u_0=0.18$ for the given disturbance of $\xi_0=0.3$. This is equal to a ratio of $\rho=0.6$.}\label{4z}
\end{center}
\end{figure}

No sink points exist inside the square, only the saddle points of the chaotic saddle can be found in this region. Then, using the Sculpting Algorithm recursively on the initial set of Fig.~\ref{3z}, we find that after 12 iterations, the algorithm converges to a safe set. Moreover, the safe sets are mapped in such a way that the images are surrounded by the safe set itself, as expected. We show the final safe set in Fig.~\ref{4z}.

The Safe-Set Sculpting Algorithm of intermediate sets when applied to the Hénon map is shown in Fig.~\ref{3z}. All the iterations of the algorithm are shown. There is a value of the control parameter $u_0^{min}$ which corresponds to the smallest $u_0$ for which there is a safe set.

\renewcommand{\thesubfigure}{(\arabic{subfigure})}
\makeatletter
\renewcommand{\subfigbottomskip}{-8pt}
\makeatother

In the simulation that we have made with the H\'enon map to obtain Fig.~\ref{4z}, we have used a value of $\xi_0 = 0.3$ for the bounded disturbance. For this value, the minimum control bound (to two-digit precision) for which there is a safe set is $u_0 = 0.18$. Of course, if for the same value of the disturbance the control allowed were higher, the safe set found would be a little larger. The minimum safe ratio obtained for this particular case is $\rho=u_0/\xi_0 = 0.6$.

\subsubsection{Duffing oscillator}

\renewcommand{\thesubfigure}{(\alph{subfigure})~}

Now we consider the Duffing oscillator with this choice of
parameters:
\begin{equation}\label{Duffing_oscillator}
\ddot{x}+0.15\dot{x}-x+x^{3}=0.245\sin(t).
\end{equation}

\begin{figure}
\begin{center}
\subfigure[basins of attraction]{\includegraphics[width=0.316 \textwidth]{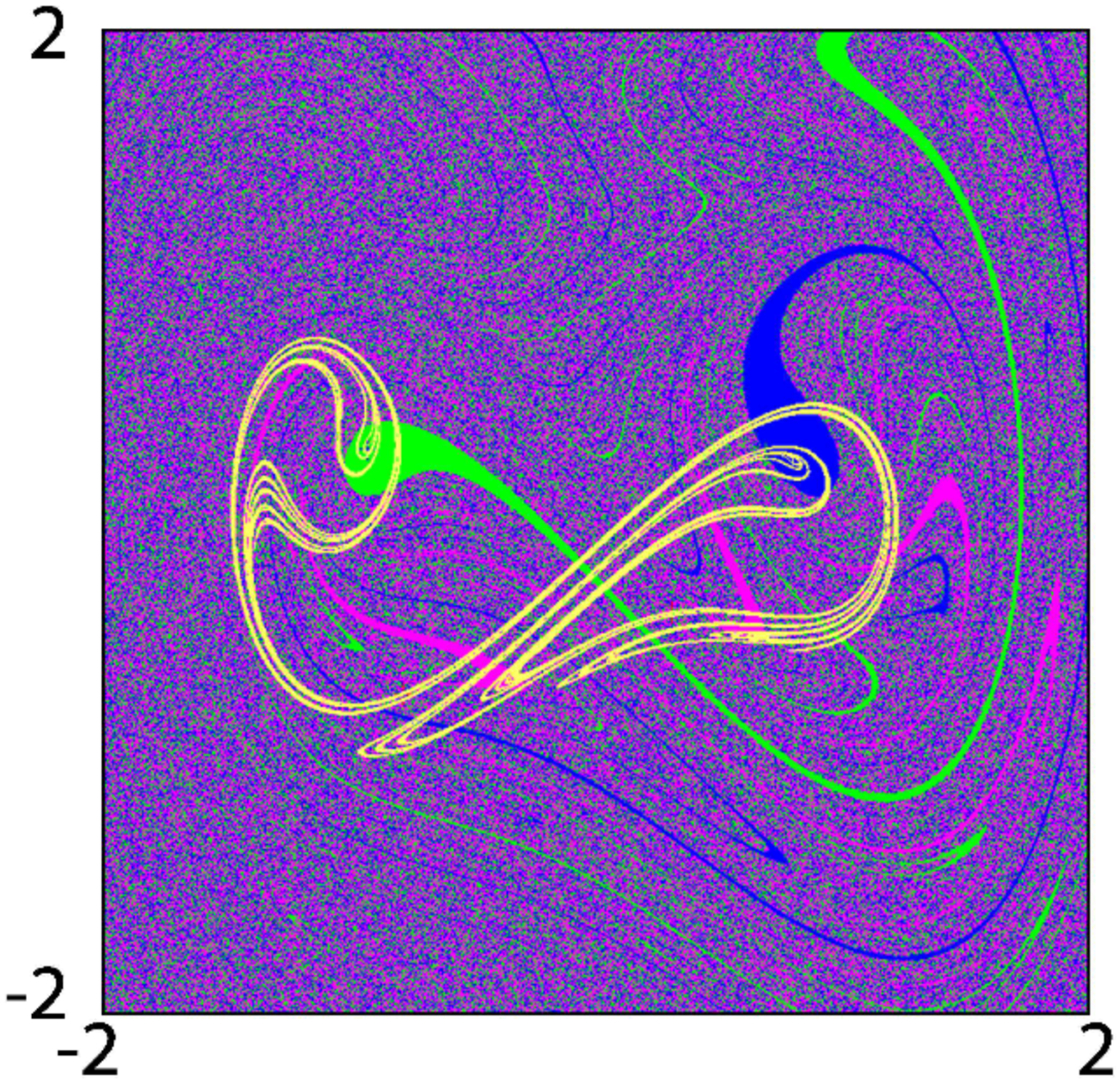}}
\subfigure[Periodic attractors]{\includegraphics[width=0.3 \textwidth]{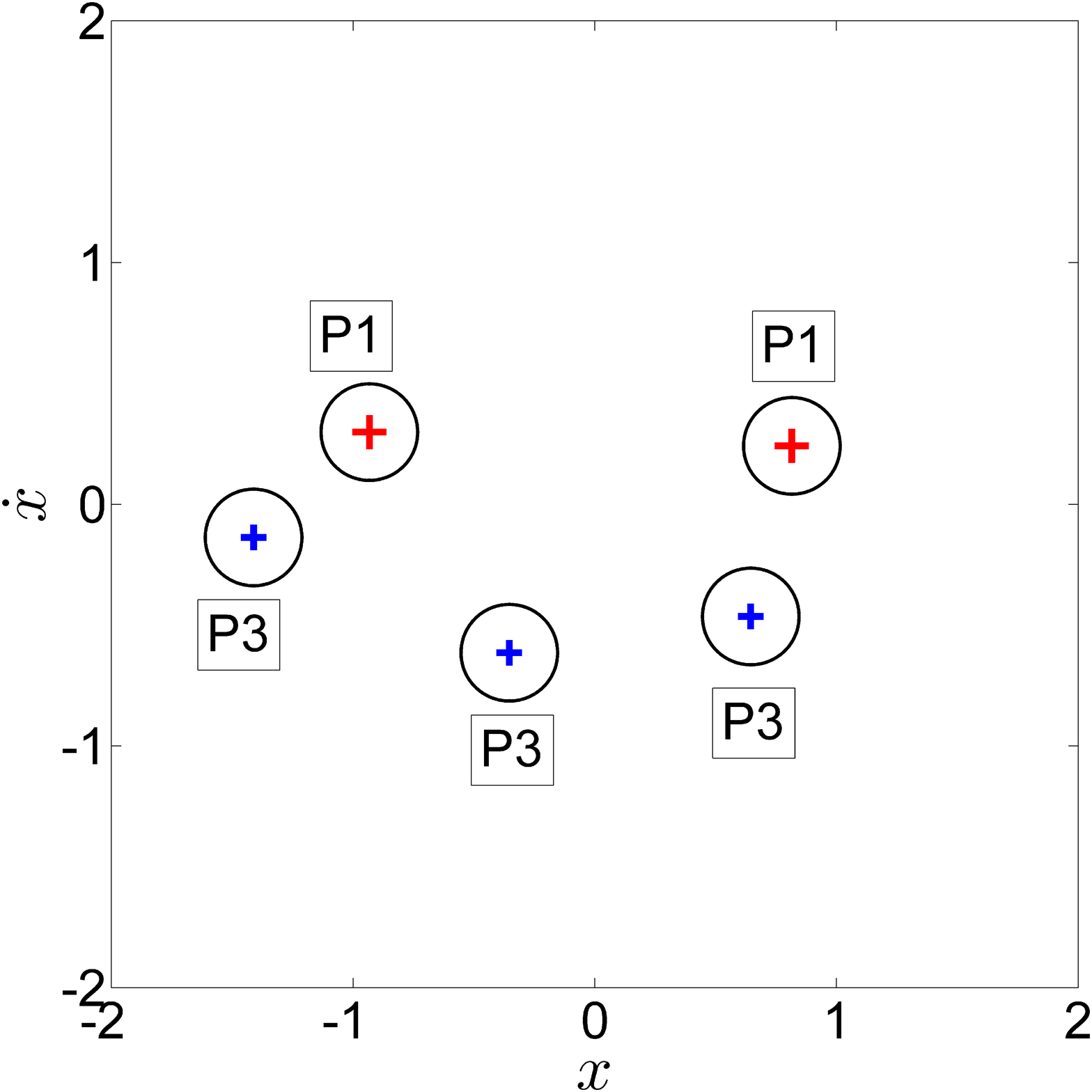}}
\subfigure[Initial region Q]{\includegraphics[width=0.3 \textwidth]{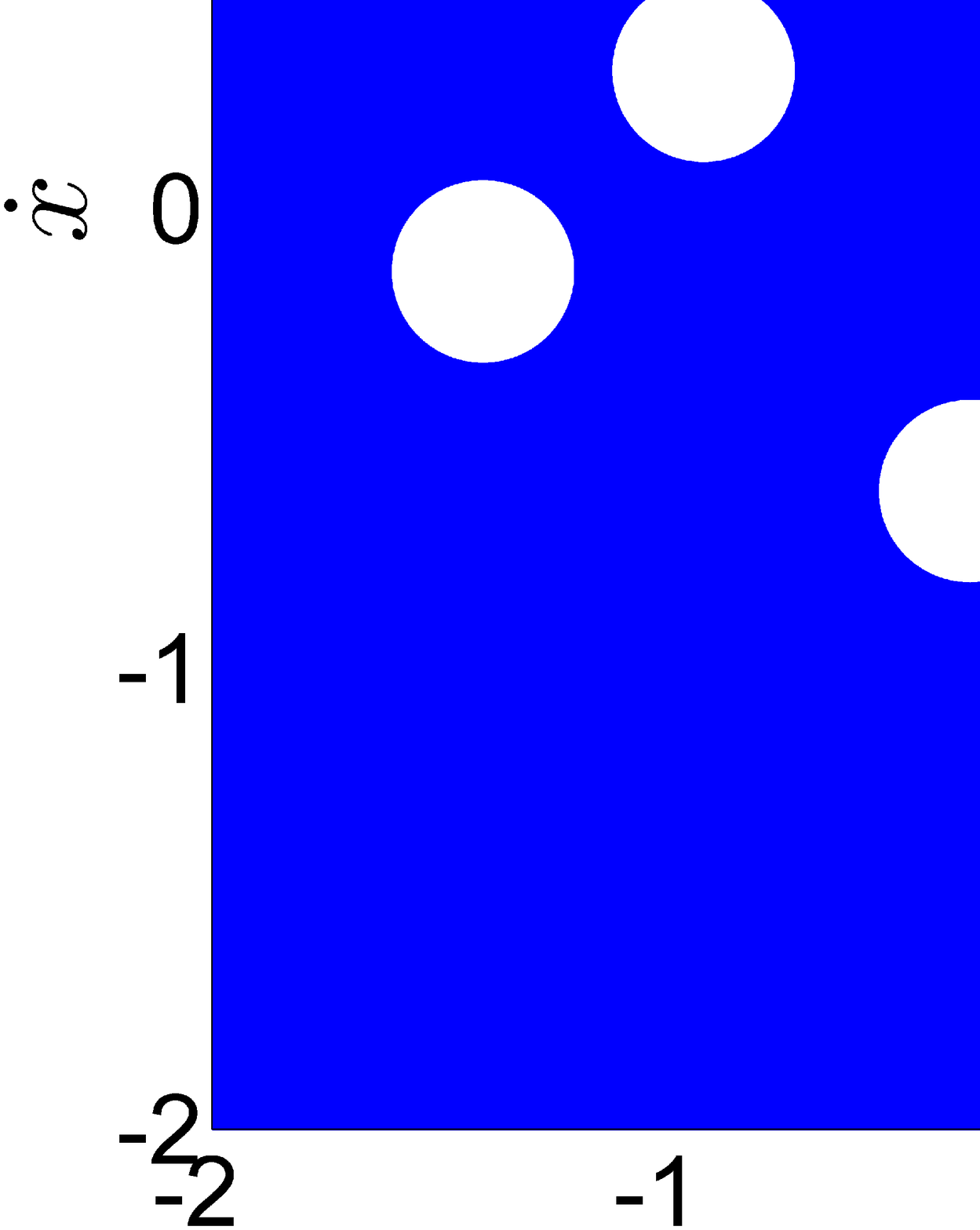}}
\\[20pt]
\caption{\textbf{Periodic attractors of the Duffing oscillator}. (a) In this figure we show the complex structure of the phase space for the Duffing oscillator $\ddot{x}+0.15\dot{x}-x+x^{3}=0.245\sin(t)$. In this system are present three different basins of attraction (magenta, blue and green)  and the system has the Wada property. The invariant unstable manifold associated to the chaotic saddle appears in yellow. (b) This figure reproduces the periodic attractors: two period-1 attractors and one period-3 attractor. We also show with circles of radius $0.2$ the region of the phase space that we want to avoid, whatever the disturbances. (c) We use a grid of $6000\times 6000$ points in the square $[-2,2]\times[-2,2]$ as our initial set, but removing the zones that we want to avoid, that is the circles. Applying the Sculpting Algorithm over several iterations, we will obtain the desired safe set. We let $\xi_0=0.08$ be the maximum size of the vector perturbation.}
\label{5z}
\end{center}
\end{figure}

With these parameters, a very interesting topological property appears here. This is the Wada property. Due to this property, every point on the boundary of any basin is also on the boundary of the other two basins
\cite{DuffingA}. This is what we see in the Fig.~\ref{5z}(a).
With this configuration, the Duffing oscillator has a region that shows a transient
chaotic behavior in the square $[-2,2]\times [-2,2]$ due to the presence of a chaotic
saddle. For this choice of parameters, the system possesses two period-1 orbits
and one period-3 orbit. We can see this in the Fig.~\ref{5z}(b).

The idea of applying the partial control method to the Duffing oscillator in this particular case is slightly different than that of using it in the H\'enon map. The region $Q$ in this case contains several attracting periodic orbits that will eventually attract almost every trajectory. Our goal here is to have the trajectories partially controlled so that they stay away from the attracting fixed points and the attracting periodic orbit of period $3$. The unperturbed, uncontrolled behavior of the system exhibits transient chaotic behavior. The orbits behave chaotically, but after some time, the orbits fall close enough to some of the stable periodic attractors. In presence of disturbances the orbit can sporadically jump to other periodic orbits arising a mixture of irregular and periodic behaviour.

The upper bound of the disturbance that we consider in this system is $\xi_0 = 0.08$. The situation changes drastically if we use the partial control method. Then it is possible to maintain the chaotic behavior indefinitely, with a control smaller than the disturbances, avoiding the orbits escape to the periodic regime. We have found that it is possible to achieve this with a ratio of control versus the disturbances of approximately $0.59$. For $u_0$ significantly smaller than $0.0475$, there is no safe set.

\renewcommand{\thesubfigure}{(\arabic{subfigure})}
\makeatletter
\renewcommand{\subfigbottomskip}{-8pt}
\makeatother

\begin{figure}
\begin{center}
\addtolength{\subfigcapskip}{-152pt}
\addtolength{\subfigtopskip}{0pt}
\addtolength{\subfiglabelskip}{-9pt}
\subfigure[]{\includegraphics[width=0.3 \textwidth]{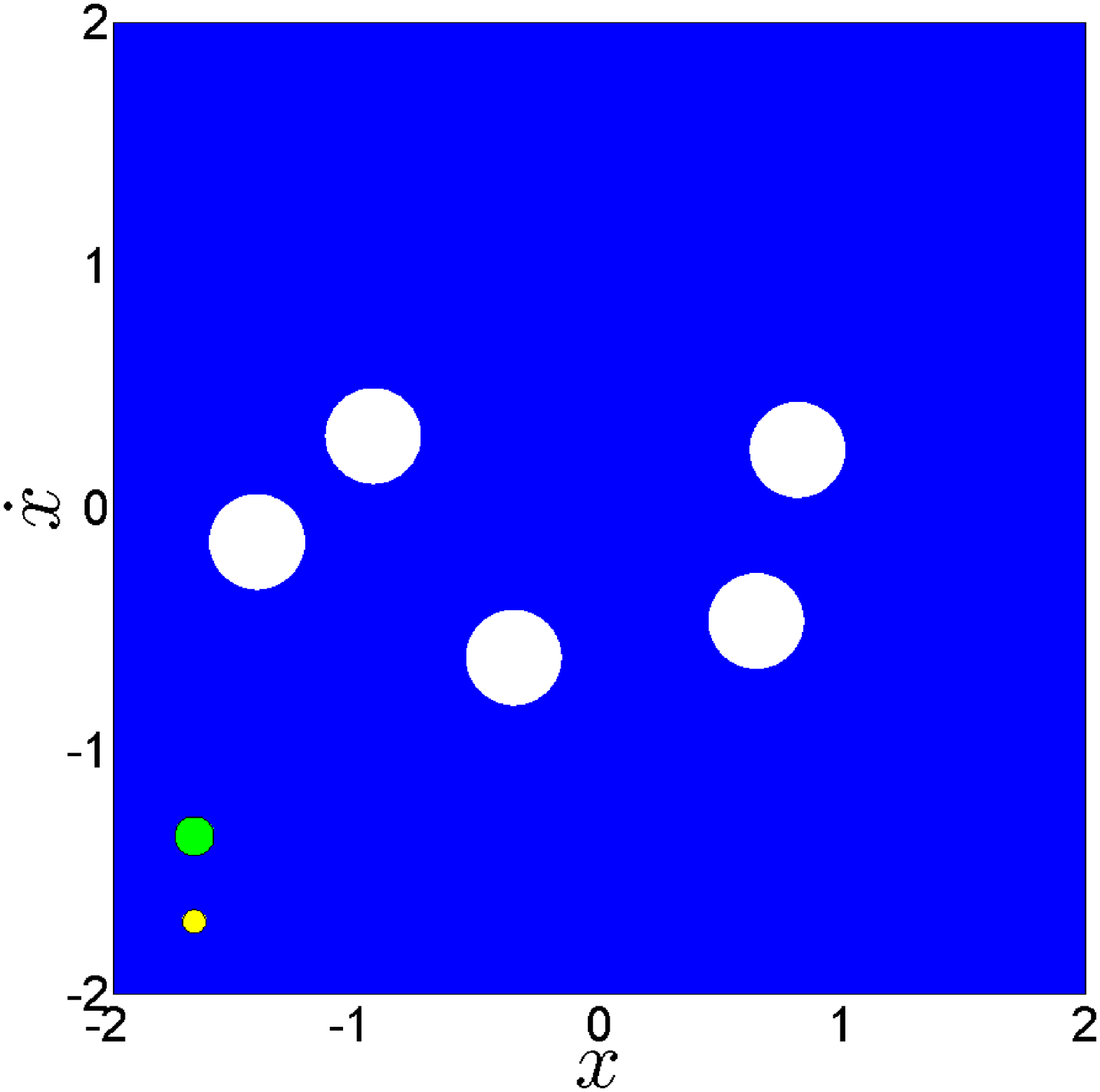}}
\subfigure[]{\includegraphics[width=0.3 \textwidth]{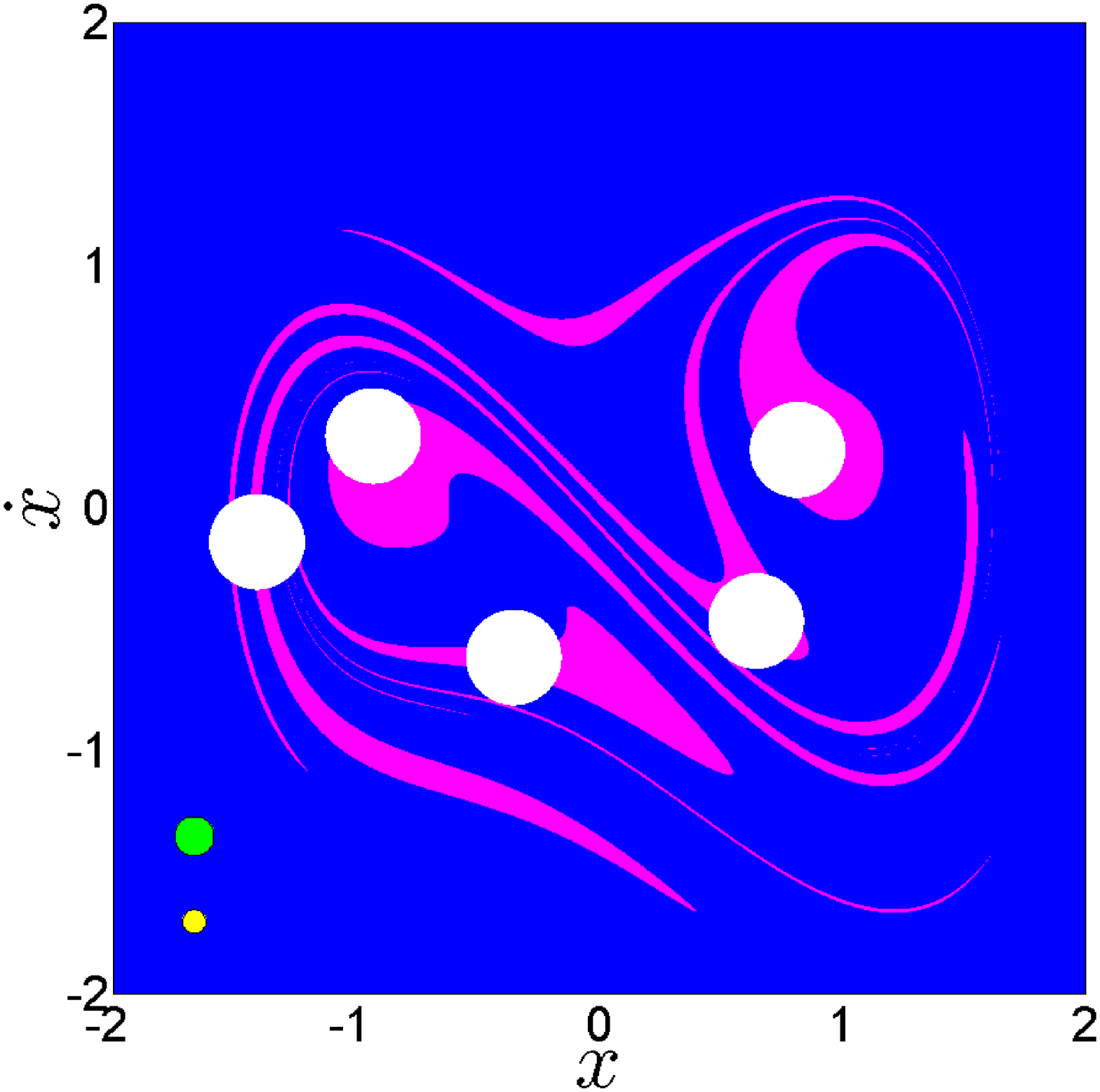}}
\subfigure[]{\includegraphics[width=0.3 \textwidth]{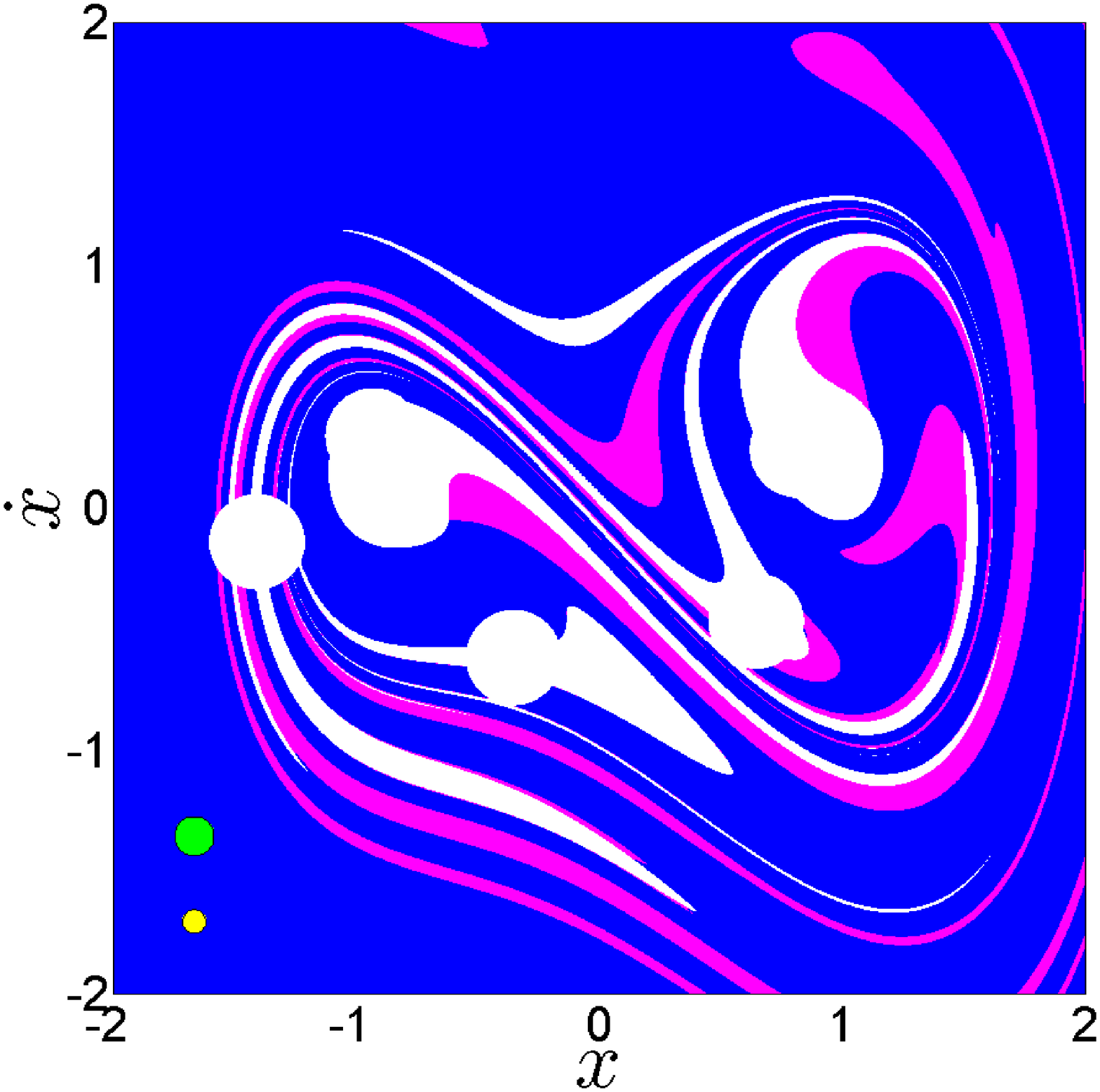}}
\subfigure[]{\includegraphics[width=0.3 \textwidth]{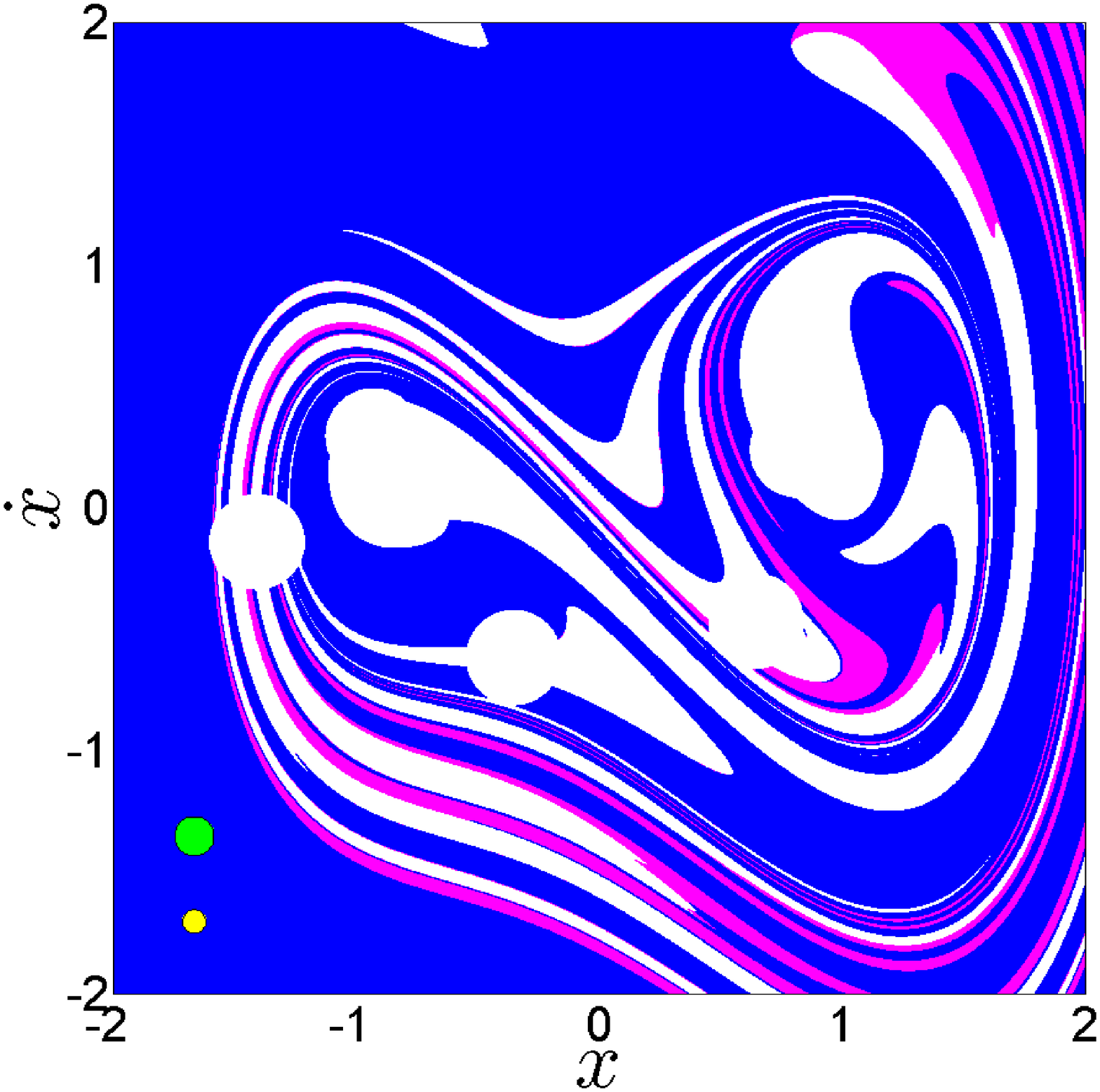}}
\subfigure[]{\includegraphics[width=0.3 \textwidth]{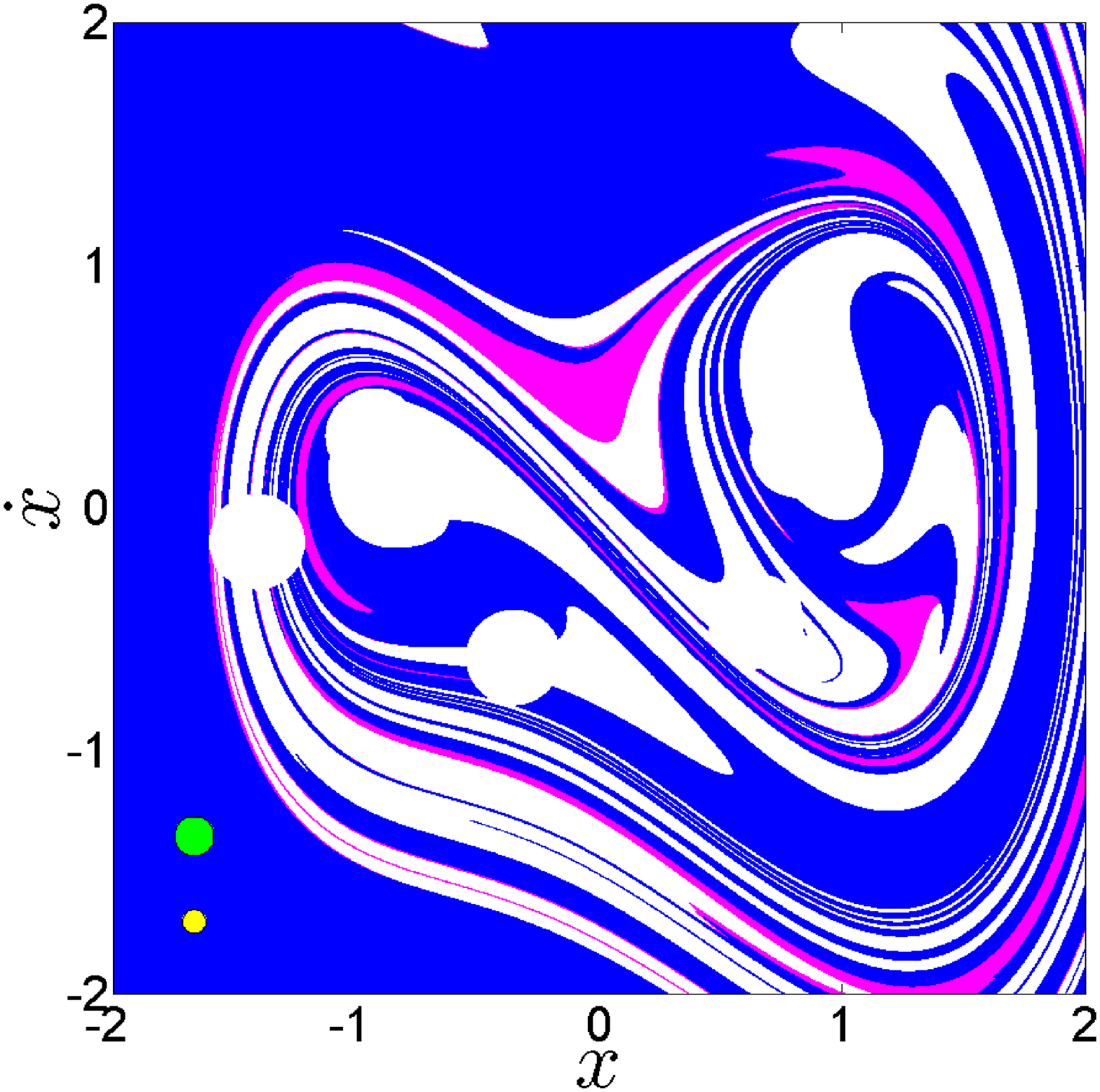}}
\subfigure[]{\includegraphics[width=0.3 \textwidth]{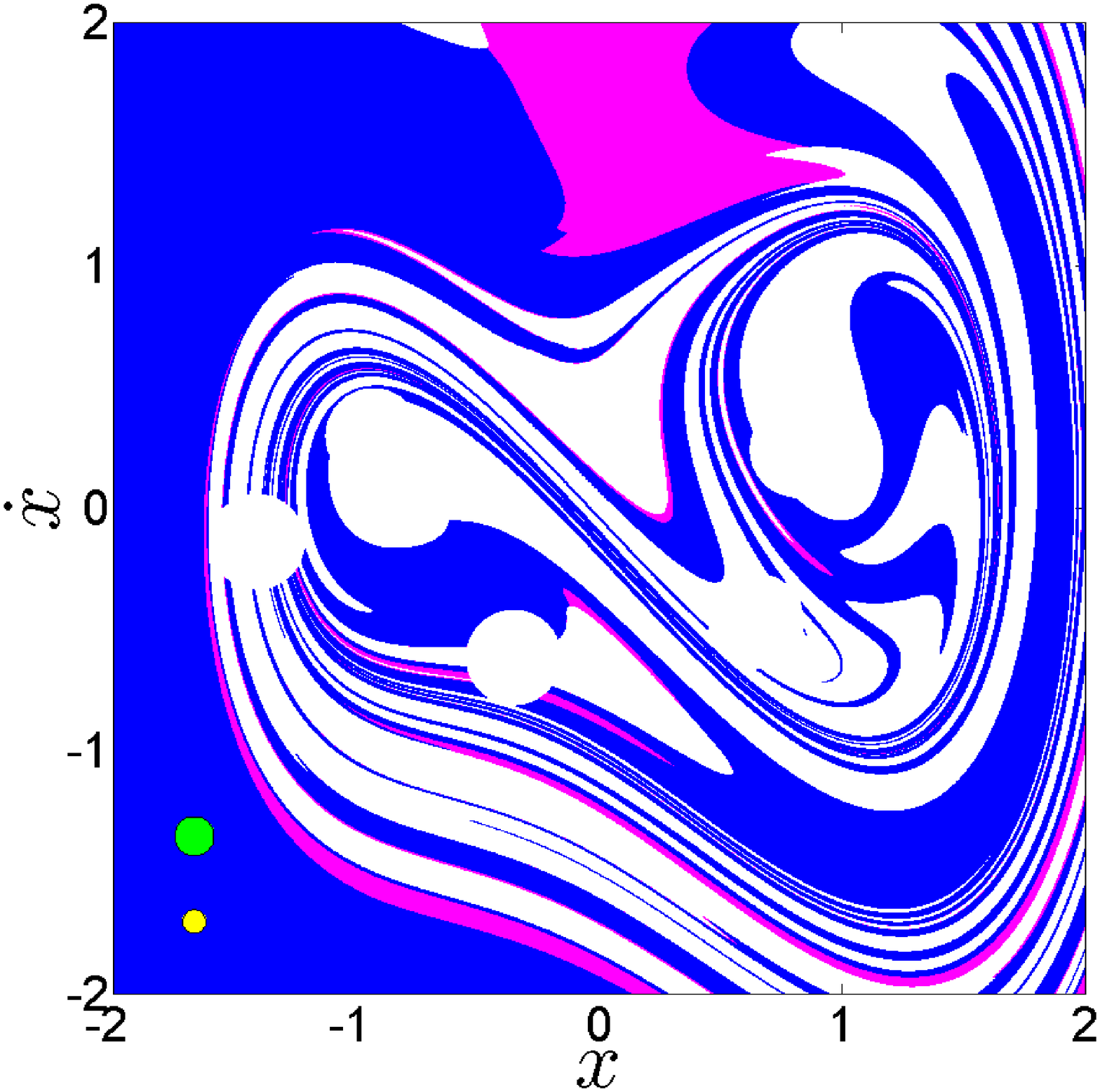}}
\subfigure[]{\includegraphics[width=0.3 \textwidth]{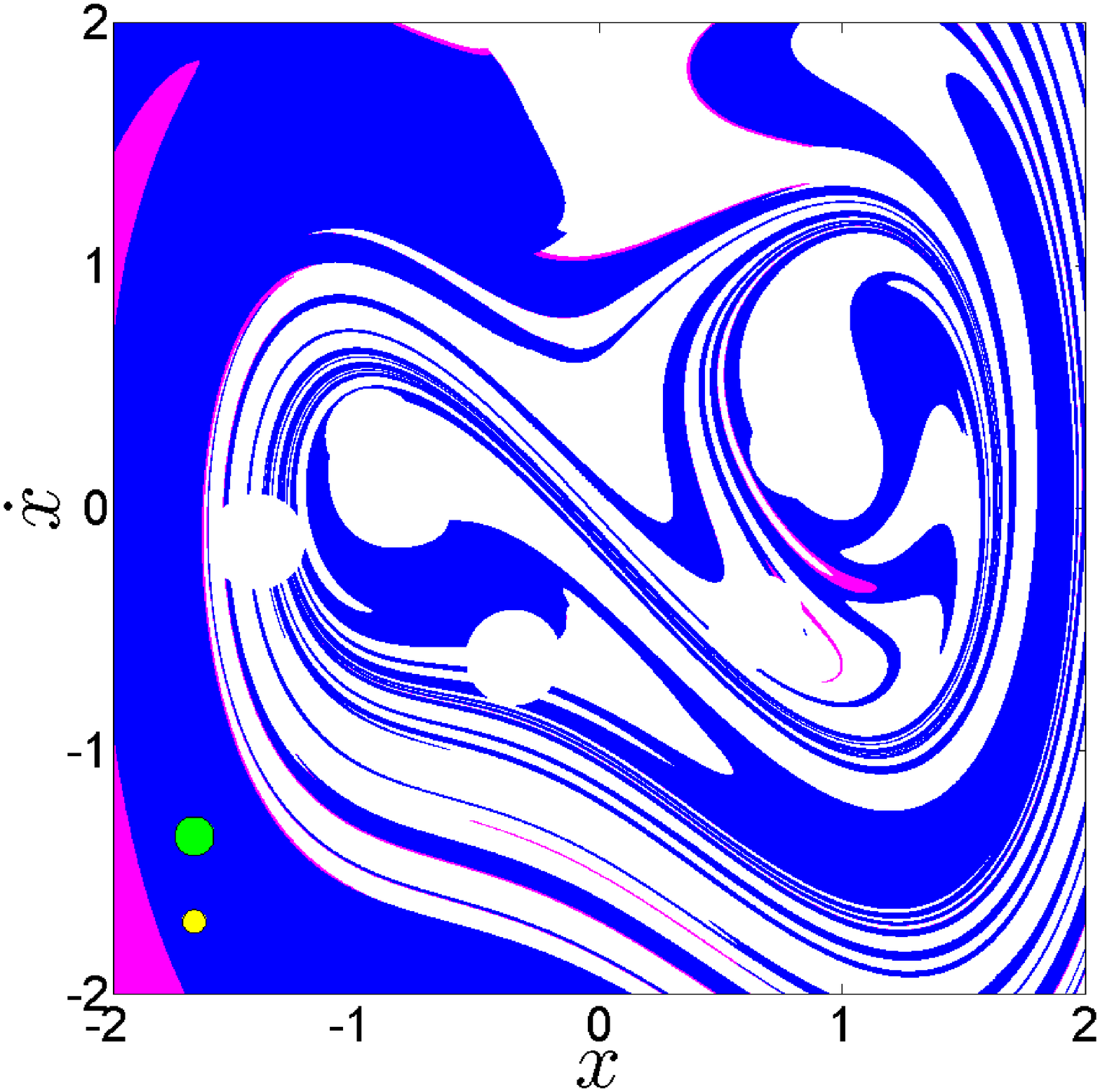}}
\subfigure[]{\includegraphics[width=0.3 \textwidth]{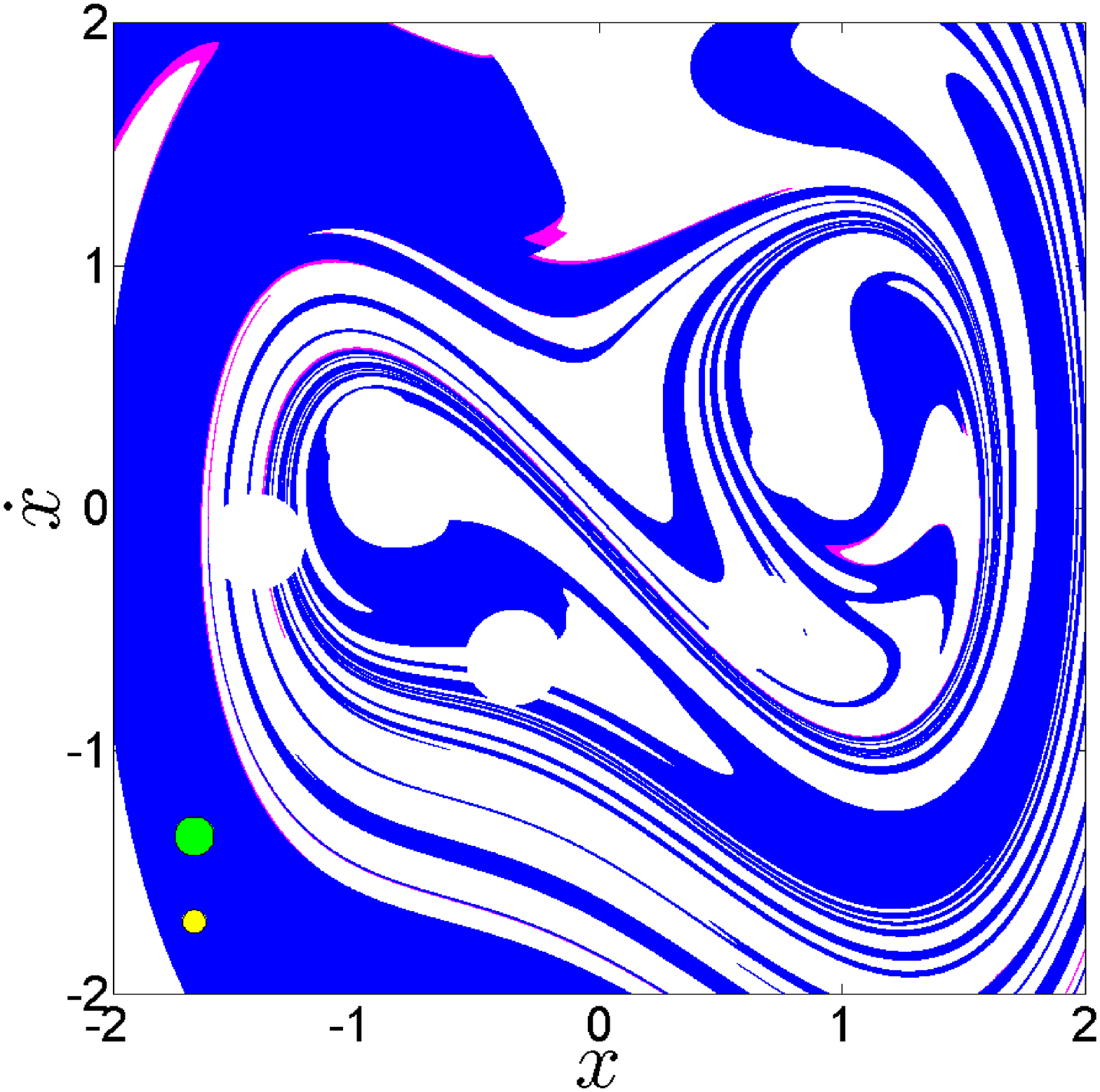}}
\subfigure[]{\includegraphics[width=0.3 \textwidth]{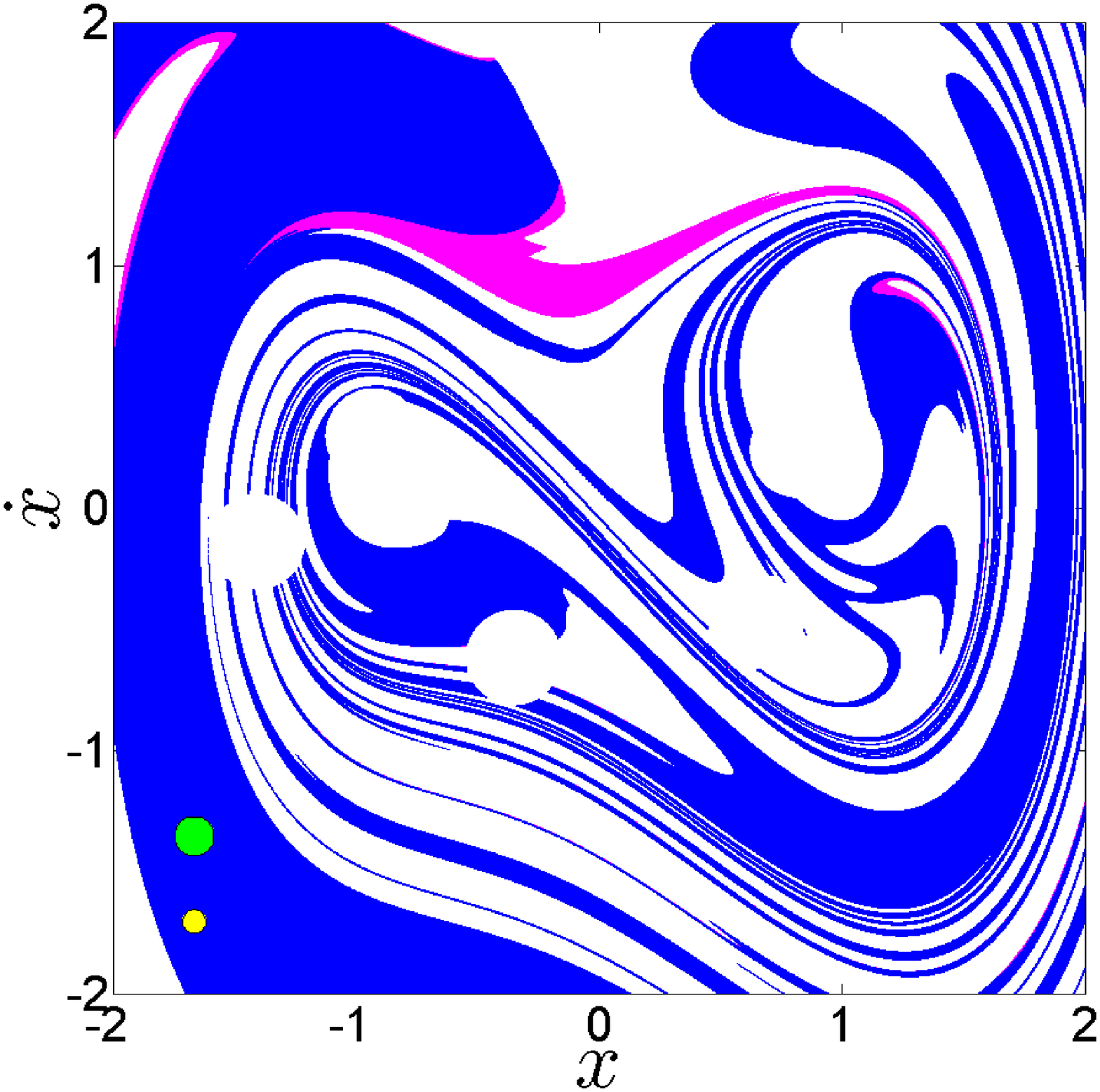}}
\subfigure[]{\includegraphics[width=0.3 \textwidth]{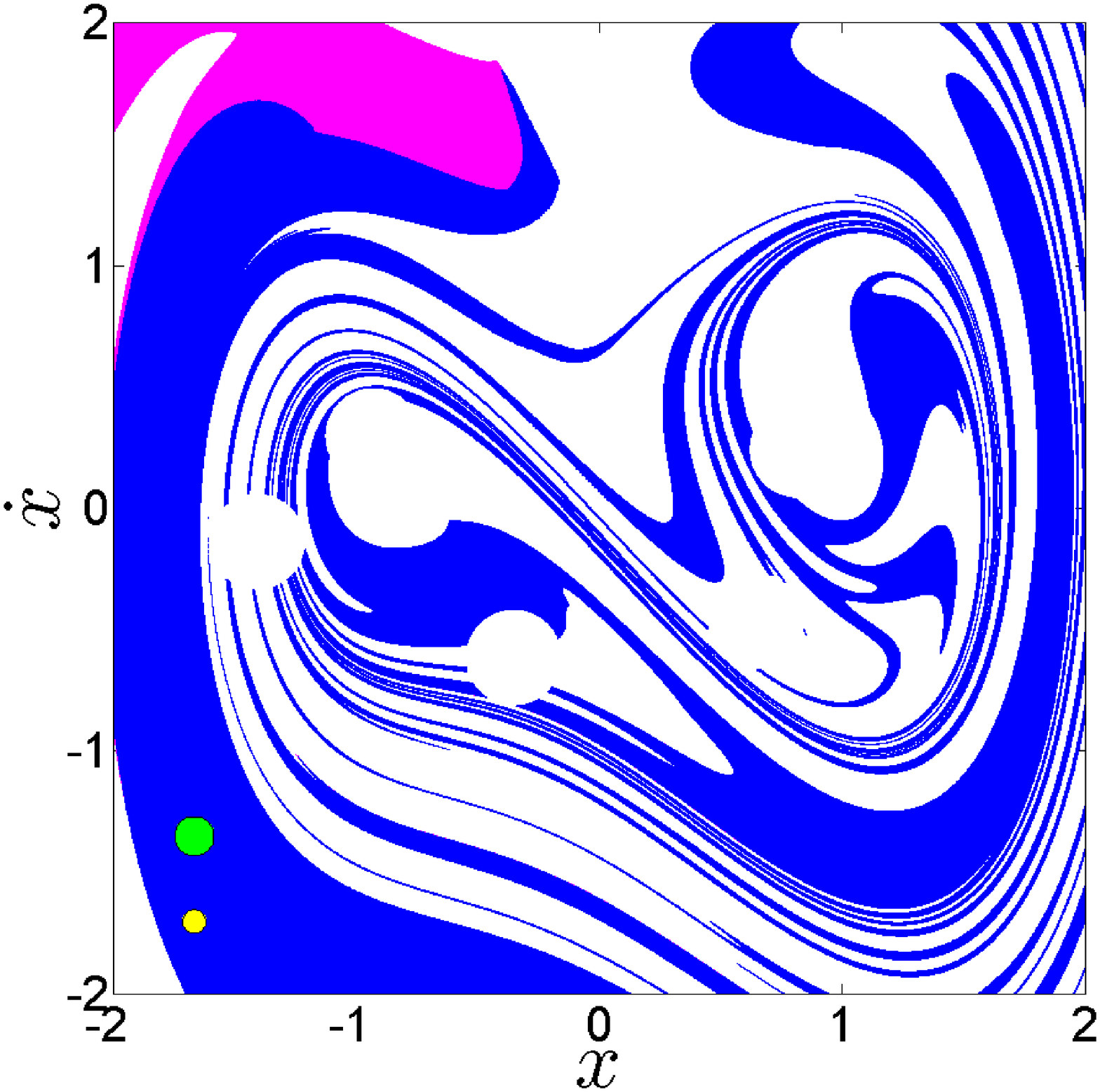}}
\subfigure[]{\includegraphics[width=0.3 \textwidth]{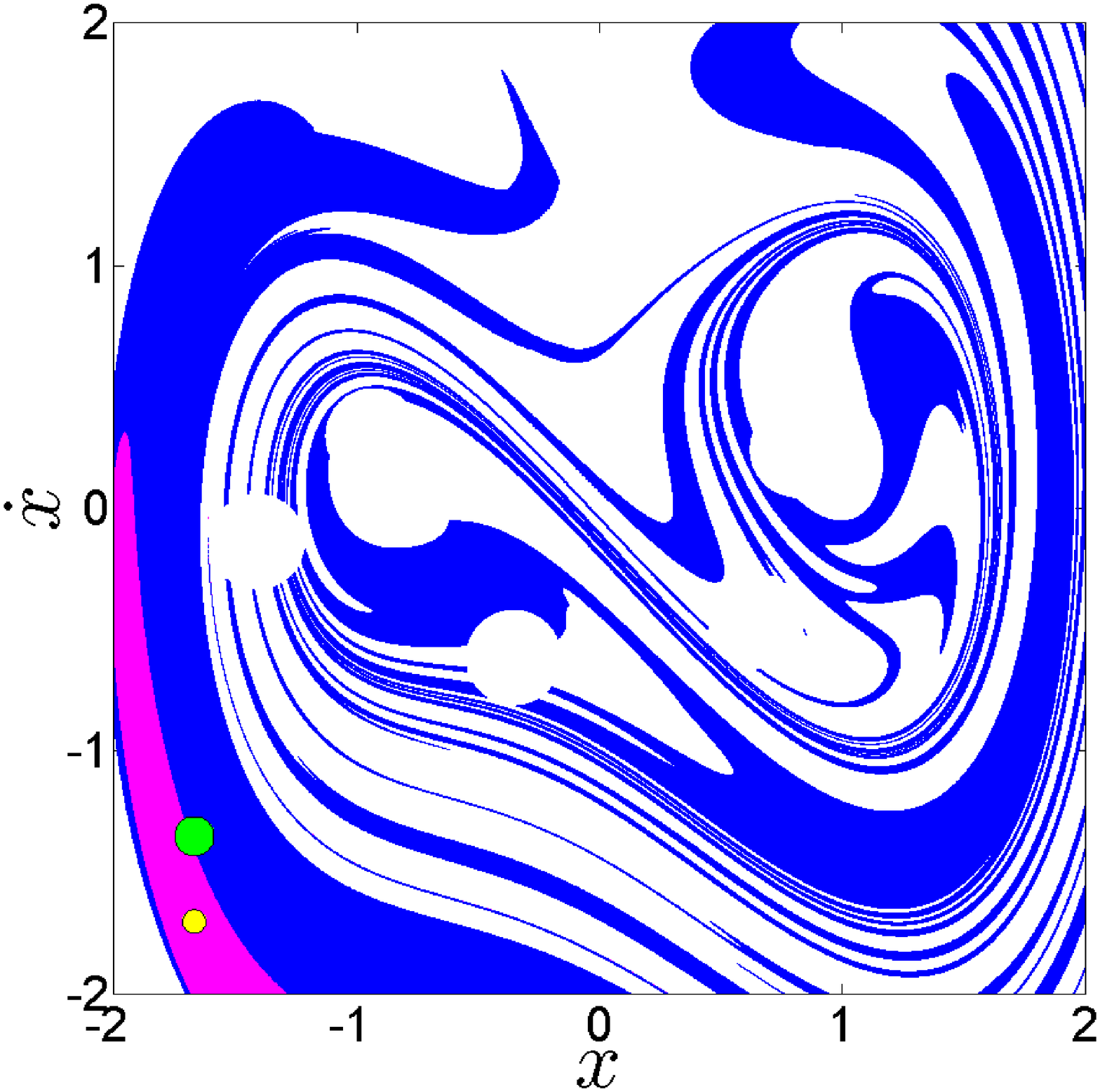}}
\subfigure[]{\includegraphics[width=0.3 \textwidth]{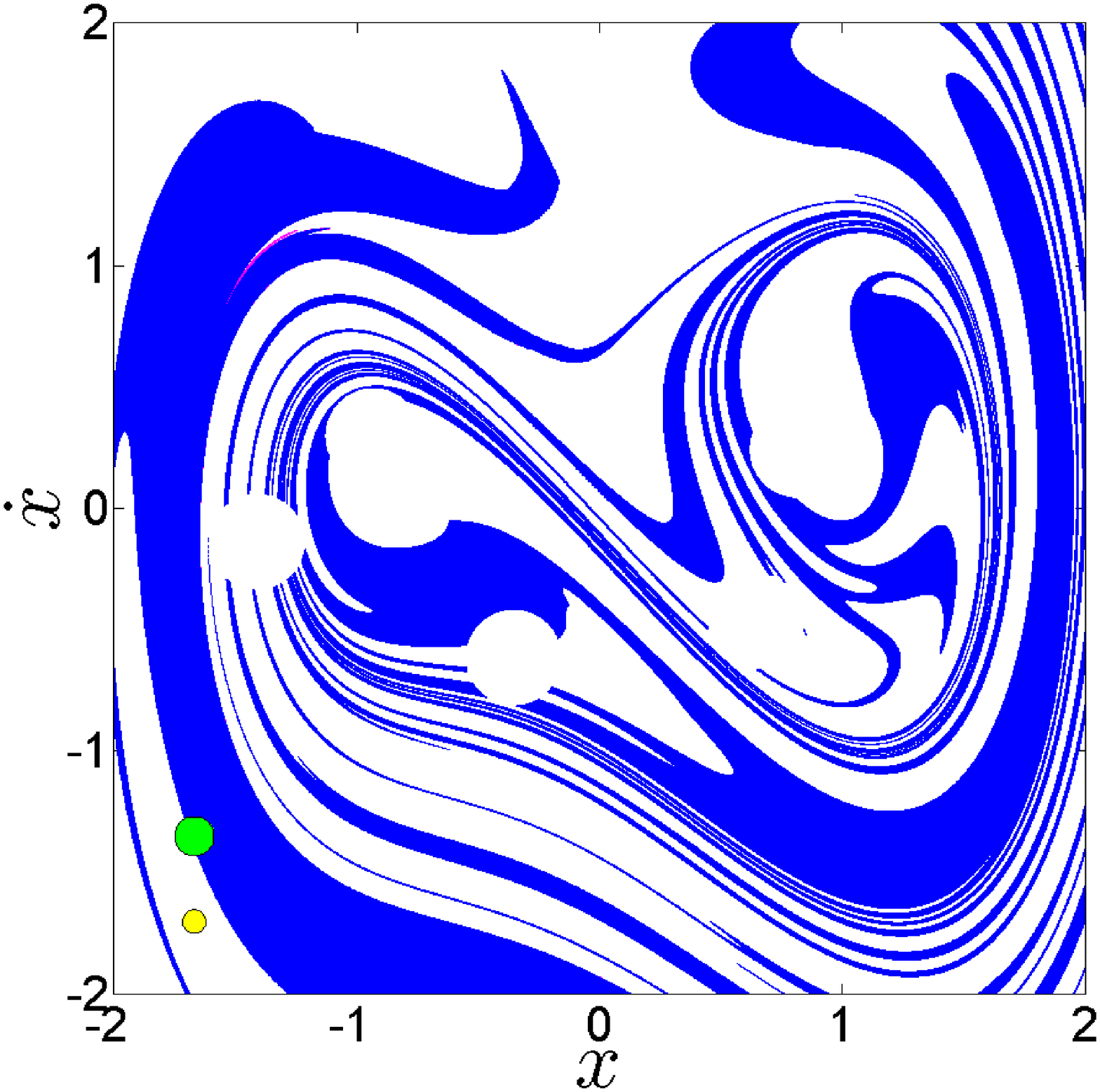}}
\caption{\textbf{Application of the Sculpting Algorithm to the Duffing oscillator}. This figure shows the sequence for creating the safe set in the Duffing oscillator. At each step, part of $Q_n$ is removed. The blue color represents the part of the set that remains and the magenta the part that is to be removed. Only 12 iterations out of the 15 appear. The small green and yellow circle represents the intensities of the control and disturbance used ($u_0=0.0475$ and $\xi_0=0.08$).
}\label{6z}
\end{center}
\end{figure}

To apply the partial control method to the Duffing oscillator, we need to clarify  the concept of escape in this case. The region from which all the trajectories escape will be the square $[-2,2]\times[-2,2]$ minus certain holes around the periodic attractors. We say there is an escape here if a given trajectory enters one of the circles or if it leaves the square. Then we use a grid of $6000\times 6000$ points in the square $[-2,2]\times[-2,2]$ as our initial set as in Fig.~\ref{5z}(c), but removing the zones that we want to avoid, that is, the circles of radius 0.2.

\begin{figure}
\begin{center}
\includegraphics[width=0.5 \textwidth]{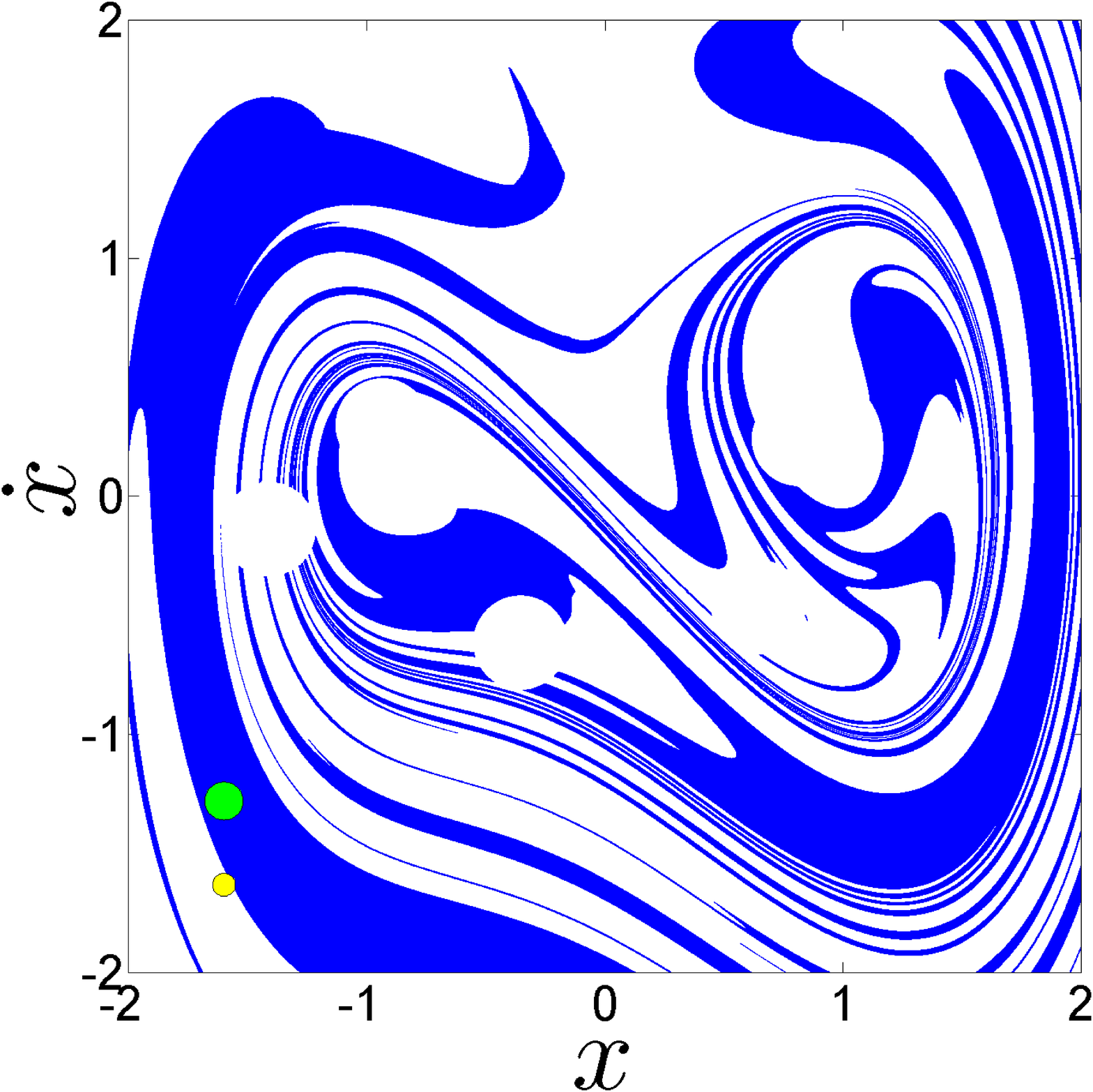}
\caption{\textbf{Safe set of the Duffing oscillator}. In this figure we can see the result of applying the Sculpting Algorithm to the Duffing oscillator
$\ddot{x}+0.15\dot{x}-x+x^{3}=0.245\sin(t)$. The safe set appears in blue. The minimum control allowed, so that it exits a safe set is $u_0=0.0475$ (yellow circle), with a maximum disturbance of $\xi_0=0.08$ (green circle). This is equal to a safe ratio $\rho\approx 0.59$.}\label{7z}
\end{center}
\end{figure}

The Safe-Set Sculpting Algorithm of intermediate sets when applied to the time-$2\pi$ map of the Duffing oscillator is shown in Fig.~\ref{6z}, though only $12$ iterations out of the $15$ appear. As it was mentioned earlier, the safe set has been computed for the parameter $u_0^{min}$ which corresponds to the smallest $u_0$ for which there is a safe set.

The final safe set obtained applying the Sculpting Algorithm to the initial set plotted in Fig.~\ref{6z}, is shown in Fig.~\ref{7z}, where the safe set appears in blue.

\subsubsection{An ecological model}

In this example, we have worked with an ecological model that describes the interaction between 3-species: resources, consumers and predators. The interest of this model lies in the fact that, for some choices of parameters, transient chaos appears involving the extinction of one of the species.  Without no control, the system evolves from a situation where the three species coexist towards a state where just two species survive, and predators get extinct.

\begin{figure}
\includegraphics [trim=0cm 0cm 0cm 0cm, clip=true,width=0.7\textwidth]{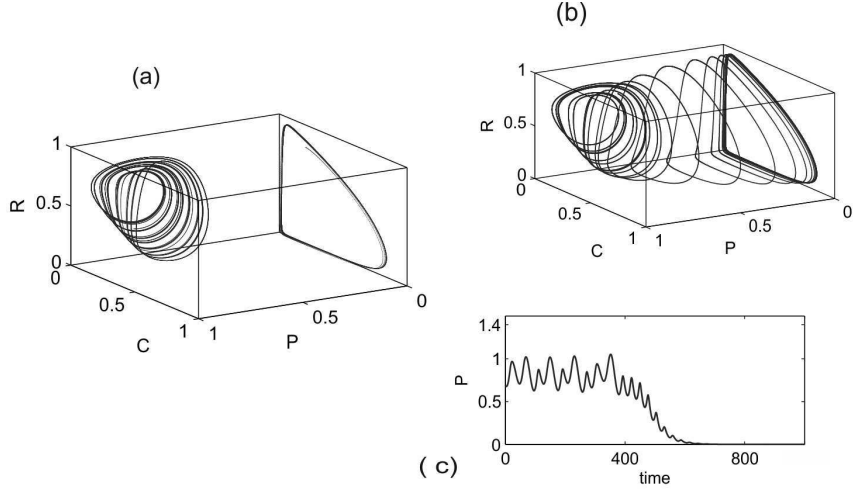}
\centering
\caption{\textbf{Dynamics of the extended McCann-Yodzis model proposed by Duarte et al from Eqs.~(3)}. Depending on the values of the parameters $(K,\sigma)$ different dynamics are possible. Fixing $K=0.99$, the boundary crisis appears at $\sigma_c=0.04166$. (a) Before the boundary crisis ($K=0.99$, $\sigma=0$), there are two possible attractors depending on the initial conditions: one chaotic attractor where the three species coexist, and one limit cycle where only the resources and consumers coexist. (b) After the boundary crisis ($K=0.99$, $\sigma=0.07$), the limit cycle is the only asymptotic attractor. (c) Time series of the predators population corresponding to the case $(b)$, where the chaotic transient before the extinction is shown.}
\label{8z}
\end{figure}

The model that we have used is an extension of the McCann-Yodzis model \cite{Yodzis} proposed by Duarte et al. \cite{Duarte}, which describes the dynamics of the population density of a resource species $R$, a consumer $C$ and a predator $P$. The resulting model is given by the following set of nonlinear differential equations:

\begin{eqnarray}
  \frac{dR}{dt}&=&R\left(1-\frac{R}{K}\right )-\frac{x_c y_c CR}{R+R_0}  \nonumber \\
   \frac{dC}{dt}&=&x_c C\left(\frac{y_c R}{R+R_0}-1\right)-\psi(P) \frac{y_p C}{C+C_0} \\
   \frac{dP}{dt}&=&\psi(P) \frac{y_p C}{C+C_0}-x_p P. \nonumber
\end{eqnarray}
Note that $R$,$C$ and $P$ are non-dimensional variables. Following McCann and Yodzis, (1995) and Duarte et al., (2009), we have fixed the ecological parameters : $x_c=0.4$, $y_c=2.009$, $x_p=x_i=0.08 $, $y_p=2.876 $, $R_0=0.16129$, $C_0=0.5$, $K=0.99$ and $\sigma=0.07$. For these values transient chaos behaviour appears, and the predators eventually get extinct (see Fig ~\ref{8z}).

\begin{figure}
\includegraphics [trim=0cm 0cm 0cm 0cm, clip=true,width=0.7\textwidth]{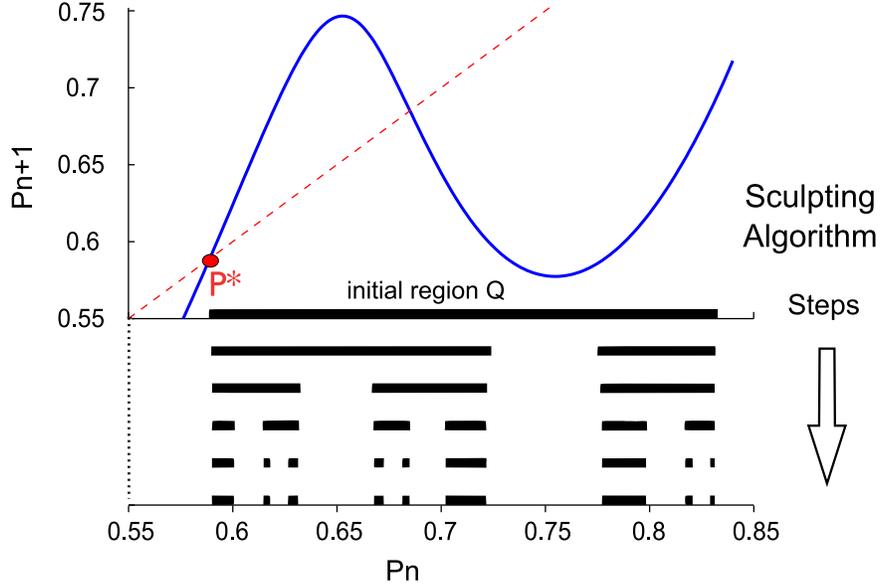}
\centering
\caption{\textbf{Return map $P_{n+1} =f(P_n)$ map obtained by using the successive local minima of the time series $P(t)$}. Notice that below $P^{*}=0.589$ the trajectory asymptotes to zero. In order to keep the trajectory in the region $Q$ indicated we compute the safe set. In the lower part it is shown the steps of the Sculpting Algorithm that converges to the final safe set. The horizontal \textit{black bars} helps us to visualize the process and represent the points $P_n$ that satisfy the condition to be a safe point at each step. In this case, the upper bound disturbance and control used are $\xi_0 =0.0114$ and $u_0  =0.0076$, respectively.}
\label{9z}
\end{figure}

With the aim to avoid the extinction, we applied the partial control method \cite{Ecology}. First we have constructed a map. Different choices are possible, but in this case we have chosen to build the map with the successive local minima $(P_n,P_{n+1})$, where $P_n$ denotes the $n$th local minimum. This set of points generates an approximately one-dimensional curve of the form  $P_{n+1}=f(P_n)$, shown in Fig.~\ref{9z}. Notice that, the iterates of any initial point for which $P_n > P^*$, follow a chaotic dynamics until they finally asymptotes to zero when it crosses a critical value $P_n < P^*$, which actually implies the extinction of the predators population. The map constructed through this way is just an approximation, so we also introduce a disturbance term $\xi_n$ into the map, in order to model potential mismatches.

  After introducing the disturbance term $\xi_n$ and the control term $u_n$ in the map, the partially controlled dynamics is given by:

\begin{eqnarray}
P_{n+1}=f(P_n)+\xi_n+u_n.
\end{eqnarray}

In our case, we want to sustain the dynamics close to the chaotic attractor, avoiding the escape produced when $P_n < P^*=0.589$, therefore we choose the initial $Q$ region to be the interval $P_n \in [0.589,0.84]$ indicated in Fig.~\ref{9z}. Then we use the Sculpting Algorithm to find the safe set. The computation of the safe set depends on the chosen values of $\xi_0$ and $u_0$. As an example, we have chosen for our simulations $\xi_0=0.0114$ and $u_0=0.0076$, where $u_0$ is very close to the minimum value for which the safe set exists. In Fig.~\ref{9z} we represent the steps of the algorithm to build the safe set.

\begin{figure}
\includegraphics [trim=0cm 0cm 0cm 0cm,width=0.7\textwidth]{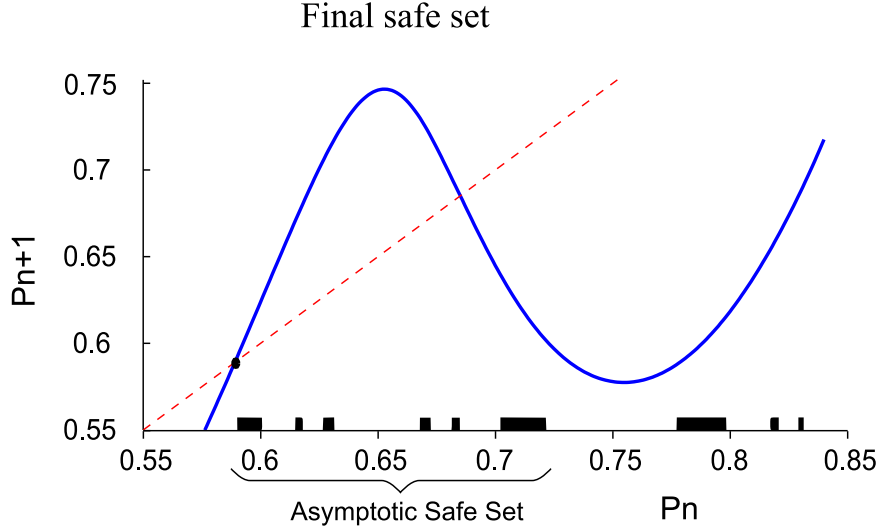}
\centering
\caption{\textbf{Final safe set}. The safe set is composed of different subsets obtained with the Sculpting Algorithm using $\xi_0 = 0.0114$ and $u_0 = 0.0076$. We also indicate the group of subsets where the dynamics remains trapped, that is, the asymptotic safe set.}
\label{10z}
\end{figure}

In Fig.~\ref{10z}, we represent the obtained final safe set that allows us to control the map constructed with the minima of variable $P$. Notice that from the point of view of real dynamics (continuous trajectories), the control is applied  every time  the trajectory crosses the set of the minima. If the value $f(P_n)+\xi_n$ is inside a safe set, we do not apply the control, and if it is outside, we relocate it inside the nearest safe point, resulting the new safe point $P_{n+1}=f(P_n)+\xi_n+ u_n$. The criterion to control the point to the nearest safe set is only an option, since in most cases there are other possible points belonging to the safe set which we can reach without exceeding the upper bound of control. From an ecological point of view, this flexibility allows us to choose the better option  considering our specific needs. For example, depending on our ease to stocking or harvesting individuals we can make the choice which involves the smallest effort.

When carrying out the numerical simulations, after some iterations, it is possible that the controlled trajectory does not visit certain regions of the safe set. This subset of the safe set is called the \textit{asymptotic safe set} (see Fig.~\ref{10z}), and it appears typically when the system is dissipative. In Hamiltonian systems, the asymptotic safe set and the safe set overlap \cite{Asymptotic}.

\begin{figure}
\includegraphics [trim=1cm 0cm 0cm 0cm,width=1\textwidth]{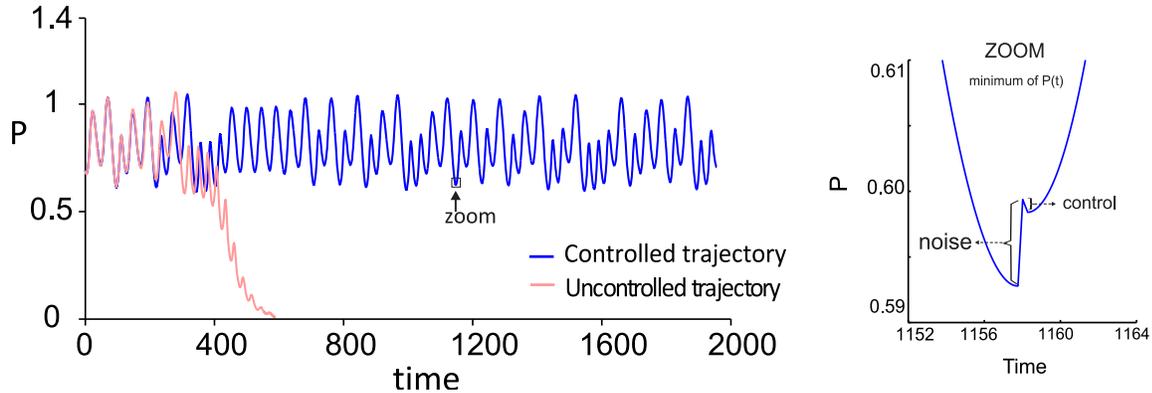}
\centering
\caption{\textbf{Controlled time series controlled}. Red line: Time series of the predators population without control exhibiting a escape towards zero, what implies the extinction of the predators. Blue line: Controlled time series of the predators population where the extinction is avoided. At every minimum, the value of $P$ is evaluated and if necessary a small control is applied. This time series corresponds to $50$ iterations in the return map $P_{n+1}=f(P_n)$. A zoom of one of the minima of the time series of $P(t)$ is also shown on the right in order to see how the noise is introduced and how the corresponding control is applied.}
\label{11z}
\end{figure}

To control the trajectory with the safe set, we proceed in this way: when the trajectory reaches a minimum of the $P$ variable, we evaluate if this point belongs to the safe set or not. If not, we apply control to shift it to the nearest safe point. In Fig.~\ref{11z}, we represent the corresponding controlled time series of the predators population (blue line) in contrast to the uncontrolled trajectory (red line), involving the extinction. On the right of the figure we also represent a zoom of one minimum to highlight how the noise (that we call disturbance)  appears and how the control is applied. Notice that the amplitude of noise shown only represents the difference between the deterministic trajectory and the noisy one.

\subsubsection{The Lorenz system}

\begin{figure}
\includegraphics [clip=true,width=0.9\textwidth]{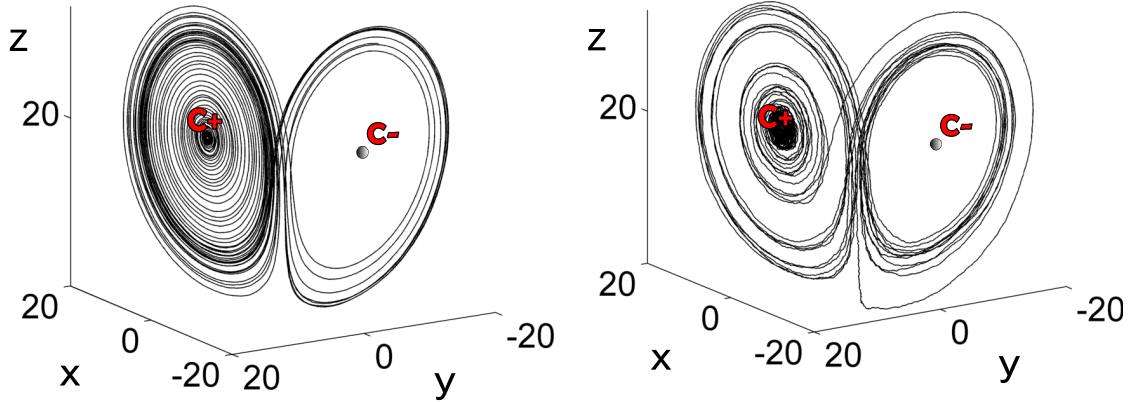}\\
\centering
\caption{\textbf{Dynamics of the Lorenz system}. We select the transient chaotic regime with $\sigma=10$, $b=8/3$ and $r=20$. On the left, the trajectory is deterministic. On the right, the trajectory is affected by some disturbances. The disturbances here, were enlarged in order to help the eye.  Almost all trajectories eventually spiral to one of the two attractors ($C^{+}$ or $C^{-}$). Here both trajectories spiral to $C^{+}$.}
\label{12z}
\end{figure}

In this section we have chosen the Lorenz system \cite{Lorenz}, which is one of the best known models in nonlinear dynamics. This system is a flow, that describes a simplified model of atmospheric convection.  The model consists of three ordinary differential equations,
\begin{eqnarray}
  \dot{x}&=&-\sigma x+\sigma y  \nonumber \\
   \dot{y}&=&-x z + rx-y \\
   \dot{z}&=&x y-b z. \nonumber
\end{eqnarray}
Depending on the parameter values $r$, $\sigma$, and $b$, the system can exhibit different dynamical behaviors, either periodic solutions, chaotic attractors or even transient chaos. Fixing $\sigma=10$, $b=8/3$, transient chaos can be found in the interval $r\in[13.93,24.06]$ as described in \cite{KaplanYorke,YorkeYorke}. For our simulations, we have chosen the value $r=20.0$. In this regime, as we show in Fig.~\ref{12z}, there are transient chaotic orbits that eventually decay towards one of the two point attractors which physically represent a steady rotation of a fluid flow, one clockwise, and the other counterclockwise. The point attractors are located in the following positions,
\begin{eqnarray}
&C^+=&(\sqrt{b(r-1)},\sqrt{b(r-1)},r-1)\approx(7.12, 7.12, 19)  \nonumber \\
&C^-=&(-\sqrt{b(r-1)},-\sqrt{b(r-1)},r-1)\approx(-7.12, -7.12, 19), \nonumber
\end{eqnarray}
In this figure, we also represent the case where some noise is present in the trajectory. The noisy trajectory behaves similarly to the deterministic one. The main difference is the time involved to reach the attractors that can be increased or reduced.

The goal of applying control here is to avoid trajectories falling in one of the attractors. To apply the control method, we need first to built a map, however we have found many interesting possibilities. In this sense we explore three of them. First, we apply this method by building a 1D map using the successive maxima of one of the variables. Next, we implement it by building a 2D map through a Poincar\'{e} section. Finally, we built a 3D map, which has the advantage of using a fixed time interval between application of the control, which can be useful for practical applications.

\begin{figure}
\includegraphics [trim=0cm 0cm 13cm 0cm, clip=true,width=0.7\textwidth]{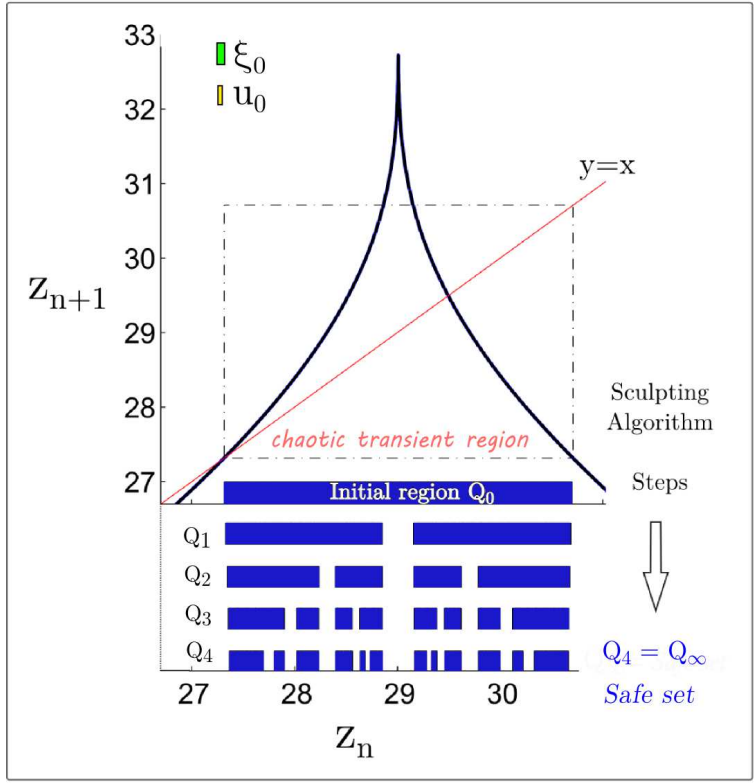}\\
\centering
\caption{\textbf{The 1D safe set}. The black curve is the 1D map built with the successive maxima of $z$. We take as initial set $Q_0$ (upper segment in blue) the region where transient chaos occurs. The map is affected by disturbances with an upper bound $\xi_0=0.080$, while we choose the upper bound of the control as $u_0=0.055$, (the bounds are the width of the bars displayed in the upper left side). The figure shows the successive steps computed by the Sculpting Algorithm, from an initial region $Q_0$ until it converges to the subset $Q_4=Q_\infty \subset Q_0$. We use a grid of $4000$ points in the interval $z_n \in [26.8,30.8]$, that corresponds to a resolution of $0.001$.}
\label{13z}
\end{figure}

  The 1D map:

  As shown by Lorenz \cite{Lorenz}, when plotting the pairs $(z_n,z_{n+1})$, one gets (approximately) a function $f$ where $z_{n+1}\approx f(z_n)$. We can see this clearly in Fig.~\ref{13z}. Knowing a local maximum of $z$ is $Z$, allows one to estimate $|x|$ and $|y|$ with considerable precision. Transient chaos can be observed in the interval $z_n \in [27.3,30.7]$, so we have chosen this interval as the set $Q_0$. We have taken $\xi_0=0.080$ and the control bound $u_0=0.055$ ($u_0<\xi_0$). This control value is approximately the minimum value for which a safe set exists. Then, we have obtained the safe set by using the recursive Sculpting Algorithm. In Fig.~\ref{13z}, we can see how the algorithm sculpts the initial region $Q_0$ until it finds $Q_4$ where it converges, so $Q_4=Q_\infty$ is the safe set. For this computation we have used a grid of $4000$ points in the interval $z_n \in [26.8,30.8]$, so the grid resolution is $0.001$.

\begin{figure}
\includegraphics [trim=0cm 0cm 0cm 0cm, clip=true,width=0.7\textwidth]{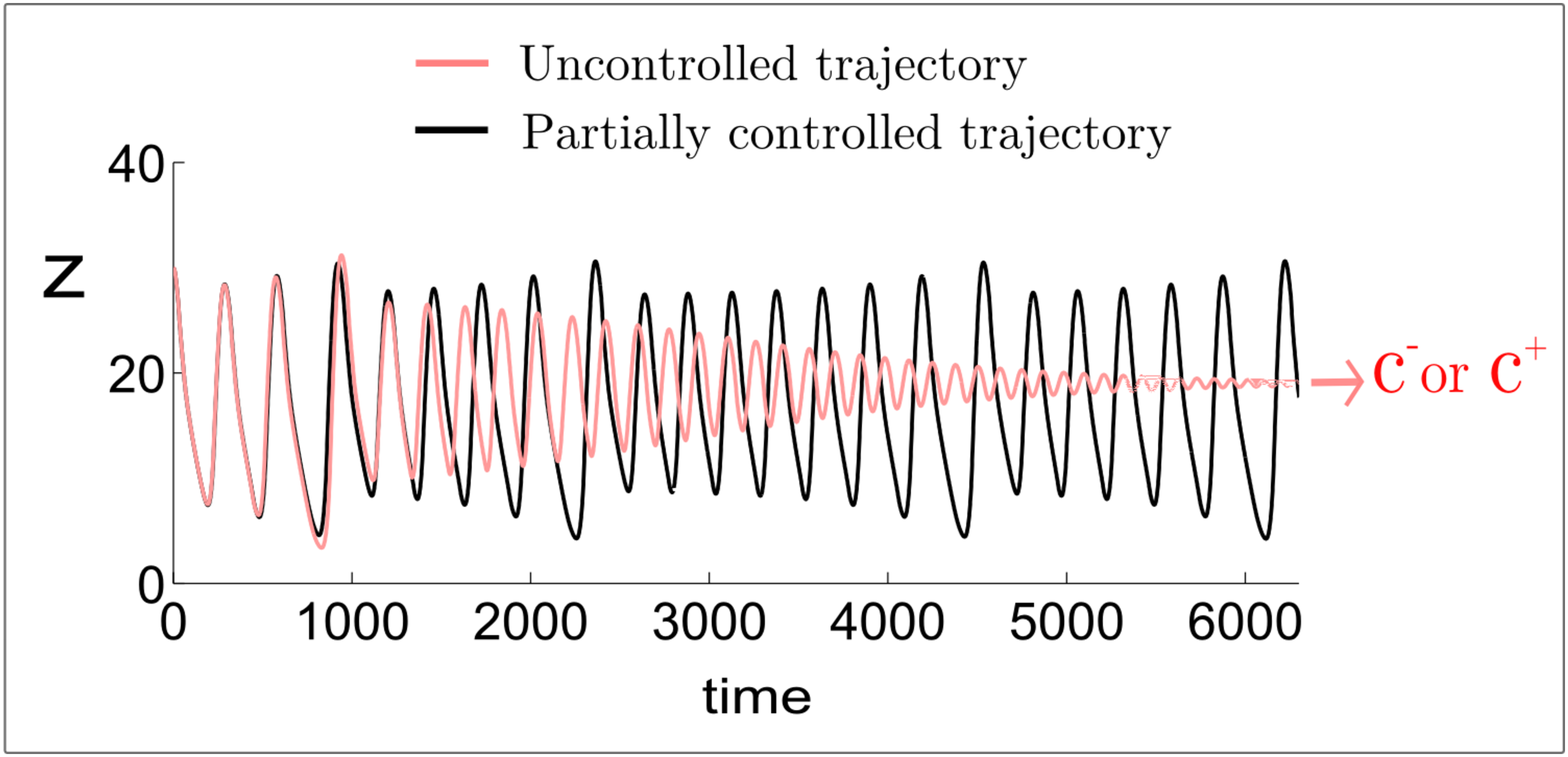}\\
\centering
\caption{\textbf{Time series of the variable $z$ for the Lorenz system with $r=20$.} The figure shows a comparison between an uncontrolled trajectory that escapes from chaos (red line) and a partially controlled trajectory (black line). Starting with the same initial condition, the uncontrolled trajectory eventually decays to $C^{+}$ or $C^{-}$, which physically means a steady rotation of the fluid flow. On the other hand the partially controlled trajectory is maintained in the chaotic transient regime, that is, the rotation of the fluid flow remains chaotic forever.}
\label{14z}
\end{figure}

  In Fig.~\ref{14z} we show a controlled time series of the $z$ variable in contrast with an uncontrolled trajectory. We can see that chaos is sustained by applying small perturbations in the maxima of the variable $z$.  Although the map only contains the variable $z$, the control in the original phase space must be applied in the three variables. The reason is because each local maximum of $z$ is described by $3$ coordinates $(x_m,y_m,z_m)$, and the coordinates $x_m$ and $y_m$ vary from maximum to maximum.  Sometimes, if this variation is negligible, it is not necessary to apply control in these coordinates as in the case of the ecological model example where we have applied control to one of the variables (the predators species). The main advantage of this 1D approach is that the computation of the safe set is easy and fast. This kind of map is useful when the disturbed trajectories mainly spread out along the expanding direction of the chaotic saddle, as it occurs in the case of stochastic noise or uncertainties in the application of the controls.

\begin{figure}
\includegraphics [trim=0cm 0cm 0cm 0cm, clip=true,width=0.6\textwidth]{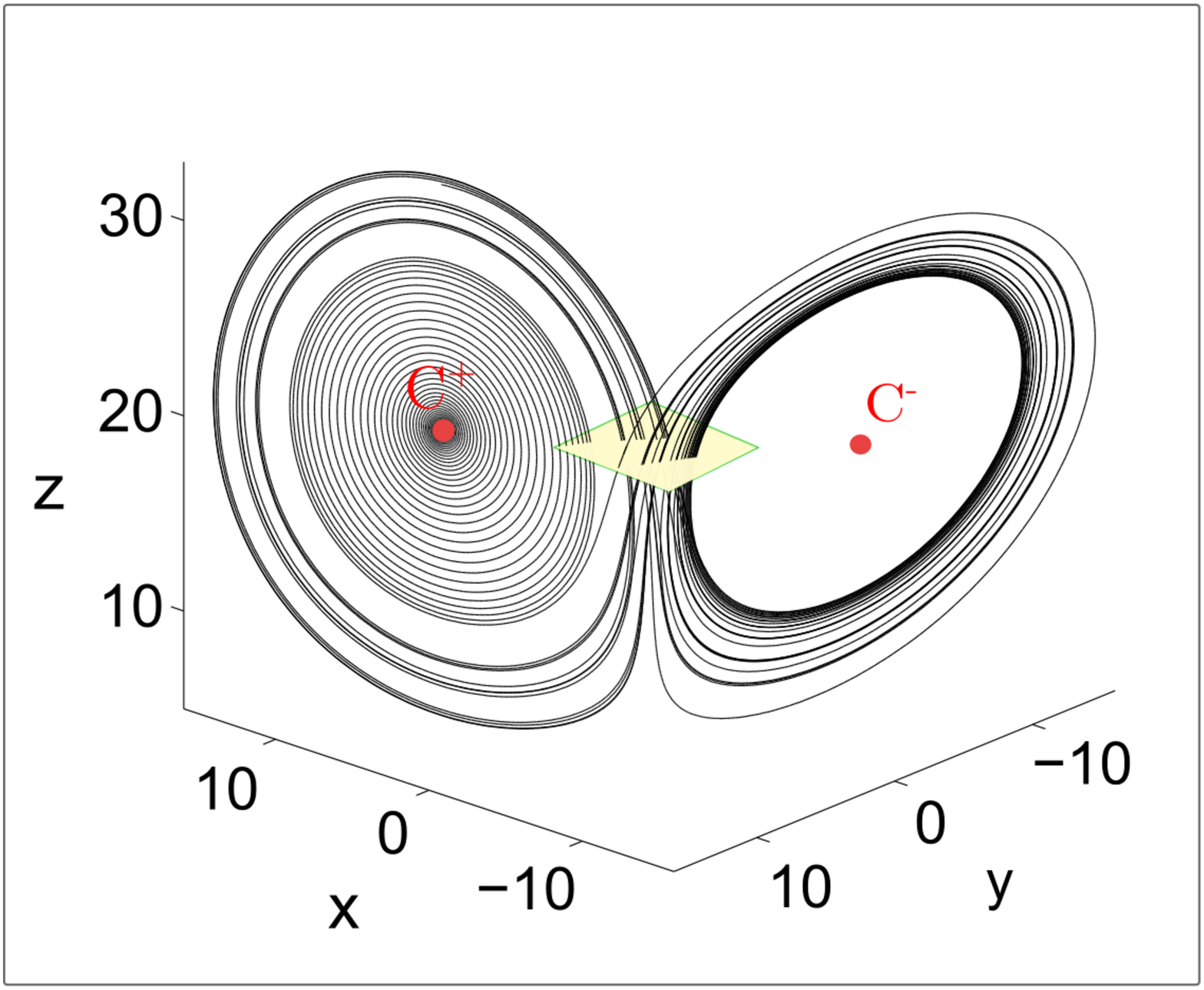}\\
\centering
\caption{\textbf{The Lorenz system with $r=20$ (transient chaos).} The figure shows an uncontrolled trajectory in phase space crossing a square with $x\in[-3,3]$ and $y\in[-3,3]$ in the plane $z=19$. To built the map, we use a grid of initial conditions in the plane, and evaluate the images of the trajectories when they cross again the plane. The goal of the control will be to keep the trajectories in this plane, avoiding the escape to one of the attractors $C^+$ or $C^-$, placed outside.}
\label{15z}
\end{figure}

  The 2D map:

  It is straightforward to build a 2D map taking a Poincar\'{e} section that intersects the flow. For our purpose, we have chosen the plane $z=19$ with the ranges $x\in[-3,3]$ and $y\in[-3,3]$, as shown in Fig.~\ref{15z}. The trajectories that cross this plane are in the transient chaotic regime, while the attractors $C^+=(7.12, 7.12, 19)$ and $C^-=(-7.12, -7.12, 19)$ that we want to avoid, are situated outside this plane (see the location in Fig.~\ref{15z}). For this reason, we have taken as $Q=Q_0$, the square $x\in[-3,3]$ and $y\in[-3,3]$ in the plane $z=19$. Then we have used the Sculpting Algorithm to find the safe set $Q_\infty \subset Q$, designed to avoid the eventually decay to the attractors.

\renewcommand{\thesubfigure}{(\alph{subfigure})~}

\begin{figure}
\begin{center}
\subfigure[Safe set]{\includegraphics[trim=0cm 0cm -10cm 0cm, clip=true, width=0.43 \textwidth]{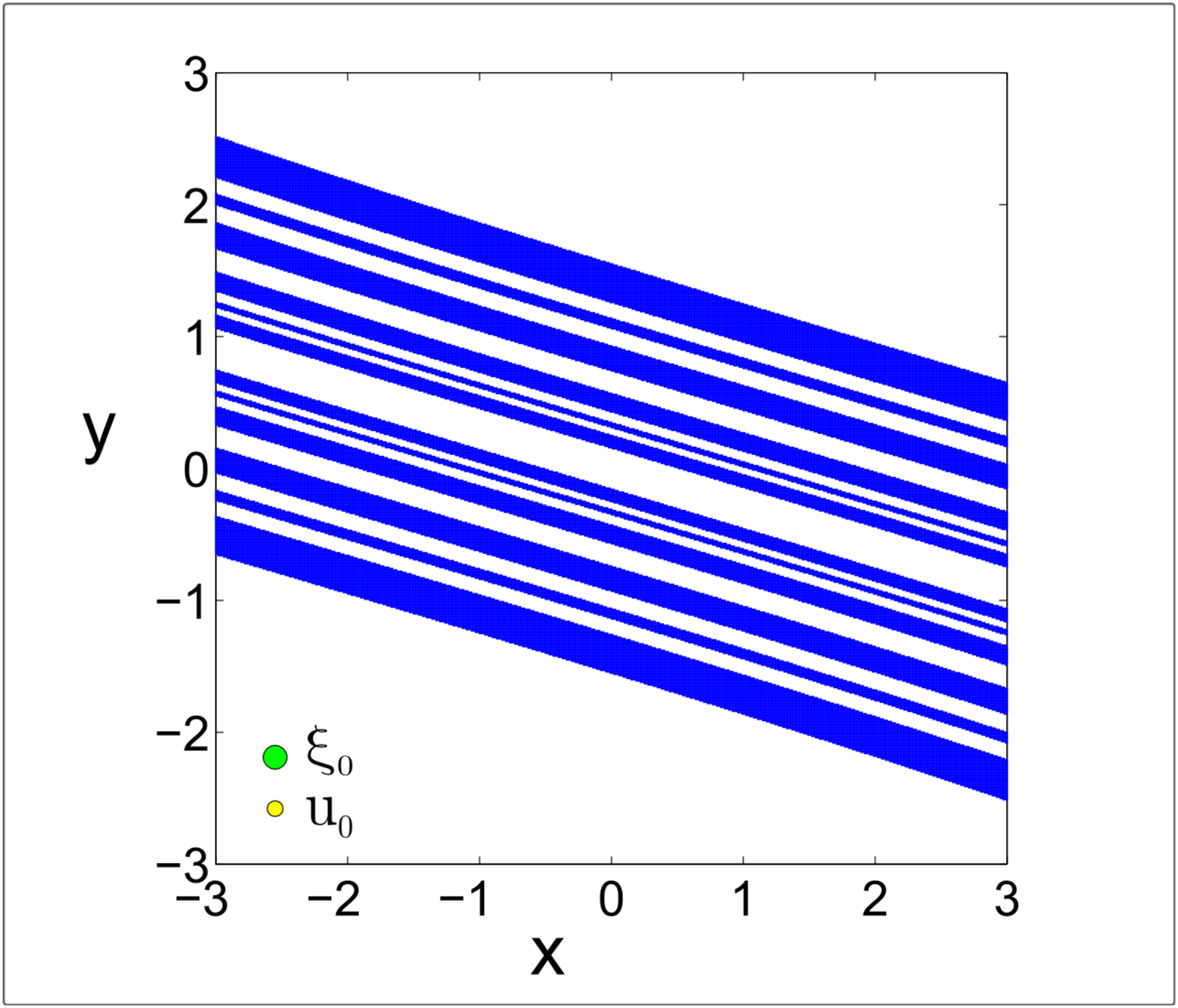}}
\subfigure[Safe set and controlled trajectory]{\includegraphics[width=0.33 \textwidth]{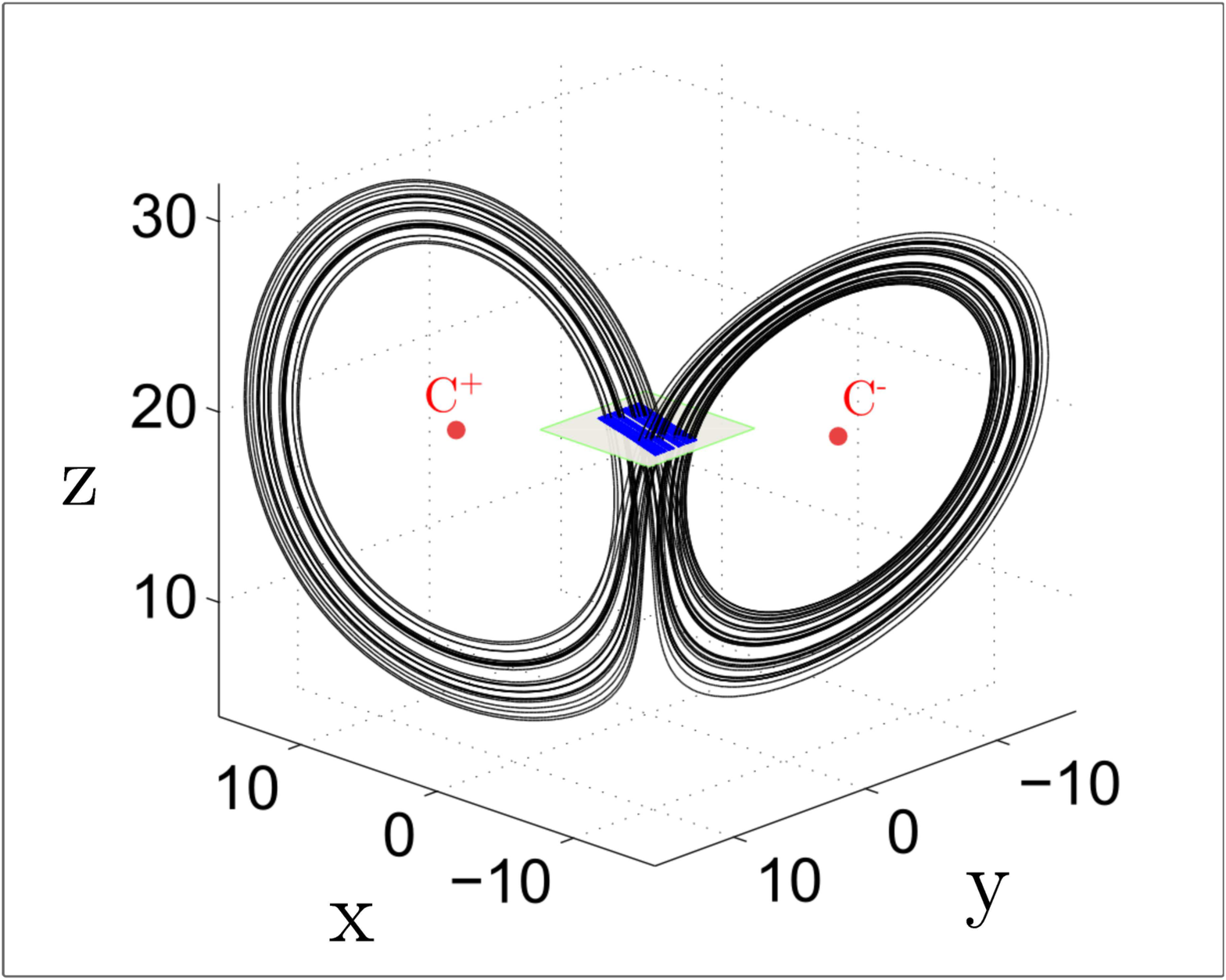}}
\subfigure[Zoom of the controlled trajectory]{\includegraphics[trim=0cm 0cm 0cm -8cm, clip=true,width=0.32 \textwidth]{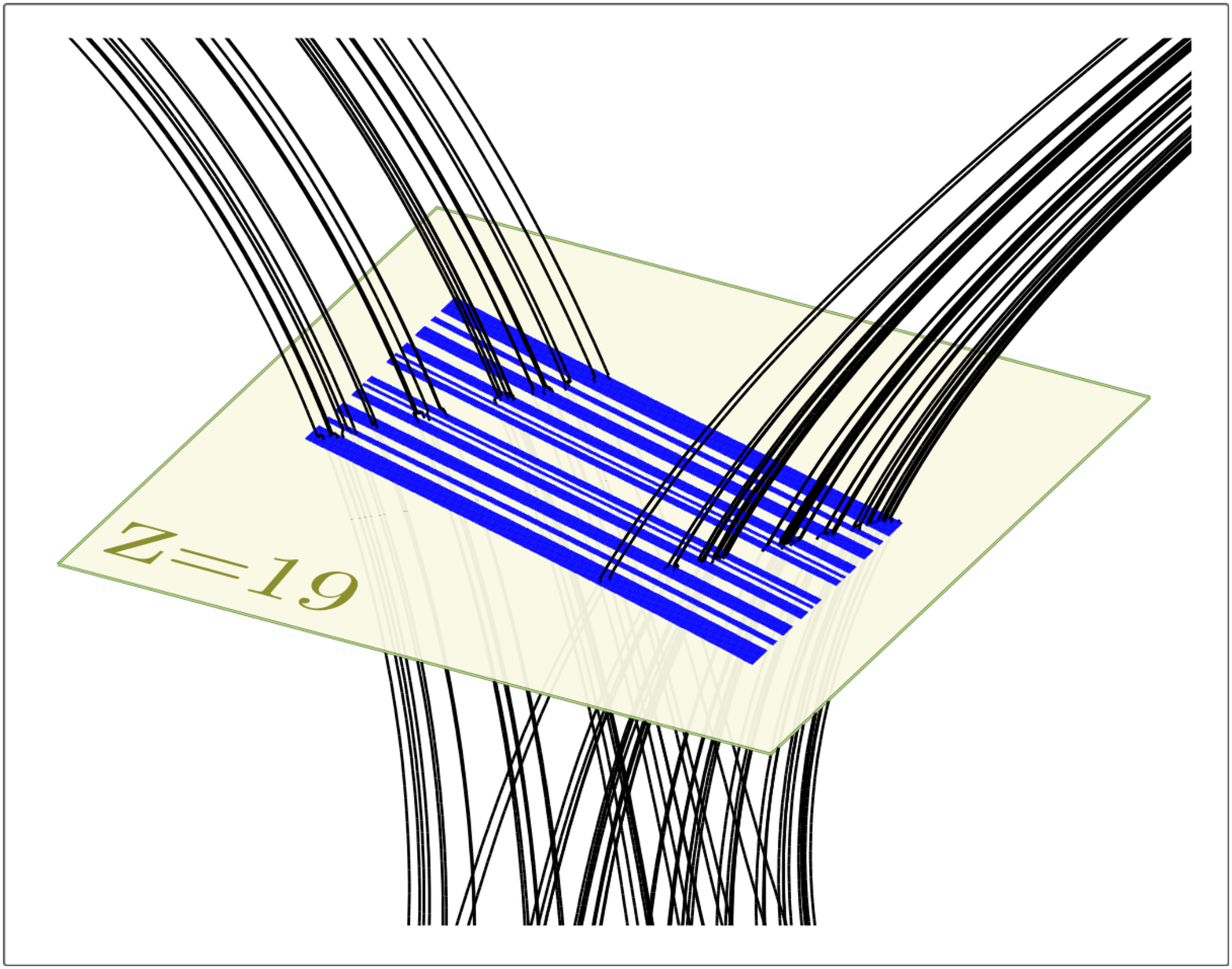}}
\\[20pt]
\caption{\textbf{The 2D safe set and how it is used to control the trajectory}. (a) The safe set obtained using the map built with the plane displayed in Fig.~\ref{15z}. We show in blue the computed safe set $Q_\infty$ for $\xi_0=0.09$ and $u_0=0.06$ ($u_0<\xi_0$). The grid size used is $1201 \times 1201$ points. The radius of the balls in the lower left side indicates the bounds of the disturbance, $\xi_0=0.09$ (green) and the control $u_0=0.06$ (yellow). (b) A partially controlled trajectory in phase space for case. Each time that the trajectory crosses the safe set plane (placed in $z=19$), the control is applied pushing the trajectory onto the set  avoiding the escape from chaos. (c) Zoom of how the control is applied in the safe set.
}\label{16z}
\end{center}
\end{figure}

 As an example, we have assumed that the map is affected by some disturbances with upper bound $\xi_0=0.09$. Applying the Sculpting Algorithm, we have found the safe set for the minimum possible value of the control, that is $u_0=0.06$ ($u_0<\xi_0$). In Fig.~\ref{16z}(a), the resultant safe set is displayed. A partially controlled trajectory is represented in Fig.~\ref{16z}(b), where we have also shown the safe set in phase space in order to see how it is used to control the system. Notice that, we are able to avoid the attractors, applying only small perturbations in the plane. A zoom of this region is shown in Fig.~\ref{16z}(c). The computation was carried out taking a grid size of $1200 \times 1200$ points, (grid resolution is $0.005$ in both variables $x$ and $y$).

The main advantage of a map is that allows to partially control systems where all the variables are affected by disturbances since the image $x_{n+1} =f(x_n)+ \xi_n$ in the Poincar\'{e} surface is a certain ellipse, and both dimensions of the surface are controlled. In addition, as opposed to the 1D map, where we have to act on the $x$, $y$ and $z$ variables to control the system, the control in the 2D map is only applied in the variables $x$ and $y$, since $z$ is constant. This can be an advantage in systems where it is difficult or expensive to apply the control in each variable.

The 3D map:

The 1D approach as well as the 2D approach, have the disadvantage of having to track the trajectory to know when it passes through the control region, where we apply the control corrections. Another strategy is to use a time discretization of the Lorenz system, by taking a suitable time interval $\Delta t$ between the current state of the system and the future state, that is, $x(t_0), y(t_0), z(t_0) \rightarrow x(t_0+\Delta t),y(t_0+\Delta t),z(t_0+\Delta t)$. By computing the time-$\Delta t$ image of each point of a 3D grid that covers the phase space, we can obtain the 3D map. The choice of $\Delta t$ is important. The topological explanation for this, is that the flow is acting like a pastry transformation which takes some time to be completed. Once this time is reached, the safe set appears. For our Lorenz system, there are safe sets for values of $\Delta t \geq 1.2$.

\begin{figure}
\includegraphics [trim=0cm 0cm 0cm 0cm, clip=true,width=1\textwidth]{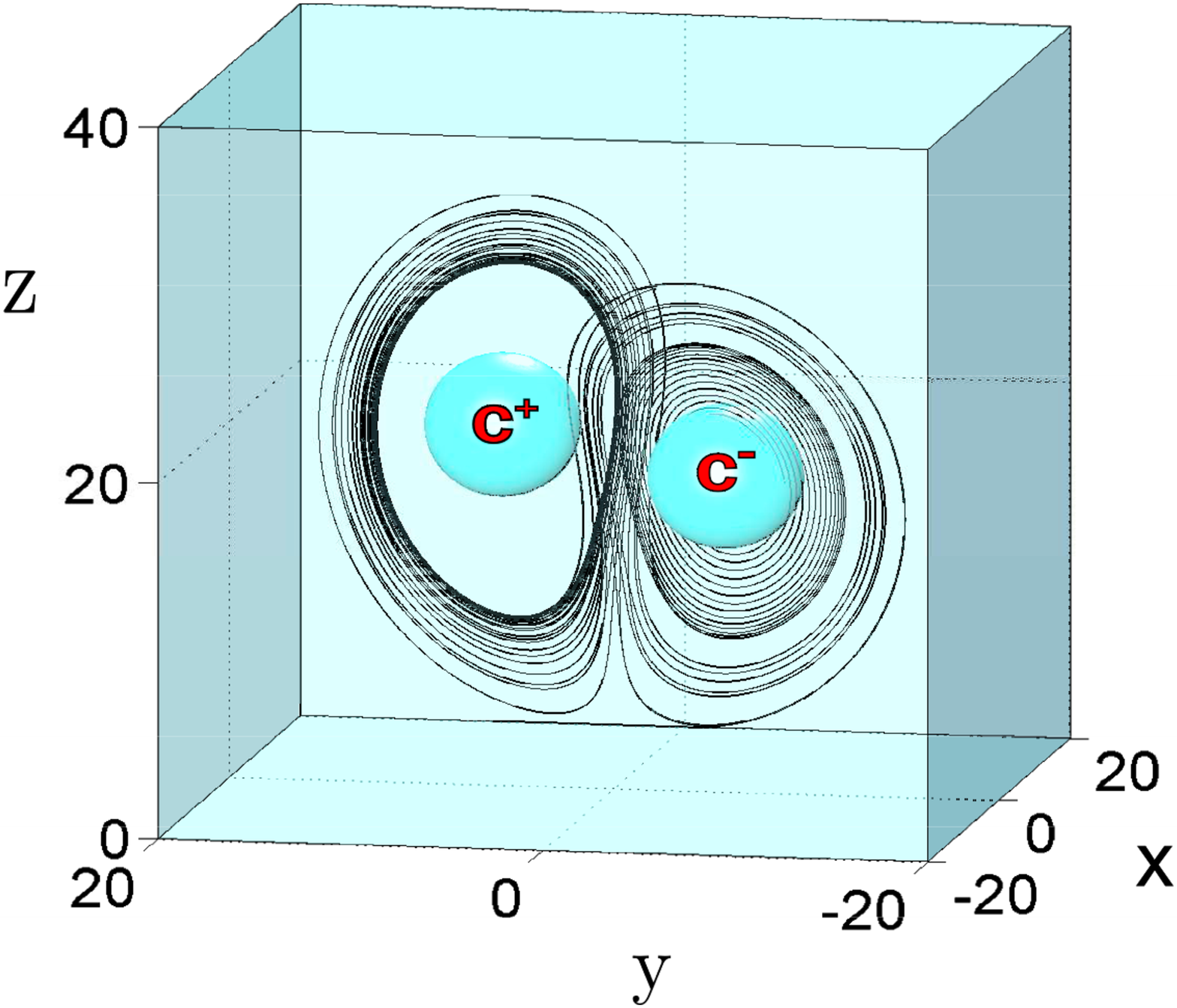}\\
\centering
\caption{\textbf{A choice of 3D set $Q$}. The 3D set $Q$ is the cube $x\in[-20,20]$, $y\in[-20,20]$, $z\in[0,40]$ except that the balls of radius $4$, centered in $C^+=(7.12, 7.12, 19)$ and $C^-=(-7.12, -7.12, 19)$ are removed from $Q$. We want trajectories to stay in $Q$ and not fall to these attractors. A trajectory is plotted to show the chaotic transient behavior in this region.}
\label{17z}
\end{figure}

\renewcommand{\thesubfigure}{(\alph{subfigure})~}

\begin{figure}
\centering
\subfigure[Safe set with $\Delta t=1.2$]{\includegraphics[trim=-1.3cm 0cm -1cm 0cm, clip=true,width=0.4\textwidth]{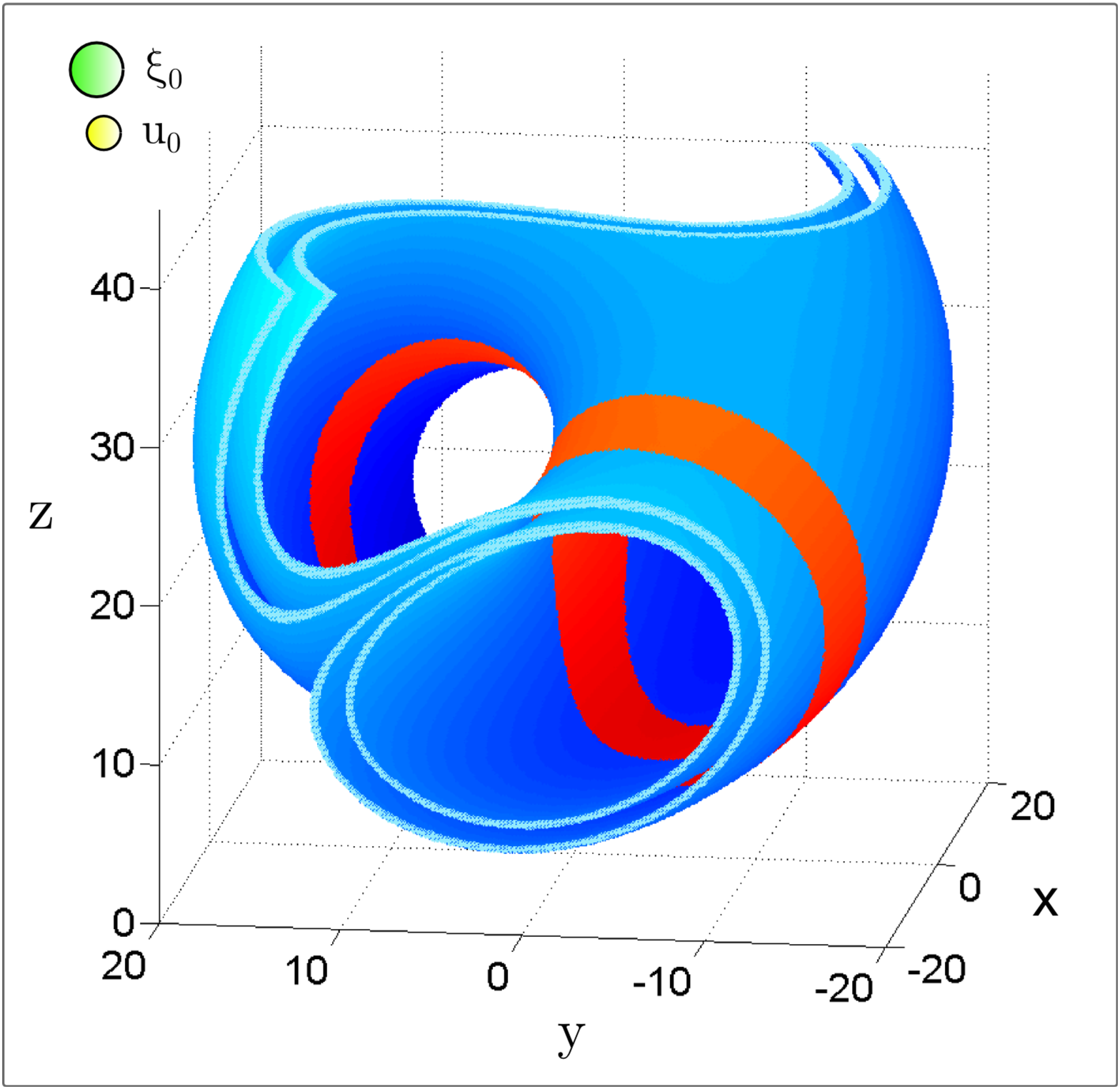}}
\subfigure[Asymptotic safe set]{\includegraphics[trim=-2cm 0cm 0cm 0cm, clip=true,width=0.4\textwidth]{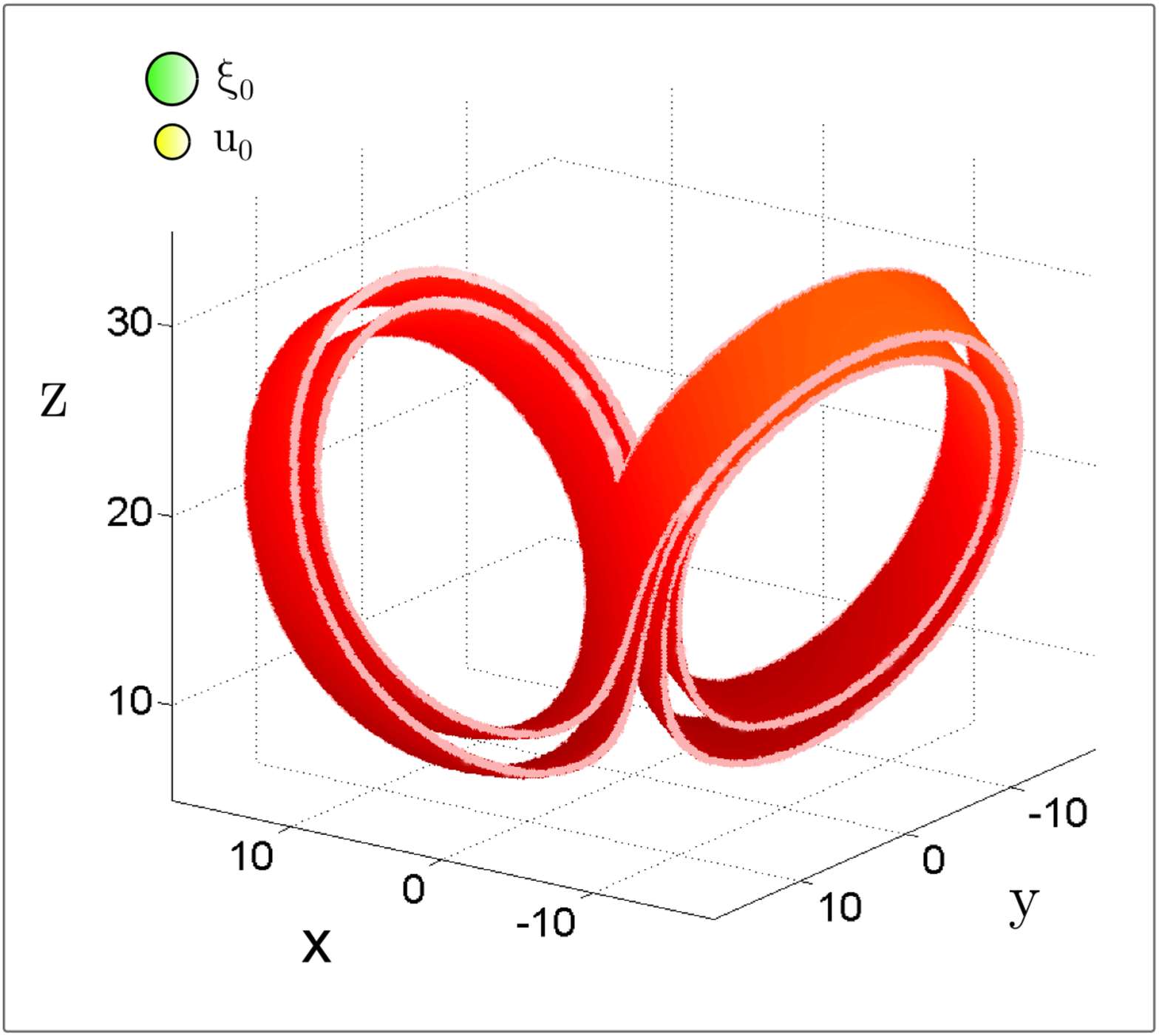}}
\subfigure[Controlled trajectory]{\includegraphics[trim=0cm 0cm -3cm -6cm, clip=true,width=0.41\textwidth]{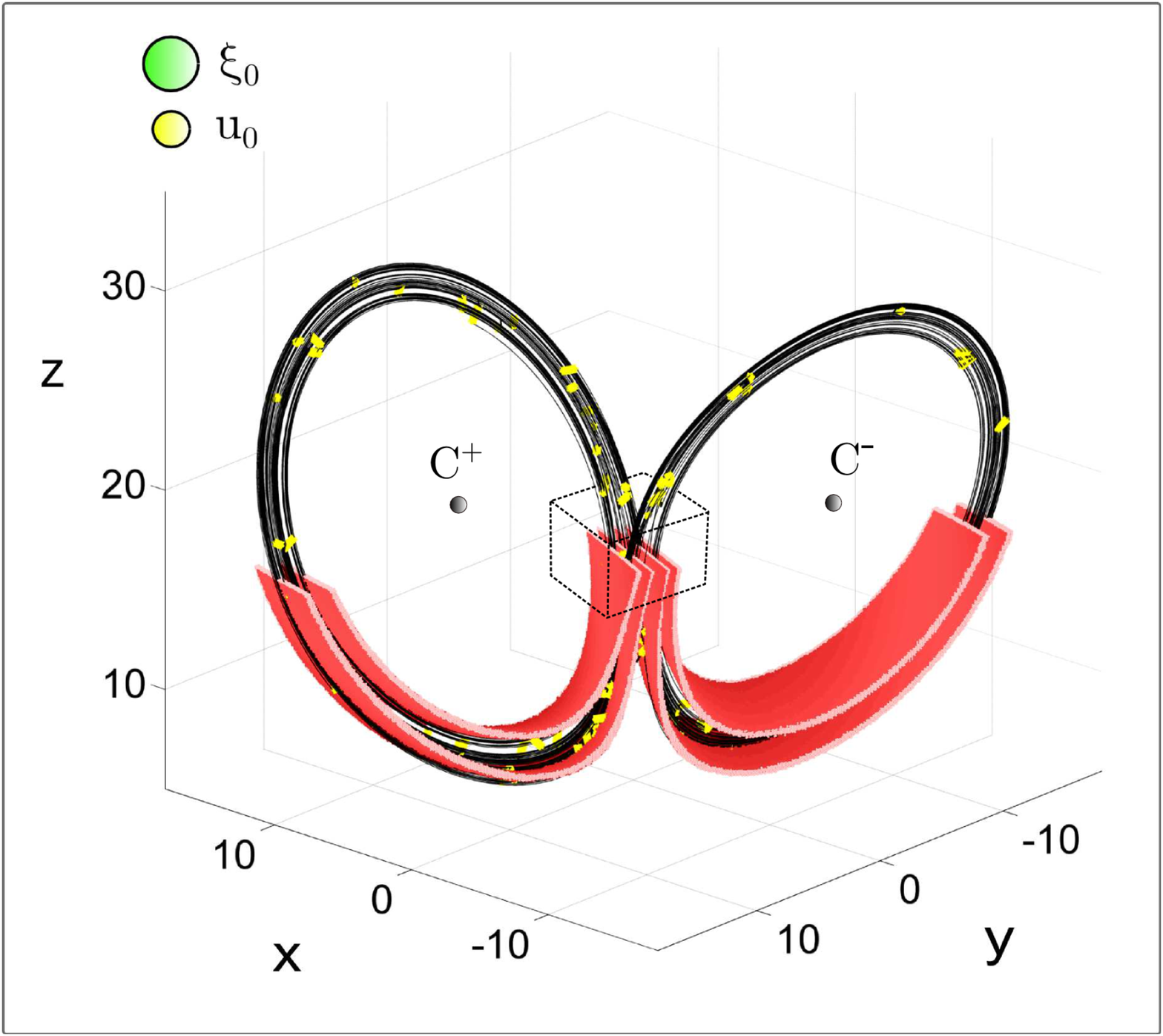}}
\subfigure[Zoom of the controlled trajectory]{\includegraphics [trim=-1cm 0cm 0cm -6cm, clip=true,width=0.365\textwidth]{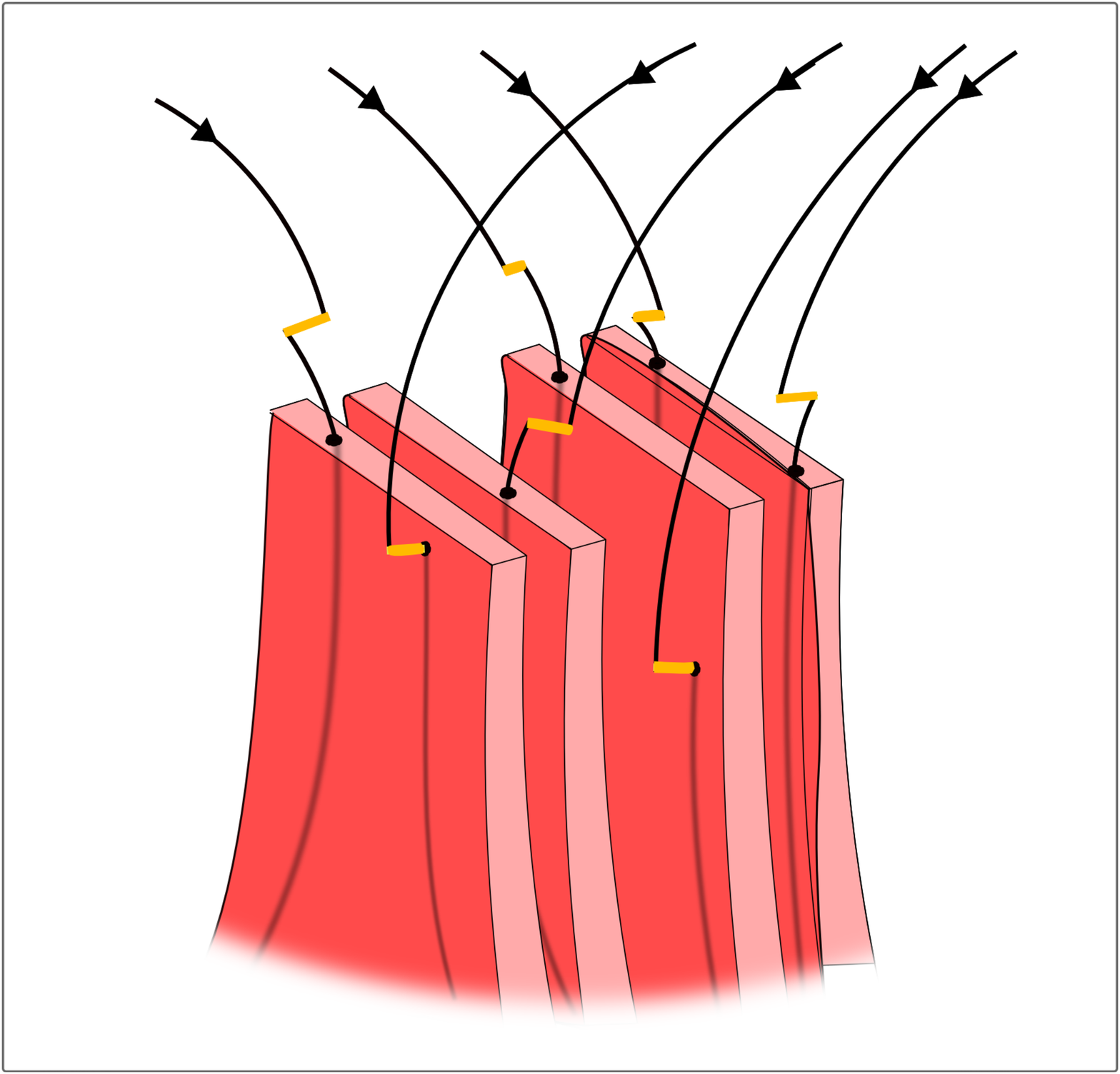}}
\\[20pt]
\caption{\textbf{The 3D safe set and how it is used to control the trajectory}. (a) In blue the 3D \emph{safe set} $Q_\infty$ for Fig.~\ref{17z}, obtained after applying the Sculpting Algorithm. We set $\Delta t=1.2$, $\xi_0=1.5$ ($\xi_0=$ radius of the green ball) and $u_0=1.0$ ($u_0=$ yellow ball's radius). In red the asymptotic safe set which is a subset of the safe set. This is the region in which the controlled trajectories eventually lie. (b) The asymptotic safe set alone. Partially controlled trajectories converge rapidly to this region. (c) A cut-away section of the asymptotic safe set in order to see a partially controlled trajectory (with $\Delta t=1.2$) displayed in black.  The controls (yellow segments inserted in the trajectory) are applied every $\Delta t=1.2$. As a result, the trajectory is kept in the chaotic region and the attractors $C^+$ and $C^-$ are avoided. (d) Zoom in the small cube displayed in Fig.~\ref{18z}(c). Only few lines are displayed for a better visualization. The controls (yellow segments) are applied to move the trajectories (in black) into the asymptotic safe set (in red).}
\label{18z}
\end{figure}

For a 3D example, we take the domain with $x\in[-20,20]$, $y\in[-20,20]$, $z\in[0,40]$, with a grid size of $400 \times 400 \times 400$, so the grid resolution is $0.1$ for each variable. In this region the transient chaotic trajectories eventually decay to the attractors $C^+=(7.12, 7.12, 19)$ and $C^-=(-7.12, -7.12, 19)$.  In order to avoid  $C^+$ and $C^-$, balls centered in these attractors are removed. See the region $Q$ and a transient chaotic trajectory in Fig.~\ref{17z}. To obtain the map, we have computed the image of each point of $Q$ with $\Delta t=1.2$. Then, as an example, we have taken the value $\xi_0=1.5$ and $u_0=1.0$  (note $u_0 < \xi_0$). After applying the  Sculpting Algorithm, the safe set shown in Fig.~\ref{18z}(a) is obtained.


To describe the controlled dynamics in the 3D map we write $q_n$ for the controlled trajectory at time $n \Delta t$. To obtain a particular trajectory we choose $\xi_n$ at random with $|\xi_n| \leq \xi_0$. Then we choose the convenient $u_n$, that places  $q_{n+1}=q_n +\xi_n + u_n$ in the safe set. In each case, $\xi_n$ represents the disturbance accumulated by the trajectory in the time interval $\Delta t$, while the control is always applied at a discrete time. In this case, we apply the minimum control, however other criterion is possible as long as the constraint $|u_n| \leq u_0$ is respected.

One interesting feature of the partial control method is that the controlled trajectories converge towards a certain region of the safe set, which is called the \textbf{asymptotic safe set} In Figs.~\ref{18z}(b) the asymptotic safe set  was drawn alone and in Fig~\ref{18z}(c) a half section of it to visualize the partially controlled trajectory inside. Notice that the trajectory does not leave the asymptotic safe set once they reach it, (unless the control is turned off). Once the dynamics converges, it is sufficient to use the asymptotic safe set to control the trajectories. The controls, represented as yellow segments distributed along the trajectory, are applied every $\Delta t=1.2$. We show this fact with a zoom in Fig.~\ref{18z}(d). As a result, the trajectories never fall into the attractors $C^+$ or  $C^-$, keeping the dynamics in the chaotic region forever.

\begin{figure}
\centering
\subfigure[Safe set with $\Delta t=1.8$]{\includegraphics[trim=-1.3cm 0cm -1cm 0cm, clip=true,width=0.42\textwidth]{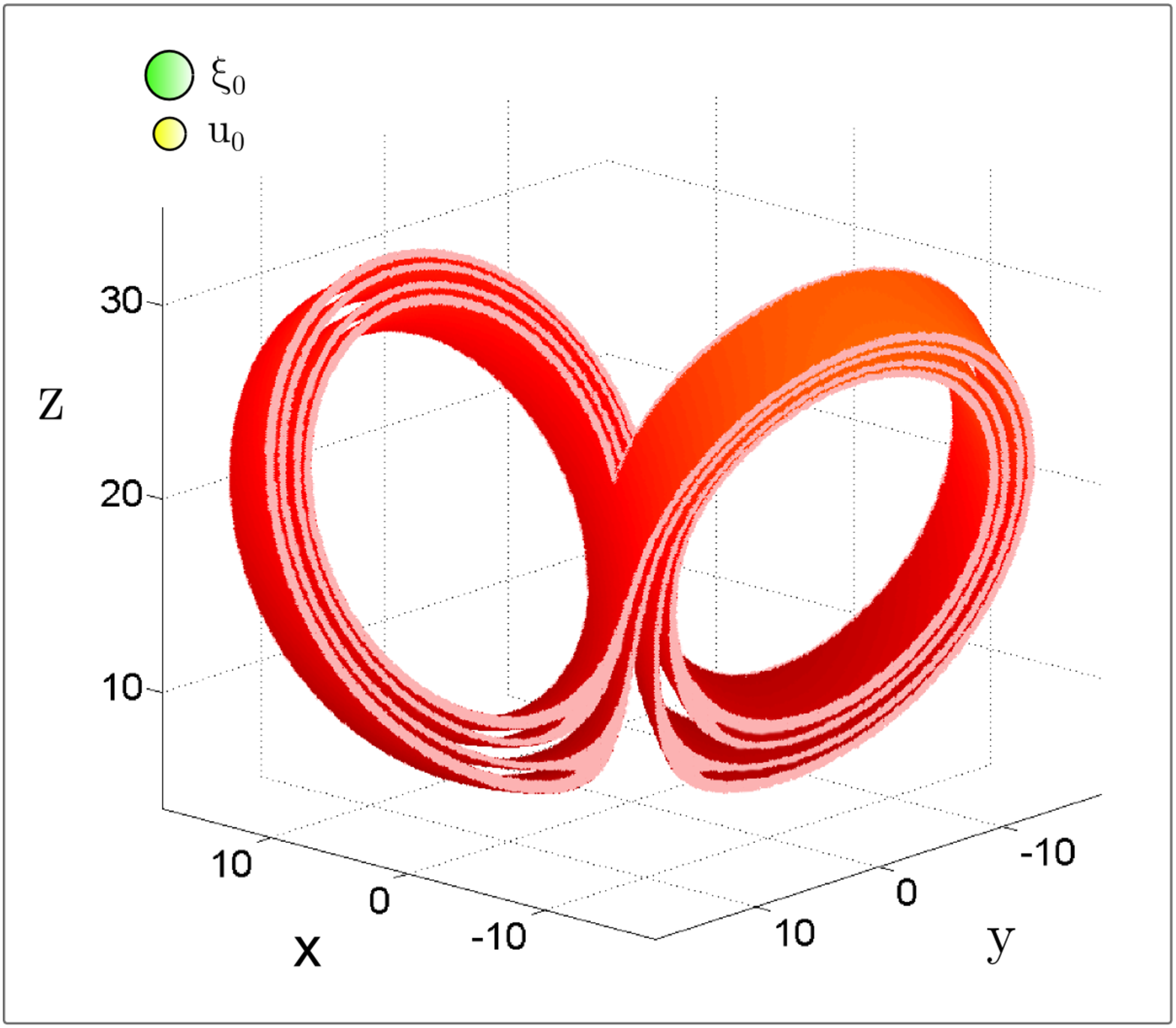}}
\subfigure[Asymptotic safe set]{\includegraphics[trim=-2cm 0cm 0cm 0cm, clip=true,width=0.4\textwidth]{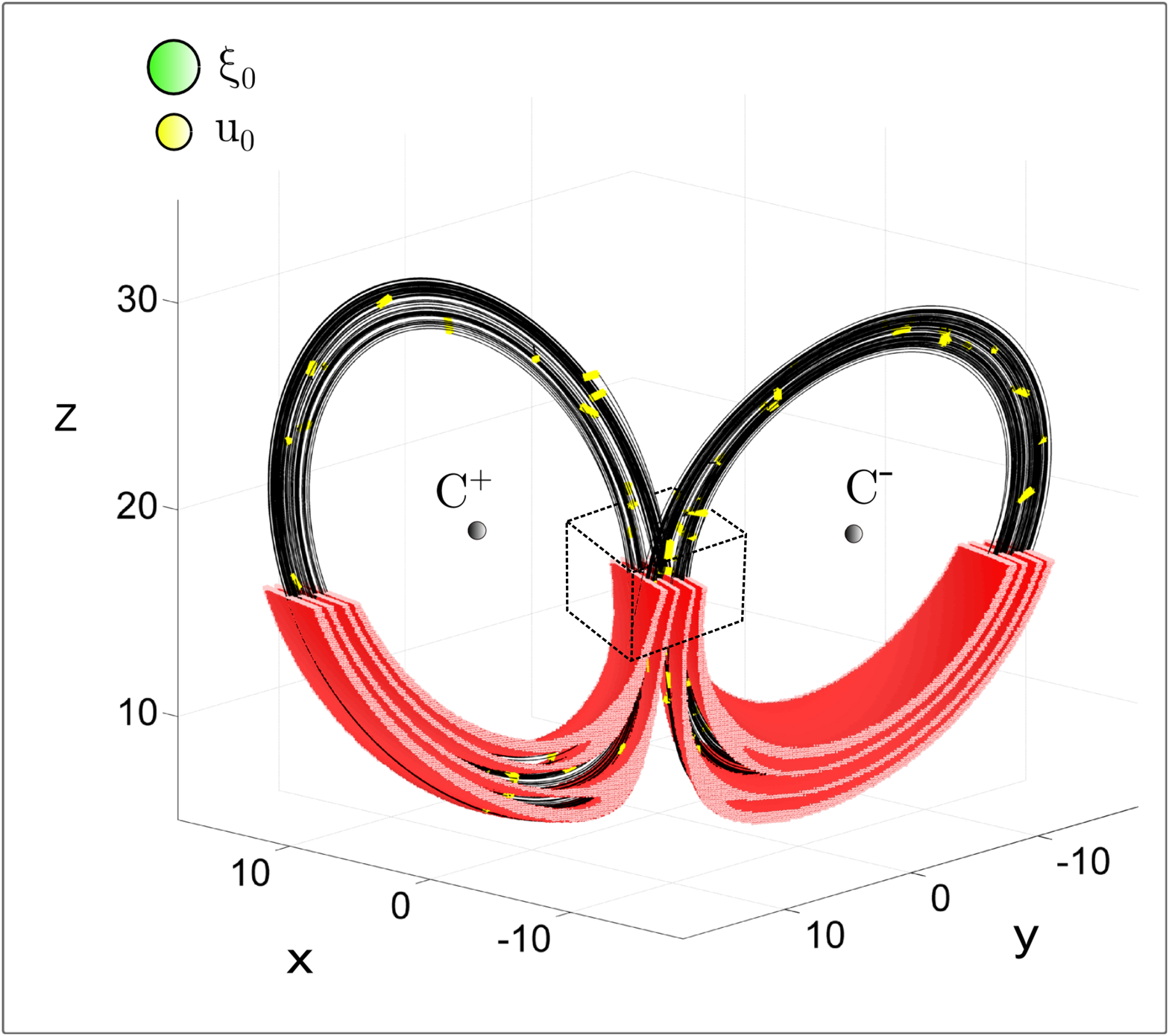}}
\\[20pt]
\caption {\textbf{Different safe set for different values of $\Delta$ time}. (a) The asymptotic safe set computed for $\Delta t=1.8$. To compute this set we have taken $\xi_0=1.5$ (green ball) and $u_0=1.0$ (yellow ball).(b) A half section of the asymptotic safe set (red) and a partially controlled trajectory (in black). In this case the controls (yellow segments inserted in the trajectory) are applied every $\Delta t=1.8$.}
\label{19z}
\end{figure}

As we have mentioned, the safe set appears for values of $\Delta t \geq 1.2$, so it is possible to adapt the control frequency to our specific requirements, taking other $\Delta t$ values. Figure~\ref{19z}(a) shows the asymptotic safe set for $\Delta t=1.8$ , and with $\xi_0$ and $u_0$ unchanged. With this set we could control the system applying a control every $\Delta t=1.8 $ (see Fig.~\ref{19z}(b)) instead of $\Delta t=1.2 $ as in the previous case. However taking a longer $\Delta t$ has a downside since in most scenarios that the cumulative effect of disturbances grows exponentially with time due to chaos, and therefore the needed $u_0$ increases as well.

\renewcommand{\thesubfigure}{(\alph{subfigure})~}

\begin{figure}
\includegraphics [trim=0cm 0cm 0cm 0cm, clip=true,width=0.75\textwidth]{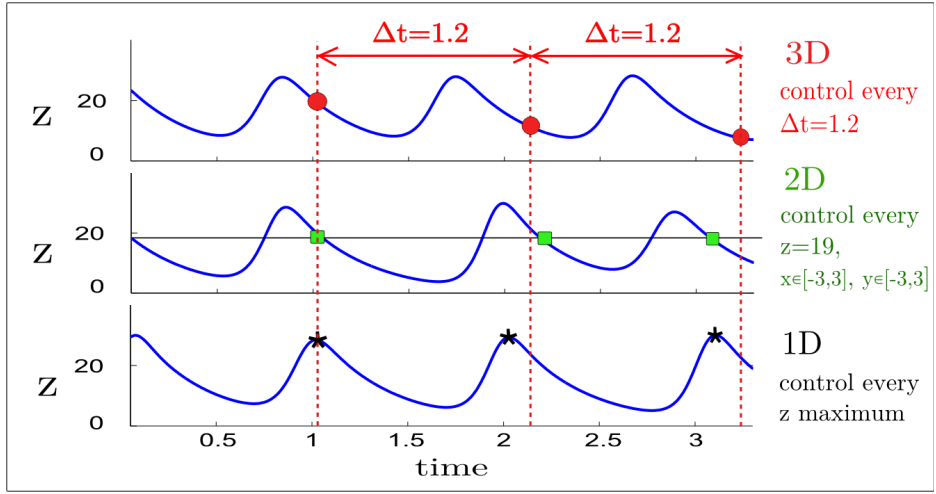}\\
\centering
\caption{\textbf{Comparison of the three controlled trajectories of the $z$ variable obtained with the 3D, 2D and 1D map respectively.} The marks indicate the moment where the control is applied. Only in the 3D case are the controls time periodic.}
\label{20z}
\end{figure}

 The use of a fixed $\Delta t$ time to discretize the dynamics can be advantageous since in some situations the application of control in periodic time intervals can be easier and more convenient. In addition, the frequency of these controls can be adapted making it very flexible. For example, in the context of medicine, a medical treatment based on the partial control method, could be applied a fixed day of the week, which supposes an easy and convenient control relationship between the physician and the patient. To highlight this feature, we compare in Fig.~\ref{20z}, three controlled trajectories obtained with the respective map (3D, 2D and 1D).

\section{Partial control applied on parameters}

  In the classical partial control method, the disturbances $\xi_n$ and the control $u_n$ were applied directly on the phase space variables of the system, that is, $q_{n+1}=f(q_n,p)+\xi_n+u_n$. In this last equation $p$ represents the parameters of the system (which are supposed to be constant over time). Here, we study a completely new control problem where the disturbances and the control terms are affecting directly some parameter of the system (instead of the phase space variables), that is, $q_{n+1}=f(q_n,p+\xi_n+u_n)$. For that reason, we call it \emph{parametric partial control}. This study is motivated by the fact that the parameters usually fluctuate from one iteration to another in most real physical systems. These kind of maps are called random maps in the literature. In the context of transient chaos, random maps are widely used to model systems where two different time scales dynamics coexist, one slow and predictable, and another with a small and fast fluctuating component. For example, this is the case in advective fluid dynamics~\cite{advective}, where the velocity field can be written as an average periodic field, plus a fluctuating component, or in some scattering processes~\cite{advection,chaos97,prl97} where the force field varies in time in a complex manner. As far as we know, the control scheme that we introduce here (parametric partial control) is the first that is able to sustain a transient chaotic dynamics in random maps.

This approach is based on the idea of the partial control method~\cite{Asymptotic} with the difference that the disturbances are introduced in a parameter of the map instead of the variables. The controlled dynamics in the region $Q_0$ where we want to keep the trajectories will be:
\begin{equation}
q_{n+1}=f(q_n,p+\xi_n+u_n),
\end{equation}
where $f$ is a function with a chaotic transient in $Q_0$, $q$ is a point of $Q_0$, $p$ is the central value of the parameter, $\xi_n$ is a bounded disturbance $\xi_n\leq\xi_0$ and $u_n$ is a bounded control, so that, $u_n\leq u_0 <\xi_0$.

\begin{figure}
\includegraphics [trim=0cm 0cm 0cm 0cm, clip=true,width=0.69\textwidth]{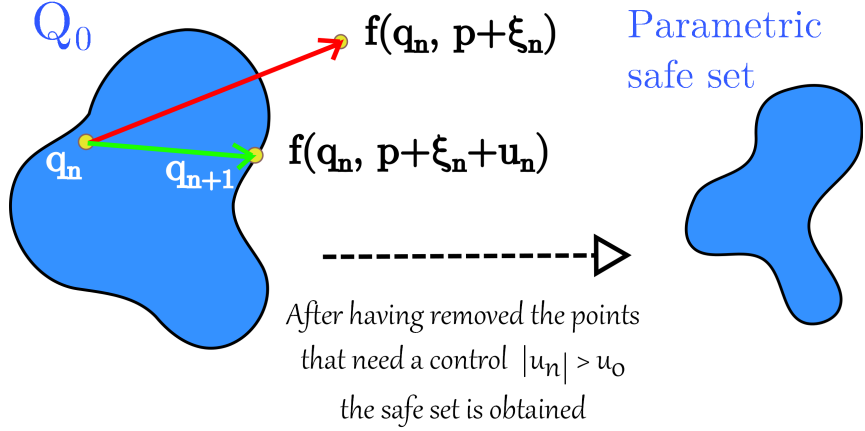}\\
\centering
\caption{\textbf{Scheme of the parametric partial control}. The red arrow shows the mapping of a point $q$, under the application of a random map in which a parameter $p$ is affected by a bounded disturbance $|\xi_n|<\xi_0$. The green arrow shows the mapping of a point $q$, once the control $u_n$ was applied to the parameter to keep the point in the blue region. Given the upper values of the disturbance $\xi_0$ and the control $u_0<\xi_0$, the partial control method removes the points of the blue region that need a control $|u_n|>u_0$ for some possible $|\xi_n|<\xi_0$. For every point we have to evaluate all possible disturbances $|\xi_n|<\xi_0$. Once the ``bad" points are removed, a new region $Q_1\subset Q_0$ is obtained. Iterating this process until it converges, we get a final region $Q_k\subset...\subset Q_1\subset Q_0$. We call this region, the \emph{parametric safe set}.}
\label{21z}
\end{figure}

We have developed an algorithm to implement the parametric safe set operator on an arbitrary set $Q$ of the phase space that has the following steps:
\begin{enumerate}
  \item  Select the region in phase space in which $f$ has a chaotic transient. We call the set of points of this region as the initial set $Q_0$. Then, we estimate the upper bound of the disturbance $\xi_0$, and we choose the upper bound of the control $u_0 < \xi_0$. Note that if the chosen $u_0$ is too small, the parametric safe set may be the empty set, and a bigger value of $u_0$ must be chosen.
  \item For every point $q \in Q_i$ ($i=0$ for the initial set), we need to check whether it is safe and can be part of an admissible trajectory or not. To do that, we compute $q_{n+1}=f(q_n,p + \xi_n + u_n)$ where the control $u_n$ is applied with the knowledge of $p+\xi_n$, to place the trajectories back in $Q_i$, if it escapes, otherwise $u_n=0$. For every point $q_n$, we have to check all possible disturbances $\xi_n$. If for all of them, the absolute value of the applied control $|u_n|$ is smaller than $u_0$, then the point $q$ is safe, otherwise, it is removed from $Q_i$.
 \item After having removed all the points that do not satisfy the control condition, a new set $Q_{n+1}\subset Q_n$ is obtained. Then, we repeat again the step 2 with the new set $Q_{n+1}$. The process is repeated until it converges, in which case $Q_{n+1}= Q_n$, and this will be the \emph{parametric safe set}. See Fig.~\ref{21z}.
\end{enumerate}

Some practical considerations have to be done. In order to compute  the parametric safe set, a finite grid covering $Q_0$ has to be used, since is not possible to compute the infinite number of points in $Q_0$. For an analogous reason, only a finite sample of disturbances $\xi_n$ can be checked for every point $q$. We will refer to the grid resolution as the distance between two adjacent points $q$, and the parameter resolution as the distance between two adjacent values of the parameter affected by different disturbances. Higher resolutions give a more accurate parametric safe set. In this sense, we have found that beyond a critical resolution of the grid over $Q$ and $\xi$, the safe set remains unchanged. For that reason and from a practical point of view, we recommend to compute the safe set with the algorithm proposed with increasing resolutions until finding the critical value for which the shape of the safe set found remains unchanged. That one will be a very good approximation of the real safe set.

In order to show how the parametric partial control approach works, we have considered three well known models, the 1D logistic map, the 2D Hénon map and the Duffing oscillator, all of them for a choice of parameters where transient chaos is present. In all cases we consider that the parameter is affected by a disturbance with a uniform probability distribution $|\xi_n|\leq \xi_0$. But any other distribution is possible, provided that it is bounded.

\subsection{The logistic map}

\renewcommand{\thesubfigure}{(\alph{subfigure})~}

\begin{figure}
\centering
\subfigure[Uncontrolled trajectory]{\includegraphics[trim=0cm 0cm 0cm 0cm, clip=true,width=0.45\textwidth]{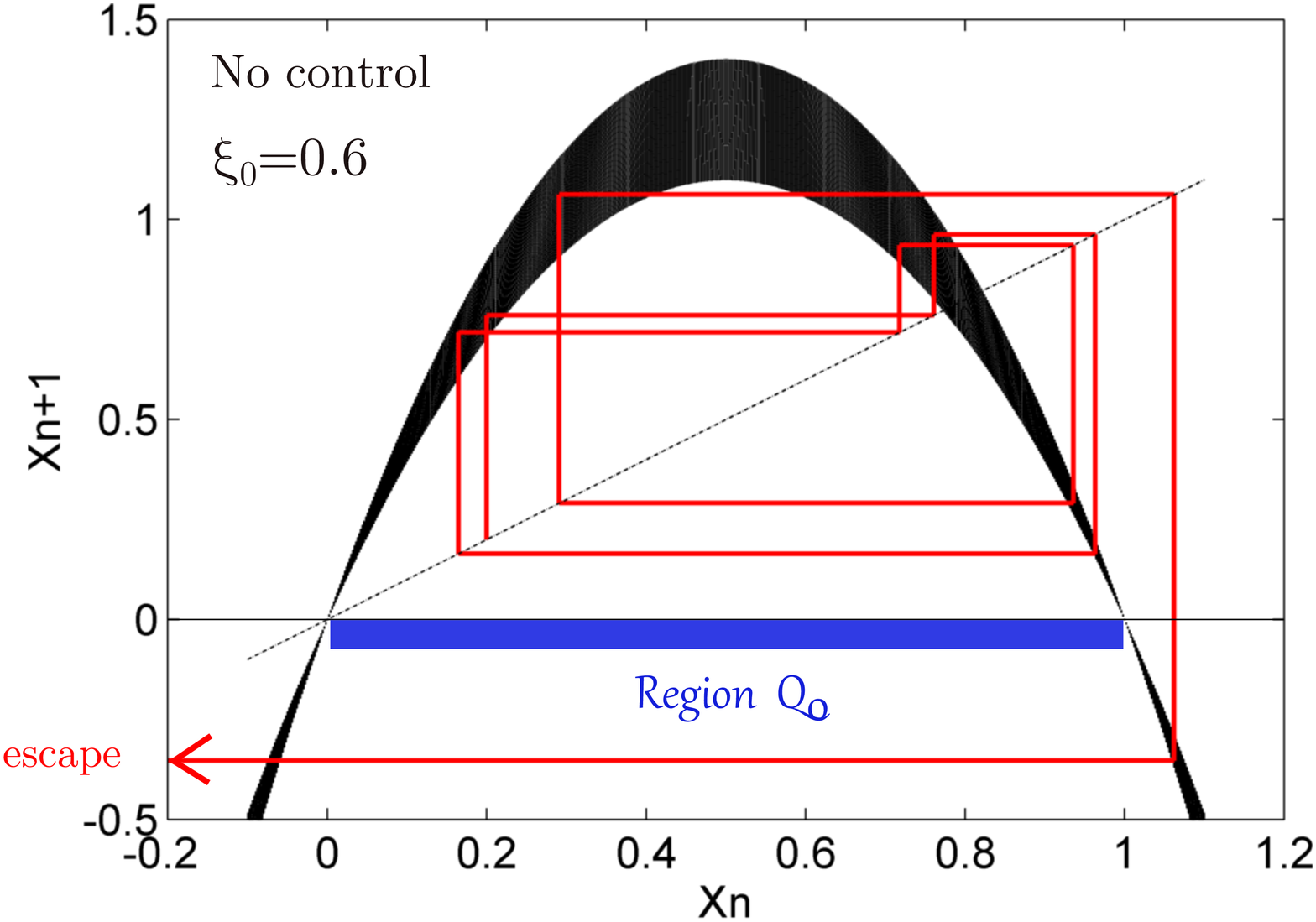}}
\subfigure[Controlled trajectory]{\includegraphics[trim=-2cm 0cm 0cm 0cm, clip=true,width=0.47\textwidth]{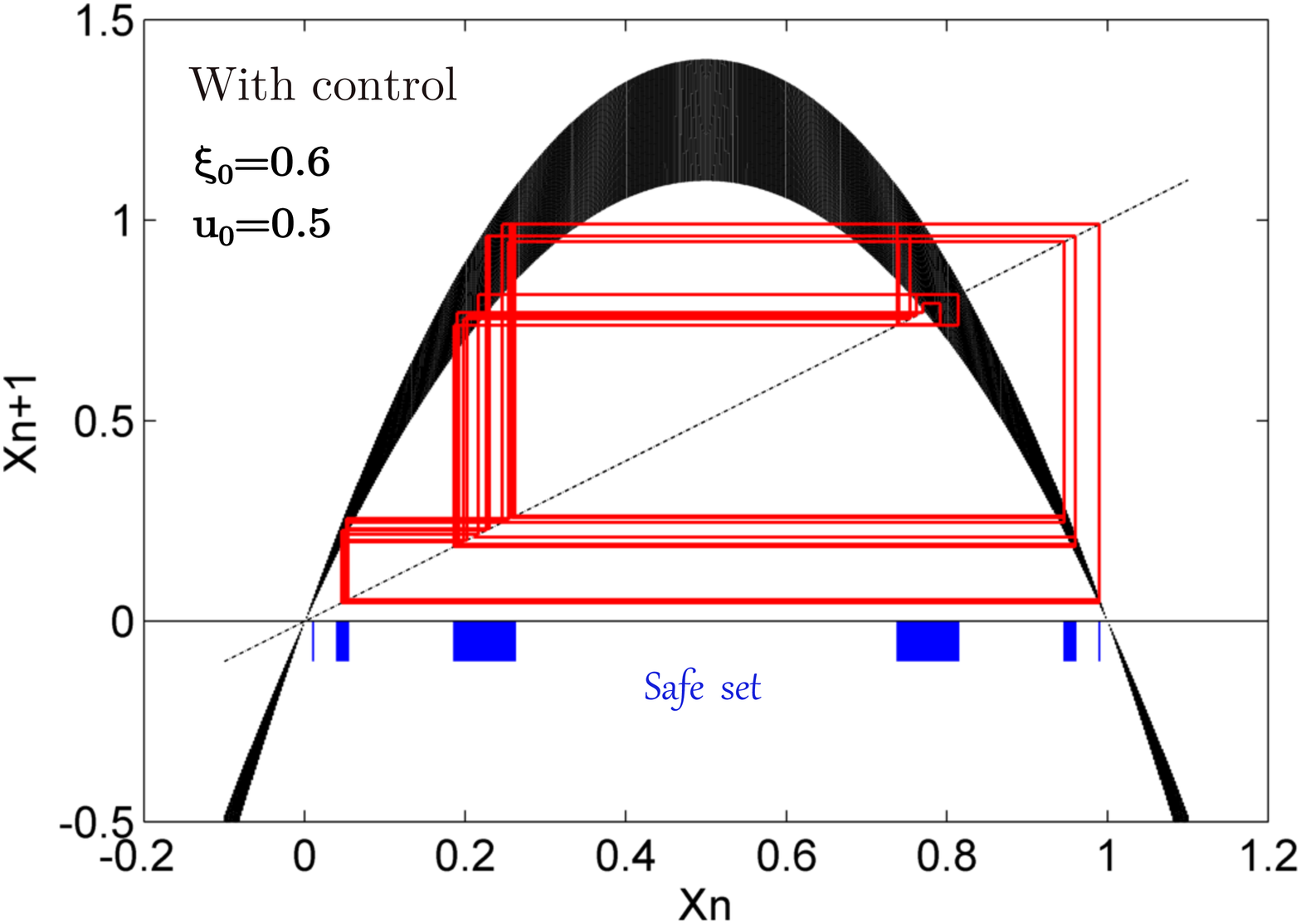}}
\\[20pt]
\caption{\textbf{Logistic map where the parameter $r$ is affected by disturbances}. (a) Logistic map $x_{n+1}=r x_n(1-x_n)$ where the parameter $r=5$ is affected by disturbances with upper bound $\xi_0=0.6$. The black wide curve is obtained for all possible values of the parameter, $r\in[5-\xi_0,5+\xi_0]$, of the logistic map. In red, we show an example of an uncontrolled trajectory that after a chaotic motion in $Q_0$, escapes to minus infinity. (b) We apply the partial control method to the logistic map, with $\xi_0=0.6$ and $u_0=0.5$ and a grid resolution of $0.001$, to obtain the parametric safe set which is shown with the wide blue segments to help the visualization.  The orbits starting in this set, remain there after applying a control $u_n\leq 0.5$ every iteration. In red, we show an example of a partially controlled trajectory. We are plotting only 50 iterations.}
\label{22z}
\end{figure}

The logistic map is a 1D map and is defined as follows:
\begin{equation}
\begin{array}{l}
x_{n+1}=r x_n(1-x_n).
\end{array}
\end{equation}
For a parameter value $r\in[0,4]$ the interval $x\in[0,1]$ maps to itself. However for $r>4$, the orbits starting in this interval, escape towards infinity after a chaotic motion (see Fig.~\ref{22z}(a)). With the aim of keeping the trajectories in $Q_0=[0,1]$ and assuming that the parameter is affected by some disturbances $|\xi_n|\leq\xi_0$, the parametric partially controlled dynamics for this map can be written as
\begin{equation}
\begin{array}{l}
  x_{n+1}=(r+\xi_n+u_n) x_n (1-x_n), \\
\end{array}
\end{equation}
where $|u_n|\leq u_0< \xi_0$ is the control applied. To show an example of control we have taken the values $r=5$, $\xi_0=0.6$ and $u_0=0.5$. After the computation of the algorithm described in the previous section, we have obtained the parametric safe set shown in Fig.~\ref{22z}(b). The blue wide segments represent the safe points of $x$. In this figure, it has also been displayed a partially controlled trajectory (in red), which as can be seen, remains chaotic and within $Q_0$ indefinitely.

%

\subsection{The Hénon map}

\begin{figure}
\centering
\subfigure[Uncontrolled trajectory]{\includegraphics[trim=0cm 0cm 0cm 0cm, clip=true,width=0.31\textwidth]{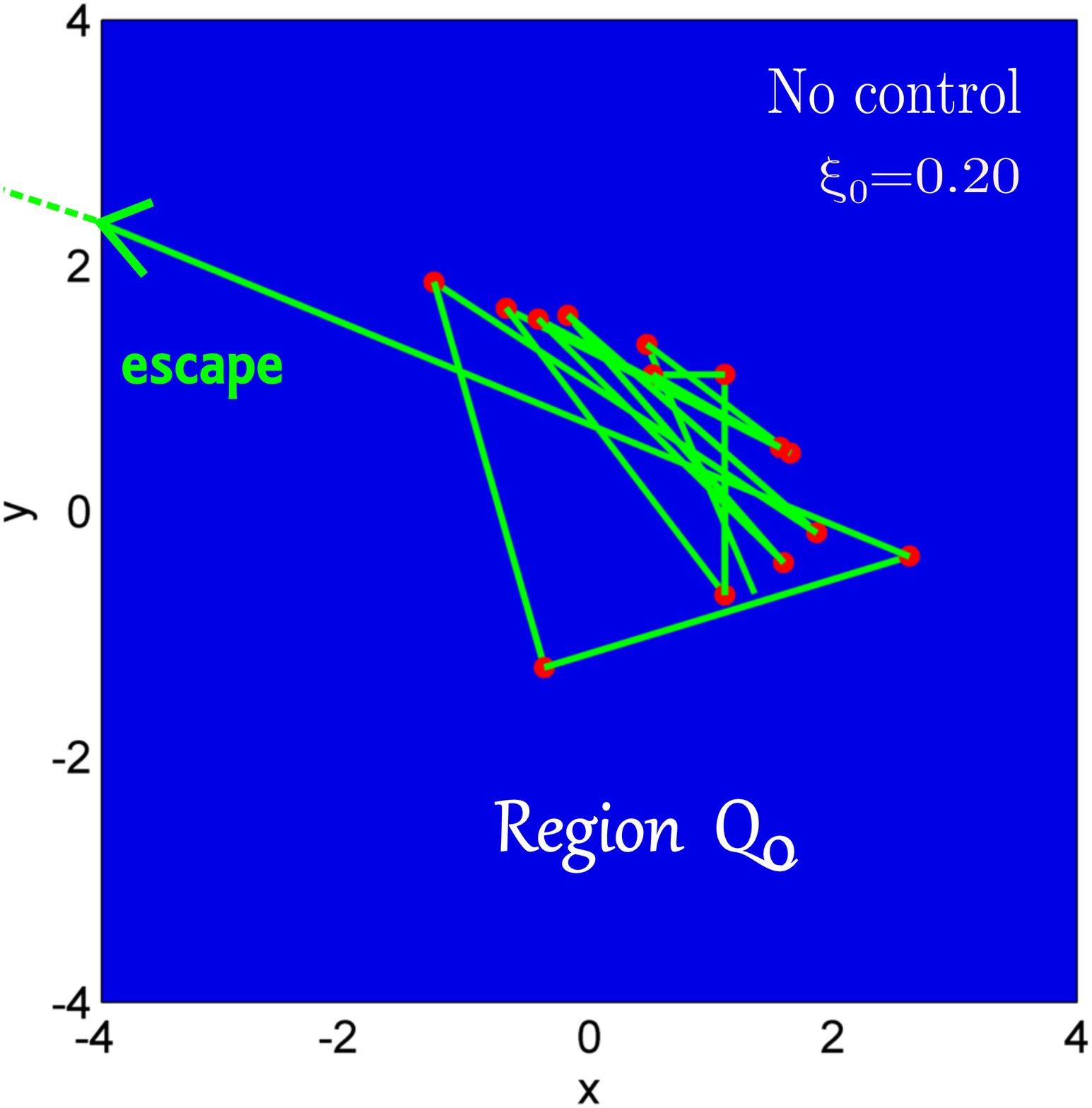}}
\subfigure[Safe set and controlled trajectory]{\includegraphics[trim=-1cm 0cm 0cm 0cm, clip=true,width=0.33\textwidth]{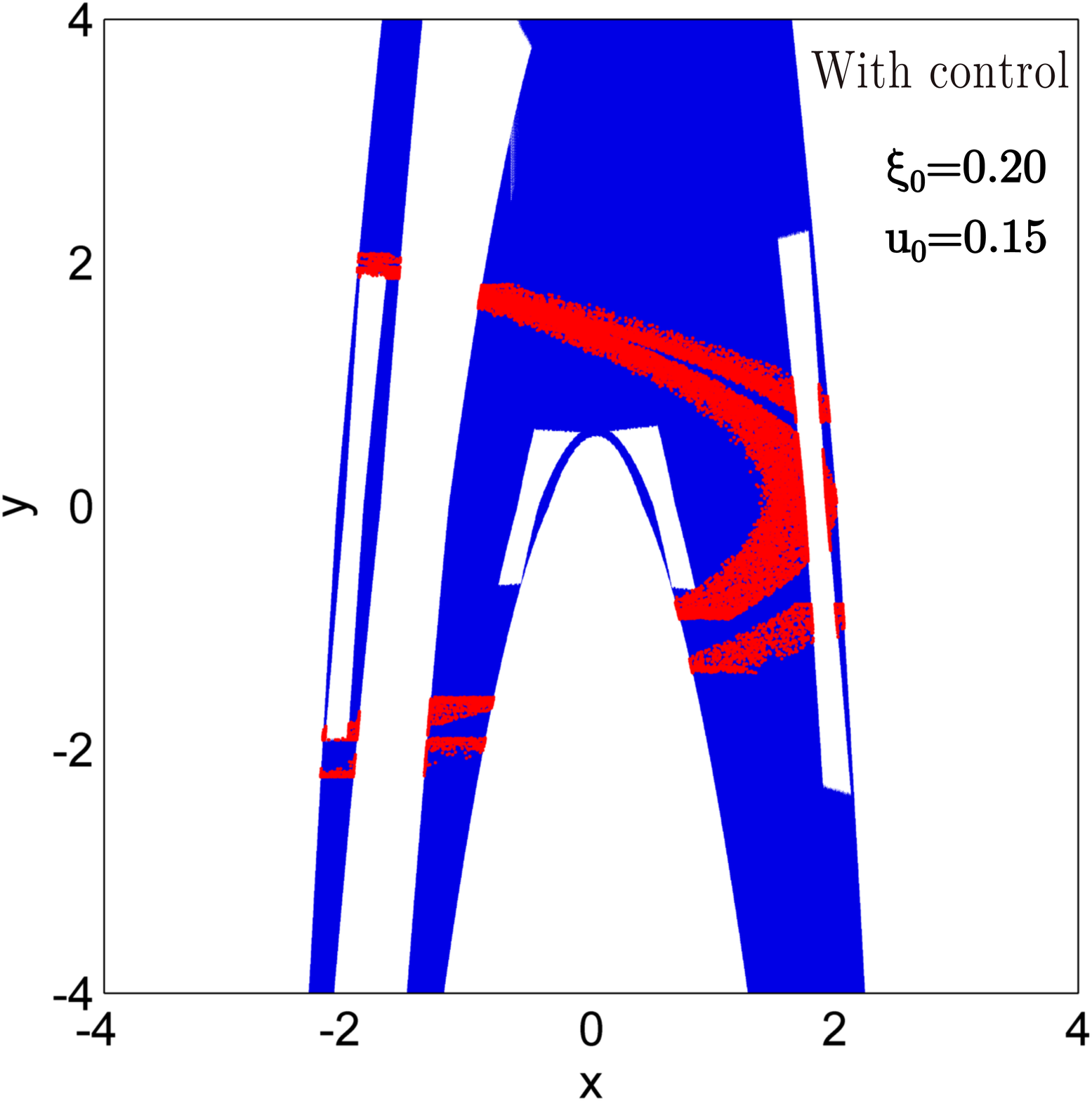}}
\subfigure[Safe set and controlled trajectory]{\includegraphics[trim=-1cm 0cm 0cm 0cm, clip=true,width=0.33\textwidth]{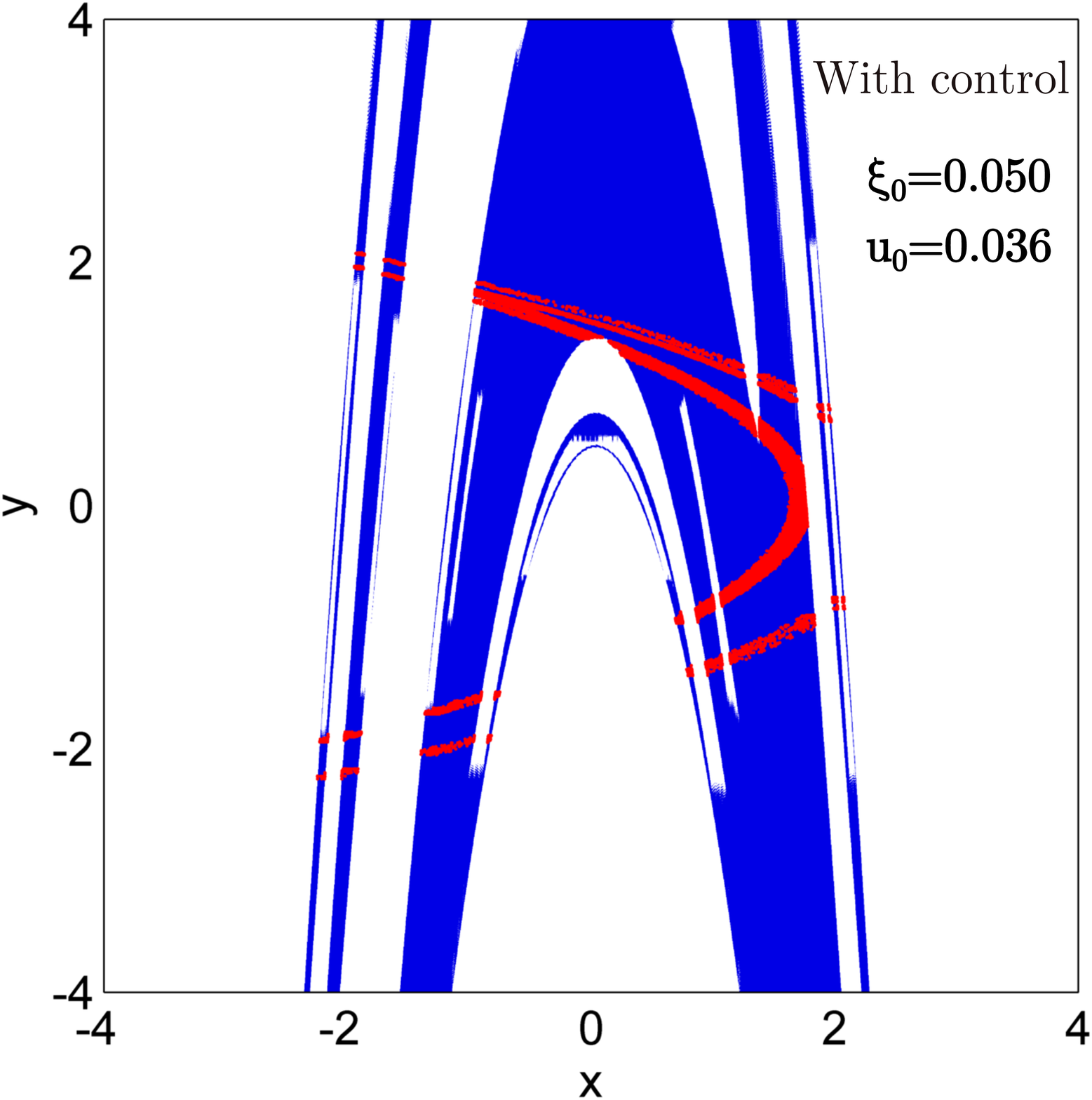}}
\\[20pt]
\caption{\textbf{ The Hénon map where the parameter $b$ is affected by disturbances}. (a) An uncontrolled trajectory in the Hénon random map with $a=2.16$ and $b=0.3$. The parameter $b$ is affected by disturbances with upper bound $\xi_0=0.20$. The blue square $[-4,4] \times [-4,4]$ is the region $Q_0$. In absence of an external control, the trajectories in $Q_0$ escape outside the square after a very short chaotic transient. An example of an uncontrolled trajectory is displayed with the red points connected by the green lines. (b) The partial control method has been applied to keep trajectories in $Q_0$ forever. The upper bound of control is $u_0=0.15$. The grid resolution taken is $0.01$ and the parameter resolution is $0.005$.  As a result, the parametric safe set (in blue) is obtained. All the orbits of the map starting in the blue set, remain there after the application of controls smaller than $u_0=0.15$. The red points display a partially controlled trajectory, where 20000 iterations of the trajectory have been plotted. (c) For this case the upper value of control is $u_0=0.036$, the grid resolution used is $0.001$ and the parameter resolution $0.0005$. In we compare it with the previous figure, we see that the appearance of the parametric safe set is more complex, due to fact that the disturbance value is smaller.}
\label{23z}
\end{figure}

The Hénon map is a 2D map defined by
\begin{equation}
\begin{array}{l}
  x_{n+1}=a-b y_n- x_n^2  \\
  y_{n+1}=x_n.   \\
\end{array}
\end{equation}
This map shows transient chaos for a wide range of the parameters $a$ and $b$. We have chosen here the parameter values $a=2.16$ and $b=0.3$. For these values, the trajectories with initial conditions in the square $[-4,4] \times [-4,4]$ have a very short chaotic transient, before finally escaping this region toward infinity. An example of this behavior is shown Fig.~\ref{23z}(a) for a given initial condition. We consider now, a situation where the parameter $b$ is affected by some disturbance
$|\xi_n|\leq\xi_0$. To keep the orbits in $Q_0=[-4,4] \times [-4,4]$ we apply a control $|u_n|\leq u_0 <\xi_0$, so that the controlled dynamics can be described as:
\begin{equation}
\begin{array}{ l }
  x_{n+1}=a-(b+\xi_n+u_n) y_n-x_n^2  \\
  y_{n+1}=x_n.  \\
\end{array}
\end{equation}

As an example we have computed two different safe sets corresponding to the values $\xi_0=0.20$, $u_0=0.15$  and $\xi_0=0.050$, $u_0=0.036$ respectively. The safe sets obtained are  shown in Fig.~\ref{23z}(b) and Fig.~\ref{23z}(c). where it was also drawn a partially controlled orbit (red points), which remains chaotic in the square forever.

\renewcommand{\thesubfigure}{(\alph{subfigure})~}

As revealed by the Figs.~\ref{23z}(b) and ~\ref{23z}(c), as the disturbance decreases, the parametric safe set becomes more and more complex due to the fractal structure of the chaotic saddle underlying the dynamics. For this reason, a higher resolution is necessary to solve this kind of safe sets. However, we always have a finite resolution in the computation, so the value of the disturbance can never be zero.

\subsection{The Duffing oscillator}

In previous sections, the partial control method was applied to the Duffing oscillator system, where disturbances and control affected in an additive way the variables of the system. In this case we have studied the same model, with the difference that disturbances and control are now affecting some parameter of the system. In contrast with the logistic and Hénon map, the Duffing oscillator model is a flow, so a previous discretization of the dynamics is required to apply the control method.

We consider here the following Duffing oscillator equations:
\begin{equation}
\begin{array}{l}
  \ddot{x}+0.15\dot{x}-x+x^{3}=0.245\sin(t).  \\
\end{array}
\end{equation}

\begin{figure}
\includegraphics [trim=0cm 0cm 0cm 0cm, clip=true,width=0.6\textwidth]{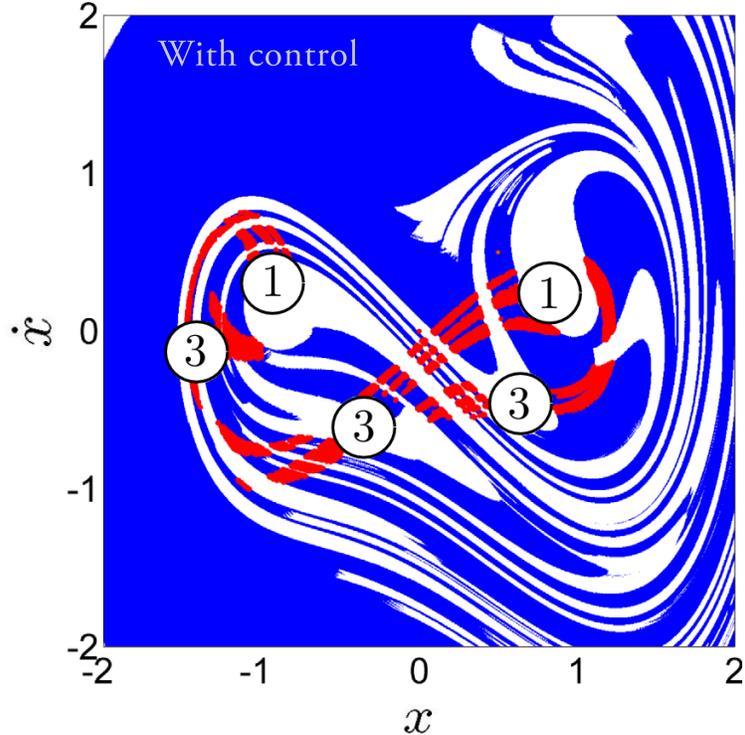}\\
\centering
\caption{\textbf{Controlled trajectory in the Duffing oscillator with $\xi_0=0.020$ and $u_0=0.014$}. Numbers indicates the three attractors of the system, two period-1 and one period-3. The aim of applying control is to avoid trajectories falling in these attractors. After removing the holes, corresponding to the attractors, the safe set (in blue) was computed with a grid of $1000\times1000$, (grid resolution $0.004$, parameter resolution $0.0002$). The red dots represent a controlled trajectory made up of $30000$ iterations in the stroboscopic map.}
\label{24z}
\end{figure}

For this choice of parameters, it is possible to find in the phase space a transient chaos behaviour of the trajectories. Due to the periodic forcing, it is suitable to build a time-$2\pi$ map, where the flow is cut every $\Delta t=2\pi$. The transient chaotic dynamics is captured in the square $[-2,2] \times [-2,2]$. Without external control, almost all initial conditions in this region, after a chaotic behaviour, fall in one of the three attractors present in the phase space. The system has two period-1 attractors and one period-3 attractor, as shown in Fig.~\ref{24z}.

With the aim of keeping the trajectories far from these attractors, we have applied the partial control method considering that the forcing amplitude is affected by some bounded disturbance $|\xi_n|\leq \xi_0$. Applying the control $|u_n|\leq u_0$ in the same parameter as well, the amplitude of the forcing vary according to $0.245+\xi_n+u_n$ every iteration.

As an example, we have computed the safe set for the  values  $\xi_0=0.020$ and  $u_0=0.014$. We have used a grid of $1000\times1000$ in the square $[-2,2] \times [-2,2]$, where the balls centered in each attractor has been removed to prevent the periodic behaviour.  The safe set obtained is shown in Fig.~\ref{24z}, where a controlled trajectory (30000 iterations in red) also appears. Notice that the partially controlled trajectory is chaotic and never fall into the attractors.

\subsection{Controlling more parameters}

   Although we have dealt with examples where the control is applied on a certain parameter, situations where more than one parameter need control are possible. The scheme of the method is easily expandable, for example, in the case of $m$ parameters $p^1,p^2,...,p^m$, the partially controlled dynamics would be described as
\begin{equation}\label{high_parametric_control}
q_{n+1}=f(q_n, (p^1 + \xi_n^1 + u_n^1), (p^2 + \xi_n^2 + u_n^2),..., (p^m + \xi_n^m + u_n^m)),
\end{equation}
with the conditions
\begin{equation}\label{high_constraints}
\sqrt{(\xi_n^1)^2 + (\xi_n^2)^2 +...+ (\xi_n^m)^2} \leq \xi_0  \hspace{0.9cm} and \hspace{0.7cm} \sqrt{(u_n^1)^2 + (u_n^2)^2 +...+ (u_n^m)^2} \leq u_0<\xi_0.
\end{equation}

The main drawback of considering the extra parameters is the considerable increase of computational time to obtain a parametric safe set due to the curse of dimensionality. However it is possible to accelerate this computation  parallelizing some parts of the Sculpting Algorithm code or also by using GPU computing techniques.

\section{The partial control applied on time-delay coordinates maps}

  \begin{figure}
\includegraphics [trim=0cm 0cm 0cm 0cm, clip=true, width=0.9\textwidth]{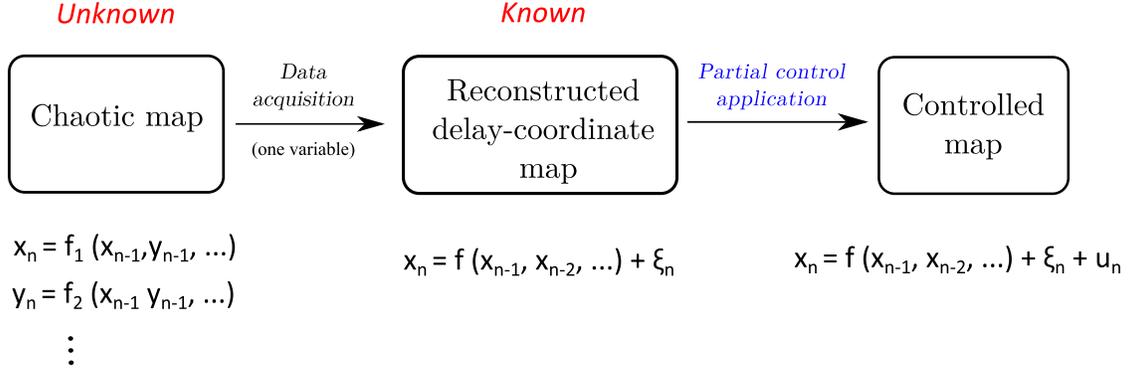}\\
\centering
\caption{\textbf{Conceptual framework.} From left to right. Step $1$: data acquisition from a chaotic system. We assume here that only one variable is observable. Step $2$: using embedding and parametric reconstruction techniques, construct a delay-coordinate map. The term $\xi_n$ represents a disturbance term that encloses all possible deviations from the real dynamics. Step $3$: apply the partial control method introducing and additive control term $u_n$ acting on the observable variable. In this work we assume that we already possess the knowledge of the delay-coordinate map.}
\label{25z}
\end{figure}

  Although in this part we describe an application of the method where the control is applied on the variables, we have considered that the case of delay-coordinate maps deserves an specific section due to the modifications needed on the control strategy to apply it (See  Ref.~\cite{Delayed}).  Delay-coordinate maps have been widely used recently to study nonlinear dynamical systems, where there is only access to the time series of one of their variables. We consider here a delay-coordinate map under external additive disturbances $f(x_n, x_{n-1}, ...)+\xi_n$, where the control can also be applied in an additive way $f(x_n, x_{n-1}, ...)+\xi_n+u_n$. This kind of framework is the one that is usually found after using the delay reconstruction method to study the phase space dynamics of a chaotic system. These maps are usually expressed in the following way:

\begin{equation}
\begin{array}{l}
x_{n}=f(x_{n-1},x_{n-2}\ldots x_{n-m}).
\end{array}
\end{equation}

We consider here the problem of controlling this kind of maps possessing a chaotic behaviour (see the  scheme of Fig.~\ref{25z}).  The main difference  with the classical partial control scheme is that the control can only be applied in the present state $x_n$, (is not physically possible to control the past states $(x_{n-1},x_{n-2}\ldots)$). Therefore we need to introduce a new approach to control the system.

Following the idea of the partial control method we consider that the system can be modelled as:
\begin{equation}
\begin{array}{l}
 x_{n}=f(x_{n-1},x_{n-2}\ldots x_{n-m})+\xi_n +u_n,
\end{array}
\end{equation}
where $\xi_n$ is the disturbance affecting the state $x_{n}$, and $u_n$ is the respective control applied, both limited by
\begin{equation}
\begin{array}{l}
|\xi_n|\leq\xi_0, ~~~~~~|u_n|\leq u_0.\nonumber
\end{array}
\end{equation}


\begin{figure}
\includegraphics [width=0.8\textwidth]{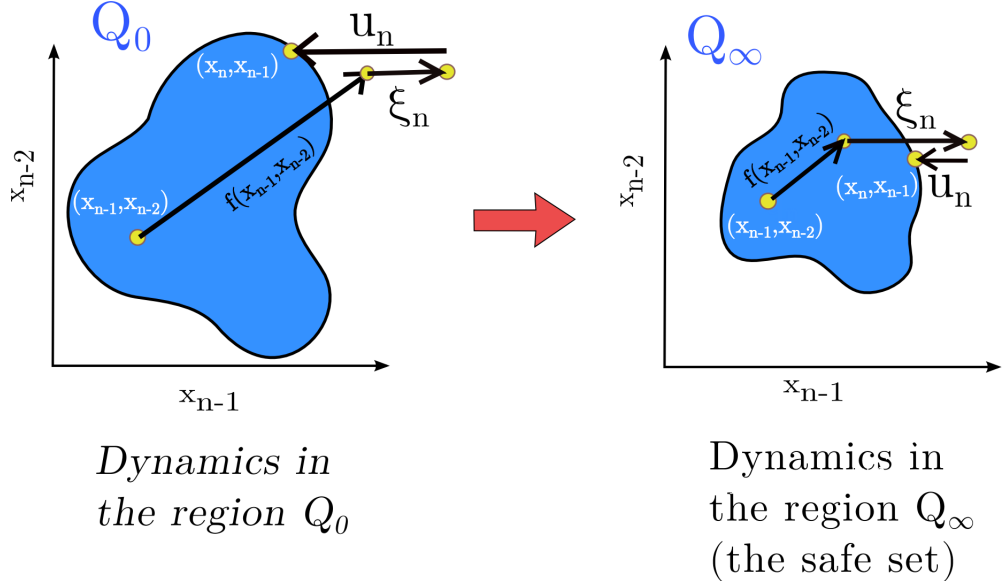}\\
\centering
\caption{\textbf{Dynamics in $Q_0$ and $Q_\infty$.} The left side shows an example of a $2D$ region $Q_0$ (in blue) in which we want to keep the dynamics described by $x_{n}=f(x_{n-1},x_{n-2})+\xi_n+u_n$. We say that $|\xi_n| \leq \xi_0$ is a bounded disturbance affecting the map, and $u_n$ is the control chosen so that $q_{n+1}$ is again in $Q_0$. Notice that disturbance and control arrows are drawn parallel to current state of the variable since only the present state is affected by them. To apply the control, the controller only needs to measure the state of the disturbed system, that is $[f(x_{n-1},x_{n-2})+\xi_n]$. The knowledge of $f(x_{n-1},x_{n-2})$ or $\xi_n$ individually is not required. The right side of the figure, shows the region $Q_\infty \subset Q_0$ (in blue), called the \emph{safe set}, where each $(x_{n-1},x_{n-2}) \in Q_\infty$ has $(x_{n},x_{n-1}) \in Q_\infty$ for some control $|u_n|\leq u_0$, which is chosen depending on $\xi_n$. Notice that the removed region does not satisfy $|u_n|\leq u_0$.}
\label{26z}
\end{figure}

Once we know the delay-coordinate-map,  all we have to do to apply the partial control method is to define the region $Q_0$ in the phase space $(x_{n-1},x_{n-2}\ldots)$ where we want to keep the trajectories, and determine the upper value of the disturbance $\xi_0$, and the upper value of the control $u_0$ used.


To compute the safe set, we have developed a modified version of the \emph{Sculpting Algorithm}~\cite{Automatic}, which evaluates the points from $Q_0$ and remove them if they do not satisfy the safety condition. The $ith$ step of this algorithm can be summarized as follows:

\begin{enumerate}
  \item  Morphological dilation of the set $Q_i$ by $u_0$ along the $x_{n-1}$ direction, obtaining the set denoted by $Q_i+u_0$.
  \item  Morphological erosion of set $Q_i+u_0$ by $\xi_0$ along the $x_{n-1}$ direction, obtaining the set denoted by $Q_i+u_0-\xi_0$.
  \item  Let $Q_{i+1}$ be the points $(x_{n-1},x_{n-2}\ldots)$ of $Q_i$, so that $f(x_{n-1},x_{n-2}\ldots)$ maps inside the set $Q_i+u_0-\xi_0$.
  \item  Return to step $1$, unless $Q_{i+1}=Q_i$, in which case we set $Q_\infty=Q_i$. We call this final region, the \textit{safe set}. Note that if the chosen $u_0$ is too small, then $Q_\infty$ may be the empty set, so that a bigger value of $u_0$ must be chosen.
\end{enumerate}

  This final set is formed by the points $(x_{n-1},x_{n-2}\ldots)$ belonging to the region $Q_0$,  where the image $x_{n}=f(x_{n-1},x_{n-2}\ldots)+\xi_n+u_n$ can be put back again on the safe set by using a control $|u_n|\leq u_0$. In Fig.~\ref{26z} we illustrate the controlled dynamics in the region $Q_0$ and the safe set $Q_\infty$.  Notice that, due to the fact that the control and disturbance affects the present state of the variable, then they are applied in the current axis direction.

  In order to show that the method can be applied on different chaotic maps, we have chosen three examples of well-known chaotic maps to illustrate it. We do not reproduce here the embedding and reconstruction model step, since is not the goal of this work. Instead of that, we have deduced by simple calculation, the expression of the delay-coordinate maps. Next, we apply the control scheme with the aim of keeping the orbits in a desirable region of phase space.

\subsection{The two-dimensional cubic map}

We consider here the system  given by:
\begin{equation}
\begin{array}{l}
x_{n}=y_{n-1}\\
y_{n}=-bx_{n-1}+ay_{n-1}-y_{n-1}^3,
\end{array}
\label{eq1}
\end{equation}
which represents the two-dimensional cubic map \cite{Holmesa}.

\begin{figure}
\includegraphics [width=0.9\textwidth]{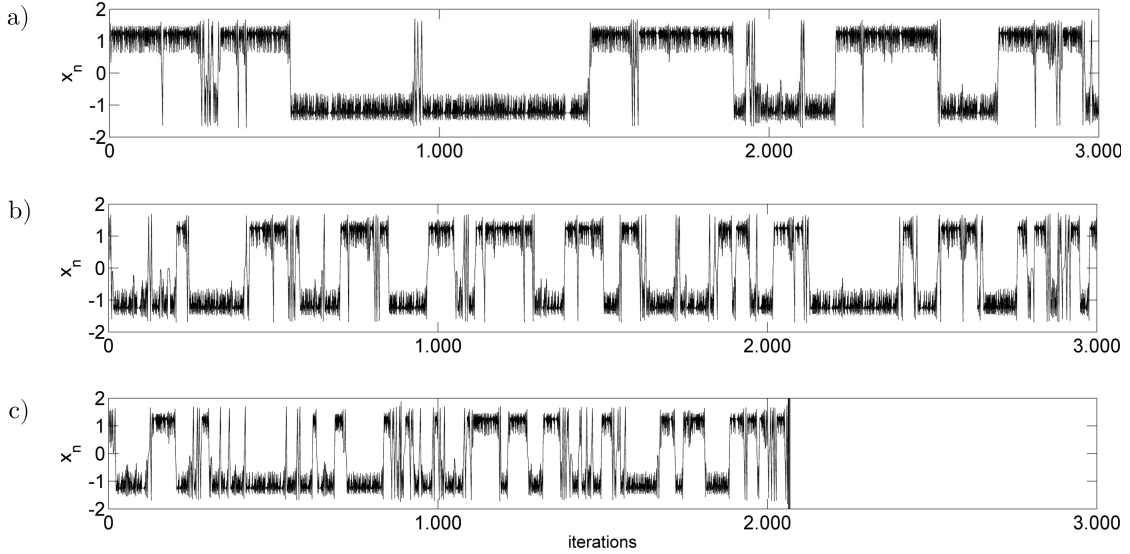}\\
\centering
\caption{\textbf{Time series of the two-dimensional cubic map for different disturbances.} (a) Time series of variable $x_n$ with no disturbance affecting it. (b) Time series with $|\xi_n|\leq\xi_0=0.02$ affecting the map. (c) Time series with $|\xi_n|\leq\xi_0=0.20$ affecting the map. After some iterations the trajectory escapes towards $-\infty$.}
\label{27z}
\end{figure}

This two-dimensional cubic map depends on two parameters and exhibits chaos for different values of them. We have selected here the values $a =2.75$ and  $b=0.2$. For this choice of parameters, we have represented in Fig.~\ref{27z}(a)  an example of the time series of the variable $x_n$ without the influence of noise. Here, we can see that the trajectories oscillate between two well differentiated regions (top and bottom), where the transitions between them occurs after some typical time. However, when we introduce additive disturbances, the frequency of the transitions increases (Fig.~\ref{27z}(b)). And for large disturbances the trajectory eventually escapes toward an external attractor due to the extra energy applied (Fig.~\ref{27z}(c)).

\begin{figure}
\includegraphics [width=0.9\textwidth]{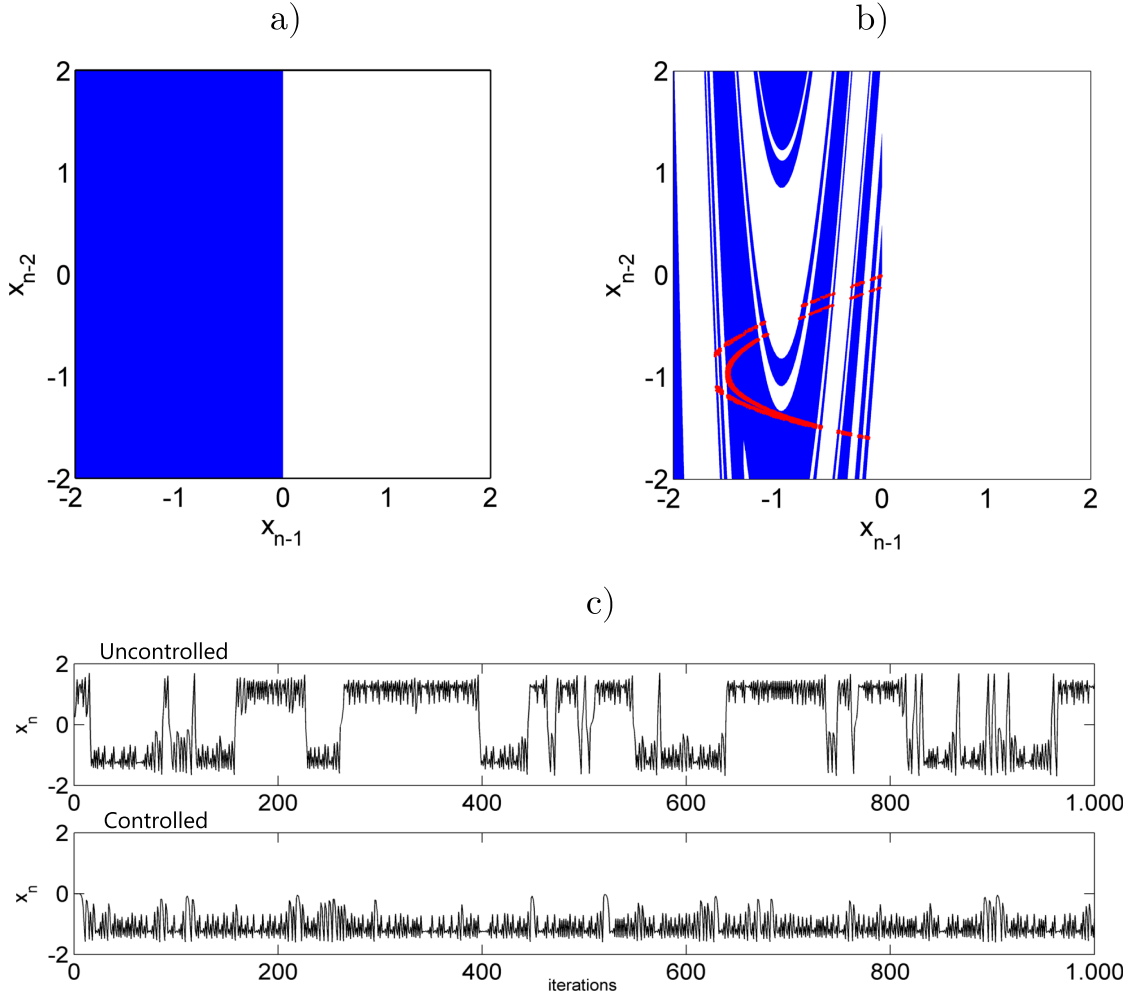}\\
\centering
\caption{\textbf{Safe set and controlled dynamics in the two-dimensional delayed cubic map ($\mathbf{x_{n}=ax_{n-1}-bx_{n-2}-x_{n-1}^3}$).} (a) In blue the initial region $Q_0$ where we want to keep the trajectories. (b) The safe set obtained with the values of disturbance $\xi_0=0.020$ and control $u_0=0.015$. A grid of $1000 \times 1000$ points has been used. The red dots represent $1000$ iterations of a partially controlled trajectory. (c) In the top it is represented an uncontrolled time series affected with $\xi_n\leq\xi_0=0.020$. In the bottom the controlled time series corresponding to the red dots shown in case b.}
\label{28z}
\end{figure}

As an example, we assume now that due to experimental restrictions we only see the dynamics of the variable $x_n$ and, with that information, we are interested in keeping the trajectory in the bottom region ($-2<x_n<0$) forever, even in presence of large disturbances.

The form of the reconstructed delay-coordinate map can be deduced by substituting  $y_{n-1}=-bx_{n-2}+ay_{n-2}-y_{n-2}^3$ into Eq.~\ref{eq1} and taking into account that $x_{n-1}=y_{n-2}$.

\begin{equation}
\begin{array}{l}

x_{n}= ax_{n-1}-bx_{n-2}-x_{n-1}^3.

\end{array}
\end{equation}

We call this map the \emph{two-dimensional delayed cubic map}. In addition, we add to the model a disturbance term $\xi_{n}$ in order to consider the potential noise present in the data acquisition or also mismatches in the reconstruction model technique.


Taking into account the disturbance and the control term $u_n$ in the system, the controlled scheme is given by:
\begin{equation}
\begin{array}{l}
x_{n}=ax_{n-1}-bx_{n-2}-x_{n-1}^3+\xi_{n}+u_n,
\end{array}
\end{equation}
with $|\xi_n|\leq\xi_0$ and $|u_n|\leq u_0$.

In order to avoid the oscillation of the trajectories, we have defined the  initial region $Q_0$ (Fig.~\ref{28z}(a)) as the interval ($-2<x_{n-1}<0$). Notice that, in this way all successive $x_n$ values remain in this interval.  The safe set (Fig.~\ref{28z}(b)) was computed  with the values $\xi_0=0.020$ and  $u_0=0.015$. The safe set obtained is used to keep the trajectories in the interval ($-2<x_n<0$), avoiding the oscillation present in absence of control. In Fig.~\ref{28z}(b) the safe set and a partially controlled trajectory (red dots) are drawn. In Fig.~\ref{28z}(c) it is represented the corresponding controlled time series, where we also show an uncontrolled trajectory in order to compare.

  \subsection{The 3-dimensional hyperchaotic Hénon map}

 In this example we explore the possibility of controlling an hyperchaotic system which involves two or more positive Lyapunov exponents. To do that we have taken the three-dimensional Hénon map \cite{HyperHenon}.

\renewcommand{\thesubfigure}{(\alph{subfigure})~}

 \begin{figure}
\centering
\subfigure[Safe set]{\includegraphics[trim=0cm 0cm 0cm 0cm, clip=true,width=0.45\textwidth]{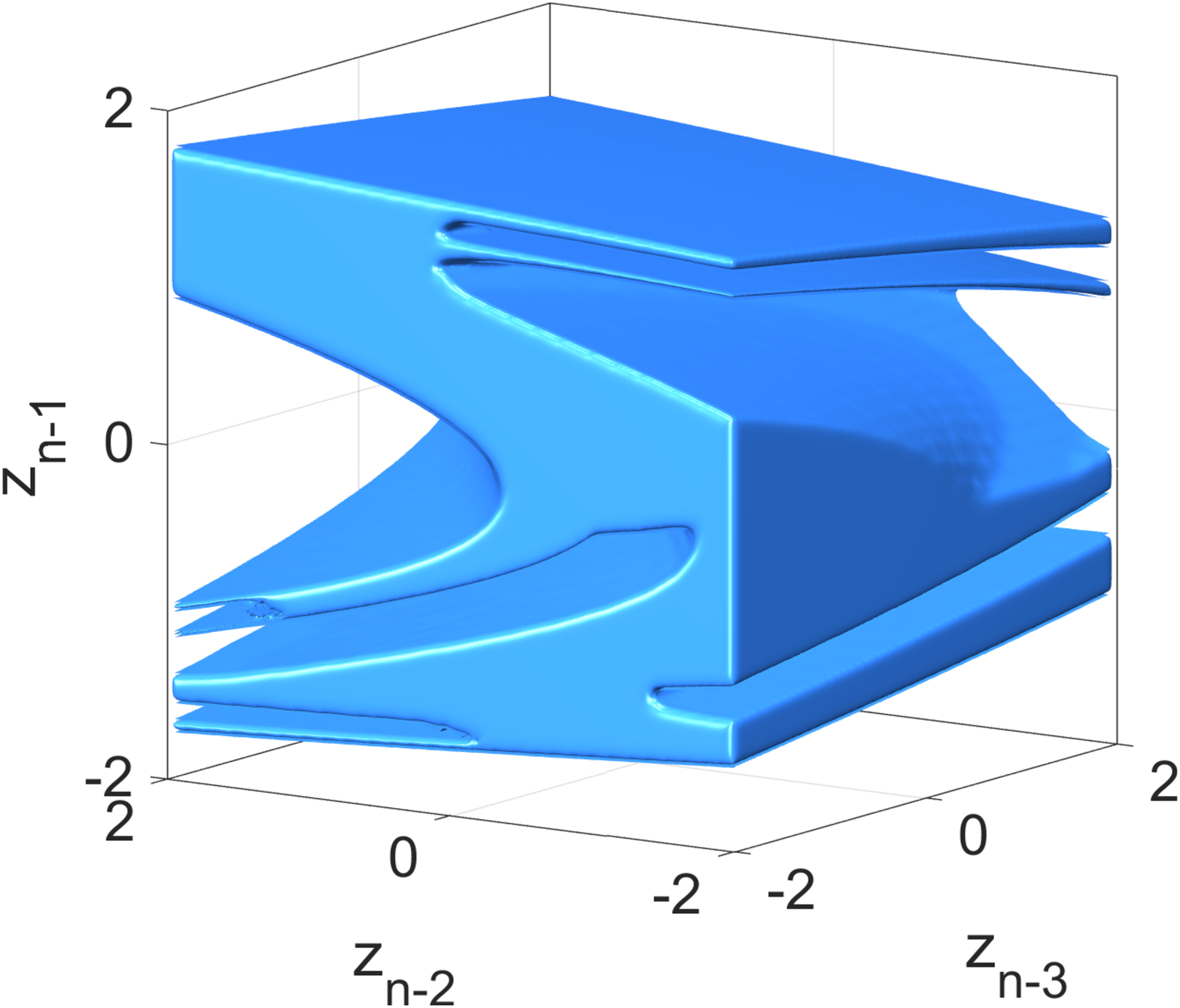}}
\subfigure[Controlled trajectory by using the safe set]{\includegraphics[trim=-1cm 0cm 0cm 0cm, clip=true,width=0.45\textwidth]{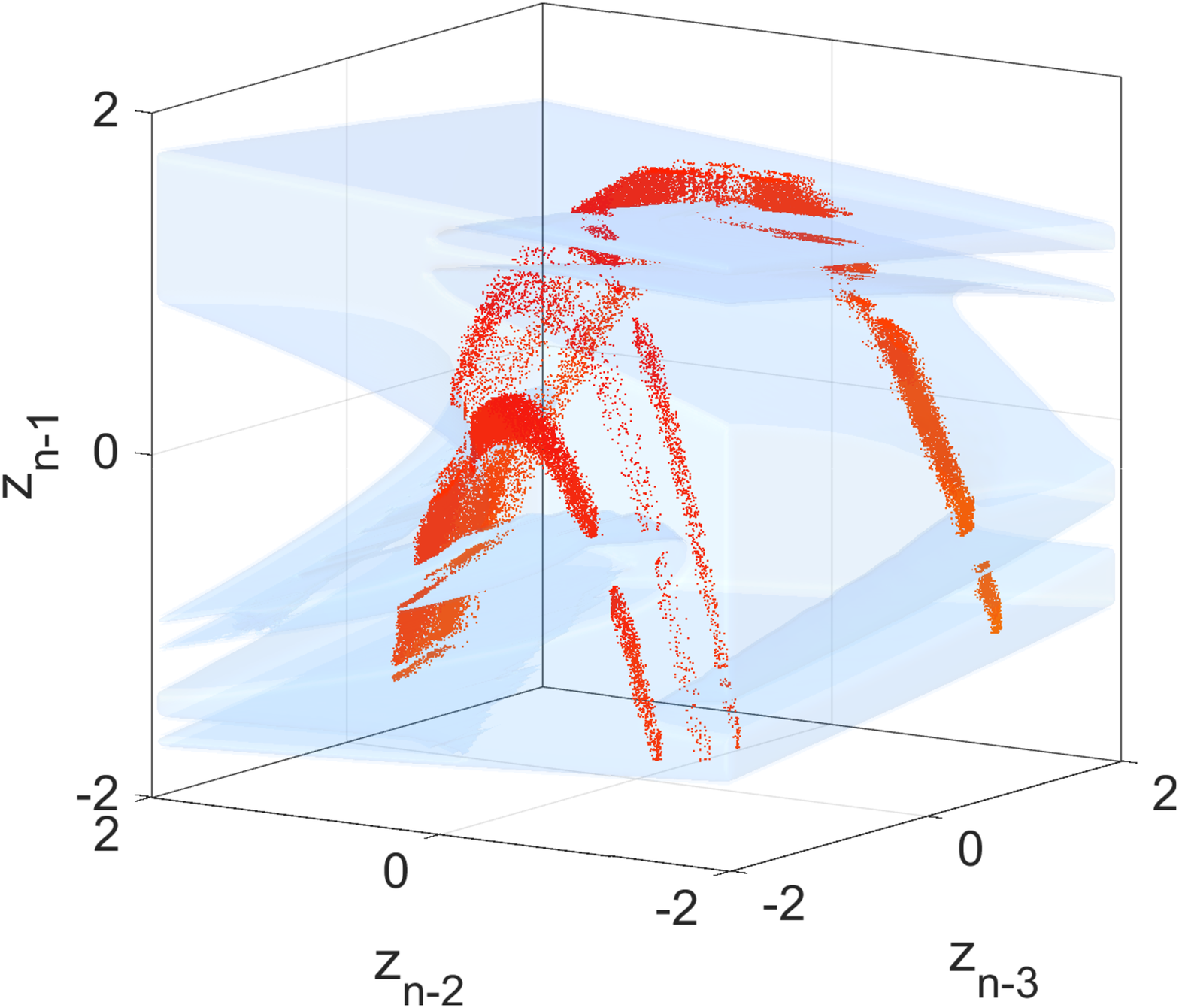}}
\subfigure[Time series of the controlled trajectory]{\includegraphics[trim=-1cm 0cm 0cm -4cm, clip=true,width=0.8\textwidth]{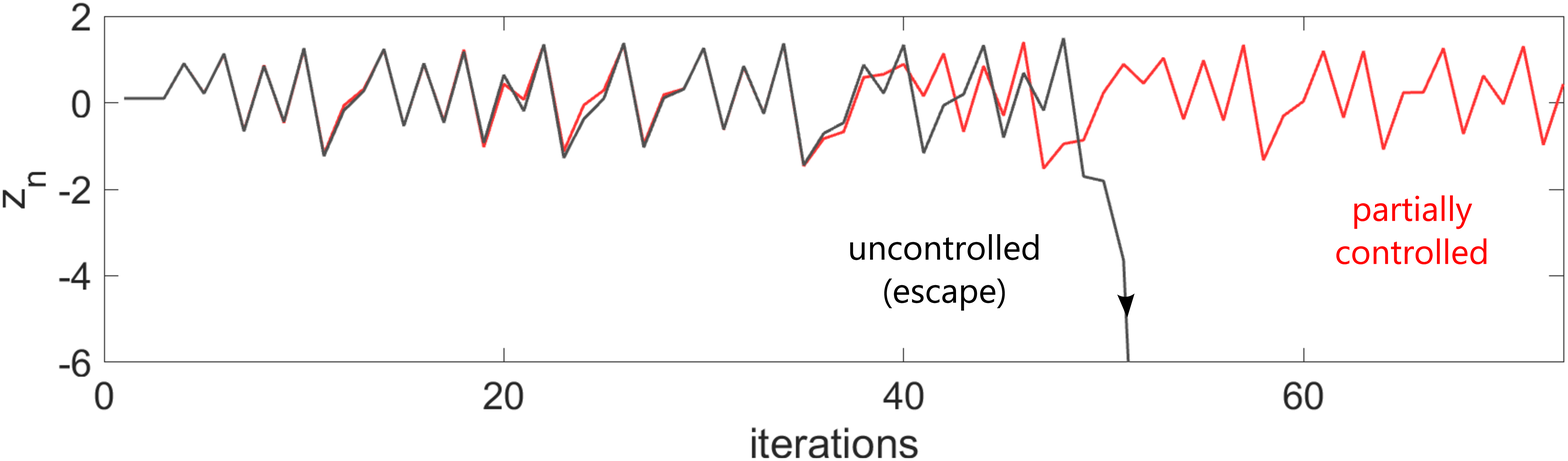}}
\\[20pt]
\caption{\textbf{Safe set and controlled dynamics for the 3D delayed Hénon map ($\mathbf{z_{n}= 1-az_{n-1}^2+bz_{n-2}-cbz_{n-3}}$)}. (a) The safe set computed for the parameter values $a=1.1$,  $b=0.3$, $c=1$. A grid of $1000\times1000\times1000$ was taken in the box $[-2,2] \times [-2,2]\times [-2,2] $ that represents the initial region $Q_0$. Taking the upper bound of the disturbance $\xi_0=0.12$ and  the control $u_0=0.08$, the safe set converges after 15 iterations. (b) The safe set is represented in transparent blue to see the controlled trajectory inside (red dots).  The  variable $z_n$ is affected by a random disturbance with upper bound $\xi_0=0.12$ and  control $u_0=0.08$. (c) Comparison between an uncontrolled trajectory and a controlled one in the 3D delayed Hénon. In black, the uncontrolled trajectory which after some iterations escapes to $-\infty$. In red, the controlled trajectory. For a fair comparison, both trajectories start with the same initial condition and are affected by the same sequence of random disturbances.}
\label{29z}
\end{figure}

  This system  is given by:
\begin{equation}
\begin{array}{l}
x_{n}=bz_{n-1} \\
y_{n}=cx_{n-1}+bz_{n-1}\\
z_{n}=1+y_{n-1}-az_{n-1}^2.
\end{array}
\end{equation}

This map shows transient chaos for a wide range of the parameters $a$, $b$ and $c$. To compute an example, we have chosen the parameter values $a=1.1$,  $b=0.3$ and $c=1$. For these values, the trajectories with initial conditions in the box $(x_n,y_n,z_n)\in[-0.5,0.5] \times [-1,1] \times [-2,2]$ have a chaotic transient, before eventually escaping from this region towards infinity.  In this case, the effect of the disturbances in the dynamics does not change dramatically the behaviour of the trajectories. It just increases or reduces the escape time in comparison with the deterministic trajectory.

Suppose now that we have collected data from the variable $z_n$ so that we were able to reconstruct a delay-coordinate map. In this case, taking three delays is sufficient to describe correctly the dynamics of the system, that is, $z_{n}=f(z_{n-1},z_{n-2},z_{n-3})$.

The form of this delay-coordinate map can be obtained by simple calculation:

\begin{equation}
\begin{array}{l}

z_{n}= 1-az_{n-1}^2+bz_{n-2}-cbz_{n-3}.

\end{array}
\end{equation}

From now on we will call this map the \emph{three-dimensional delayed Hénon map}.

In these coordinates, values of $|z_n|>2$ involve the escape to $-\infty$  of the trajectories. In order to avoid the escape, the goal is to apply control in the variable $z_n$ to keep it in the box $(z_{n-1},z_{n-2},z_{n-3}) \in [-2,2] \times [-2,2]\times [-2,2]$.

Introducing the disturbance term $\xi_n$ and the control term $u_n$, the partial control scheme is,
\begin{equation}
\begin{array}{l}

z_{n}= 1-az_{n-1}^2+bz_{n-2}-cbz_{n-3}+\xi_{n}+u_n,

\end{array}
\end{equation}
with $|\xi_n|\leq\xi_0$ and $|u_n|\leq u_0$. In order to show how the safe set changes depending on the disturbance value, we have computed the safe set taking $\xi_0=0.12$ and $u_0=0.08$. We have used a grid of $1000\times1000\times1000$ points covering $Q_0$, and then applied the modified Sculpting Algorithm to the safe set shown in Fig.~\ref{29z}(a). We have also represented in Fig.~\ref{29z}(b), 10000 iterations of a partially controlled trajectory  (red dots). Notice that the trajectory remains in the box $[-2,2] \times [-2,2]\times [-2,2]$ forever. In absence of control, the trajectory abandons this box after some iterations as it is illustrated in the time series represented in Fig.~\ref{29z}(c).

Although the variable $z_n$ was taken here as an example, in the case that the reconstructed delayed map was built with other variable $x_n$ or $y_n$, the methodology would be the same as the one presented here. The only difference would be the shape of the safe set obtained and possibly the minimum ratio $u_0/\xi_0$ achieved, since this depends on the embedded variable.

\section{A different application of partial control}

In contrast with the previous sections where the partial control method is used to keep trajectories close to the chaotic saddle and avoid an undesirable escape, the aim of the scheme proposed here, is to maintain the chaotic transient as much as we desire, before forcing an immediate escape. To do that, we use the same safe sets defined in the partial control method in a completely different way. By only using this set, we show how possible is to handle the stabilization and destabilization of the chaotic dynamics of the partially controlled system.

\begin{figure}
\fboxsep=0mm
\includegraphics [width=0.98\textwidth]{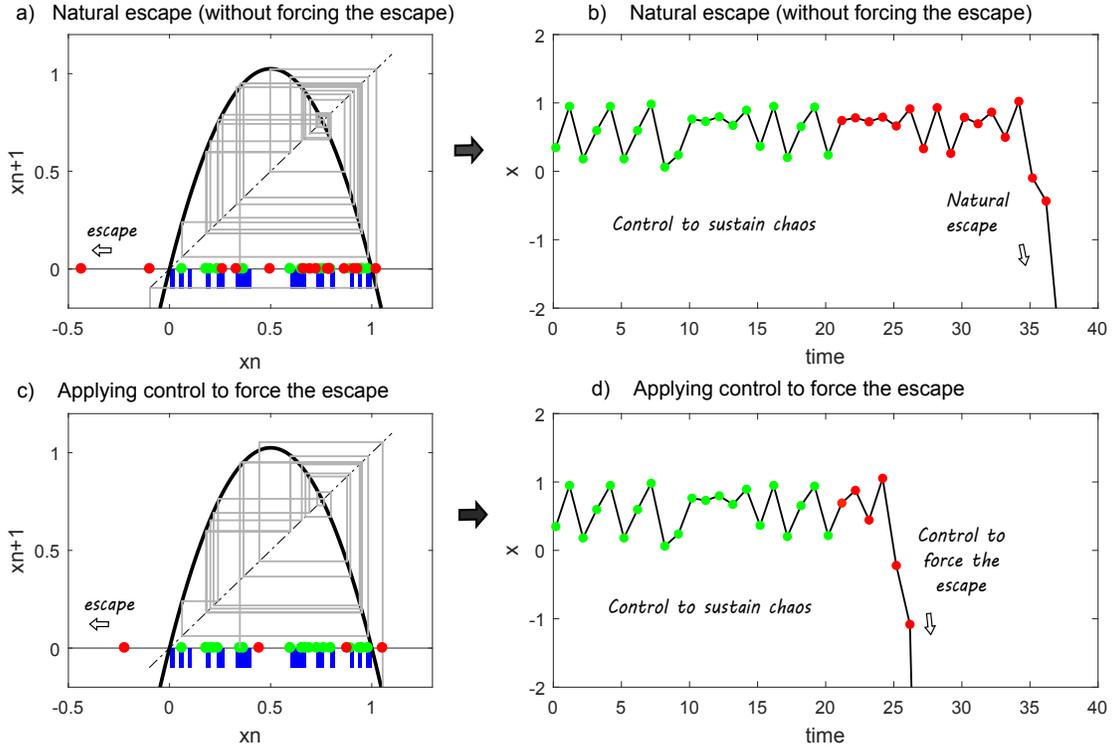}\\
\centering
\caption{\textbf{Controlled dynamics in the logistic map.} The black line in figures (a) and (c), represents the logistic map for the parameter $r=4.1$. For this value, transient chaos appears and orbits starting in the interval $[0,1]$ eventually escape to $-\infty$. In order to apply the control, the safe set was computed for the value of disturbance $\xi_0=0.03$ and control value $u_0=0.02$. The safe set is showed with thick blue bars to improve the visualization. In the first $20$ iterations (green points) the control is applied to return the orbit to the safe set. After that, in figure (a) the orbit is free to escape (no control is applied). However in figure (c) the orbit is forced to escape (red points). In figures (b) and (d) the corresponding time series are displayed. Notice that, by inducing the escape, the time to abandon the interval $[0,1]$ is greatly reduced.}
\label{30z}
\end{figure}

It is reasonable to think that if we want to force the escape of the trajectories, the simplest strategy is just stop applying the control. However, in many cases the average time between the moment in which the application of the control is stopped and the moment in which the trajectory reaches the escape may be very long. It is here where we found that the safe set can be used in another different way to speed up the escape time of the trajectory and therefore to get a higher control in the behaviour of the system. We show here that an optimal way to achieve this goal is to apply the control to drive the trajectories outside the safe set. The simplest strategy is to apply the control $|u_n|\leq u_0$ each iteration to the most far away point in the map of the safe set. As we will show in the examples, this strategy reduces significantly the average escape time, in comparison with the average escape time if no perturbations are applied.

For simplicity we use the well-known logistic map defined as follows,
\begin{equation}
\begin{array}{l}
 x_{n+1}=r x_n(1-x_n),
\end{array}
\end{equation}
where $x \in [0,1]$ and $r \in [0,4]$ to keep orbits in the interval $[0,1]$. Transient chaos appears for parameter values $r>4$. In order to compute an example, we have fixed $r=4.1$. For this value the orbits starting in the interval $[0,1]$ typically abandon the interval after a long transient.  We have also considered that these orbits are affected by disturbances with a bound $\xi_0$. The effect of this disturbance can be both, to accelerate or to slow down the escape time depending on the particular contribution of the random disturbances in each iteration of the map. To keep the chaotic trajectories in the interval $x=[0,1]$, we consider to apply the control $u_n$ bounded by $u_0$. In this way, the dynamics of the partially controlled map is

\begin{equation}
\begin{array}{l}
  x_{n+1}=r x_n(1-x_n)+\xi_n+u_n \\
 |\xi_n| \leq \xi_0, \hspace{1cm} |u_n| \leq u_0.  \\
\end{array}
\end{equation}

\begin{figure}
\fboxsep=0mm
\includegraphics [width=0.98\textwidth]{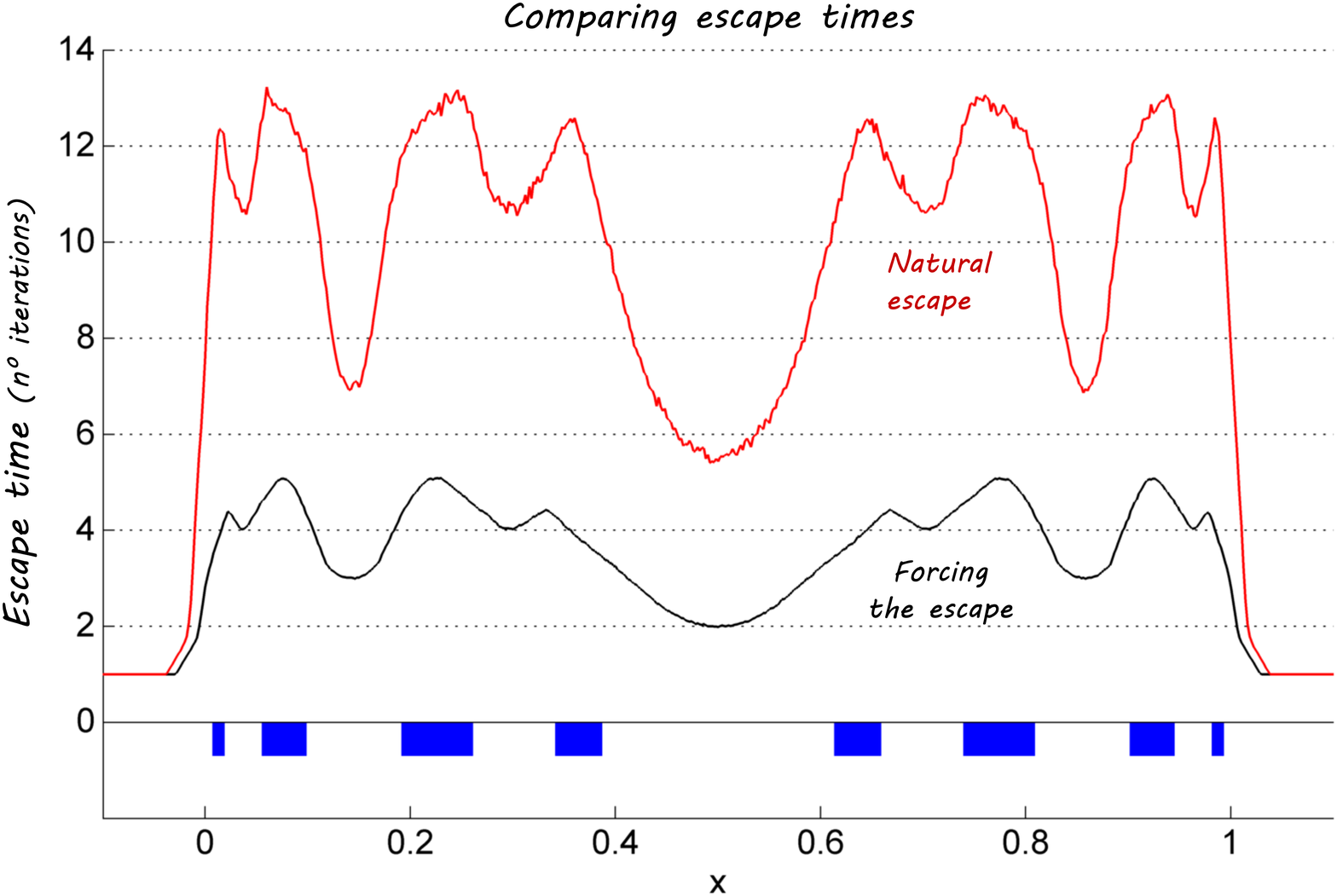}\\
\centering
\caption{\textbf{Average escape times.} The figure represents the interval $[0,1]$ and the safe set (in blue) for the same conditions as the previous figure. The upper red line shows the average escape time when the orbit abandons the interval $[0,1]$ without the application of any perturbation. The lower black line shows the average escape time when the orbits are forced to escape by applying small controls. In this way, the trajectory escapes about $2.5$ times faster than without control.}
\label{31z}
\end{figure}

For this example we have chosen the values $\xi_0=0.03$ and $u_0=0.02$. Then we compute the safe set showed in Fig.~\ref{30z}.

Imagine now that for instance, we want to keep the dynamics in the interval $[0,1]$ during $20$ iterations and then induce the escape as fast as possible. We have considered two scenarios. In Fig.~\ref{30z}(a) and \ref{30z}(b) we show the evolution of the variable $x$, where the control is applied in the first $20$ iterations to return orbits to the safe set. After that, we stop applying the control and the trajectory eventually escapes after a long time. In Fig.~\ref{30z}(c) and \ref{30z}(d) we represent the same situation with the difference that, after the first $20$ iterations, the control keeps applied with the goal of forcing its escape. As we can see in Fig.~\ref{31z}, the average escape time is much smaller when the control is applied. In addition the standard deviation of the escape time associated to the forced orbits is much smaller, which ensures that most orbits will escape very soon.

\section{Conclusions}

In this work we have presented a general overview of the partial control method. This method is used to avoid undesirable escapes or crisis in chaotic systems with minimal intervention. This scheme is based on finding a safe set in the phase space and it is specially recommended when some noise or random disturbances are present in the system, since this set minimizes the undesirable effects of uncertainties in the system. In the real time application of the control, the controller only needs to know which is the state of the system and which is the safe set. If the state of the system is in the safe set no control is applied, whereas if the state of the system is not in the safe set, a small amount of control is needed to put the system inside the safe set again.

Using the respective safe sets in each case, we have shown that is possible to control the trajectories, using a small amount of control in comparison with the disturbances affecting the system. Another remarkable feature is that the partially controlled trajectories keep the chaotic behavior of the original system, since $u_0< \xi_0$ and therefore it is impossible for the controller to completely determine the oscillatory behavior.

To show how the method works we have applied it under different situations. First, we have considered that the control was applied on the dynamical variables of the system, both in maps like the Hénon map or flows like the Lorenz system or an ecological model. Second, we have considered the situation where the control is applied in some parameter of the system considering that the parameter is also affected by random disturbances. Some applications of the method were shown using the one-dimensional logistic map or the Duffing oscillator flow. Finally, we have studied the application of the method in delay-coordinate maps, being aware of the importance of these maps in experimental data analysis. The dynamics of these systems usually depends on the present state and some past state as well. This fact involves a big challenge since only the present state is controllable in practice.  With a suitable modification of the control scheme, we aim to provide a helpful tool to control this kind of systems. The examples presented here were the two-dimensional cubic map and the 3-dimensional hyperchaotic Hénon map. In the last section we have explored a very simple but powerful modification involving the use of the safe sets. We have shown that these sets can be used also to accelerate the escape time of trajectories when it is required, increasing our control over the behaviour of the system.

Finally, we want to point out that, although we consider here mathematical models to express the maps, we believe that the method can be applied in the same way to delay-coordinate maps built from experimental time series. That would be the next step in the development of this control method. In this sense, there is still much room to continue improving the control method and develop new approaches to deal with more general problems where chaos and noise are present.

\begin{acknowledgments}
This work was supported by the Spanish Ministry of Economy and Competitiveness
under Project No. FIS2013-40653-P and by the Spanish State Research Agency (AEI)
and the European Regional Development Fund (FEDER) under
Project No. FIS2016-76883-P. MAFS acknowledges the
jointly sponsored financial support by the Fulbright Program
and the Spanish Ministry of Education (Program No. FMECD-ST-2016).

\end{acknowledgments}


\end{document}